\documentclass[twocolumn,superscriptaddress,aps,showpacs,amsmath,amstex,amssymb,citeautoscript,floatfix,pra,preprintnumbers]{revtex4-2}
\pdfoutput=1
\usepackage[english]{babel}
\usepackage{letltxmacro}
\usepackage{latexsym}
\LetLtxMacro{\ORIGselectlanguage}{\selectlanguage}
\makeatletter
\DeclareRobustCommand{\selectlanguage}[1]{%
  \@ifundefined{alias@\string#1}
    {\ORIGselectlanguage{#1}}
    {\begingroup\edef\x{\endgroup
       \noexpand\ORIGselectlanguage{\@nameuse{alias@#1}}}\x}%
}
\newcommand{\definelanguagealias}[2]{%
  \@namedef{alias@#1}{#2}%
}
\makeatother

\makeatletter
\def\maketitle{
\@author@finish
\title@column\titleblock@produce
\suppressfloats[t]}
\makeatother

\definelanguagealias{en}{english}
\definelanguagealias{English}{english}
\usepackage{graphicx}
\usepackage{amsmath}
\usepackage{amsfonts}
\usepackage{amssymb}
\usepackage{amsthm}
\usepackage{cancel}
\usepackage{mathtools}
\usepackage{bm}
\usepackage{color}
\usepackage[percent]{overpic}
\usepackage{soul} 
\usepackage{amssymb}
\usepackage{wasysym}
\usepackage{dsfont}
\usepackage{float}
\usepackage{physics}
\usepackage{hyperref}
\usepackage{comment}
\usepackage{enumitem}
\usepackage{booktabs}

\hypersetup{
  colorlinks   = true, 
  urlcolor     = blue, 
  linkcolor    = blue, 
  citecolor   = red 
}
\usepackage[mathscr]{euscript}

\newtheorem*{theorem*}{Theorem}
\newtheorem*{lemma*}{Lemma}

\usepackage{graphicx}
\usepackage{verbatim}
\usepackage[normalem]{ulem}

\newcommand{\eqs}[1]{\begin{equation}\begin{split}#1\end{split}\end{equation}}
\newcommand{\eqnref}[1]{Eq.\,\eqref{#1}}

\newcommand{\cd}[1]{c^{\dagger}_{#1}}
\newcommand{\cc}[1]{c_{#1}}

\newcommand{\MIT}{Center for Theoretical Physics, Massachusetts Institute of Technology, Cambridge, MA 02139, USA}
\newcommand{\Harvard}{Department of Physics, Harvard University, Cambridge, MA 02138, USA}
\newcommand{\ITAMP}{ITAMP, Harvard-Smithsonian Center for Astrophysics, Cambridge, MA 02138, USA}

\begin{document}
\title{
Efficiently measuring $d$-wave pairing and beyond in quantum gas microscopes
}
\preprint{MIT-CTP/5812}

\author{Daniel K. Mark}
\altaffiliation{These authors contributed equally to this work.}
\affiliation{\MIT}
\author{Hong-Ye Hu}
\altaffiliation{These authors contributed equally to this work.}
\affiliation{\Harvard}

\author{Joyce Kwan}
\affiliation{\Harvard}

\author{Christian Kokail}
\affiliation{\Harvard}
\affiliation{\ITAMP}

\author{Soonwon Choi}
\affiliation{\MIT}

\author{Susanne F. Yelin}
\altaffiliation{Corresponding author: \href{mailto:syelin@g.harvard.edu}{syelin@g.harvard.edu}}
\affiliation{\Harvard}

\begin{abstract}
Understanding the mechanism of high-temperature superconductivity is among the most important problems in physics, for which quantum simulation can provide new insights.
However, it remains challenging to characterize superconductivity in existing cold-atom quantum simulation platforms. Here, we introduce a protocol for measuring a broad class of observables in fermionic quantum gas microscopes, including long-range superconducting pairing correlations (after a repulsive-to-attractive mapping). The protocol only requires global controls followed by site-resolved particle number measurements ---capabilities that have been already demonstrated in multiple experiments--- and is designed by analyzing the Hilbert-space structure of dimers of two sites. The protocol is sample efficient and we further optimize our pulses for robustness to experimental imperfections such as lattice inhomogeneity. Our work introduces a general tool for manipulating quantum states on optical lattices, enhancing their ability to tackle problems such as that of high-temperature superconductivity.
\end{abstract}
\maketitle

Quantum technologies have seen remarkable recent developments in the control of large quantum systems for tasks ranging from quantum error correction~\cite{bluvstein2024logical,da2024demonstration,acharya2024quantum}, investigating dynamical phenomena~\cite{bernien2017probing,neyenhuis2017observation, mi2021time}, to preparing and manipulating exotic quantum many-body states~\cite{ebadi2021quantum,google2023non,iqbal2024non}. 
One promising application is the simulation of complex quantum phenomena~\cite{altman2021quantum}, such as unconventional superconductivity in interacting itinerant fermions. The Fermi-Hubbard model is a simple candidate model 
for the physics of high-temperature superconductivity~\cite{emery1987theory,lee2006doping,keimer2015quantum,bednorz1986possible,tsuei2000pairing}, but is notoriously difficult to study either theoretically and computationally, and many open questions remain~\cite{lee2006doping,keimer2015quantum,arovas2022hubbard}. Indeed, only within the past five years
have state-of-the-art numerical studies started producing predictions on the conditions under which superconductivity may arise~\cite{qin2020absence,jiang2019superconductivity,xu2024coexistence}. High-quality quantum simulations offer the ability to verify such predictions or provide new insight into uncharted parameter regimes, and therefore are highly promising avenues for resolving open problems~\cite{lee2006doping,keimer2015quantum}.

\begin{figure}[tb!]
    \centering
    \includegraphics[width=1\linewidth]{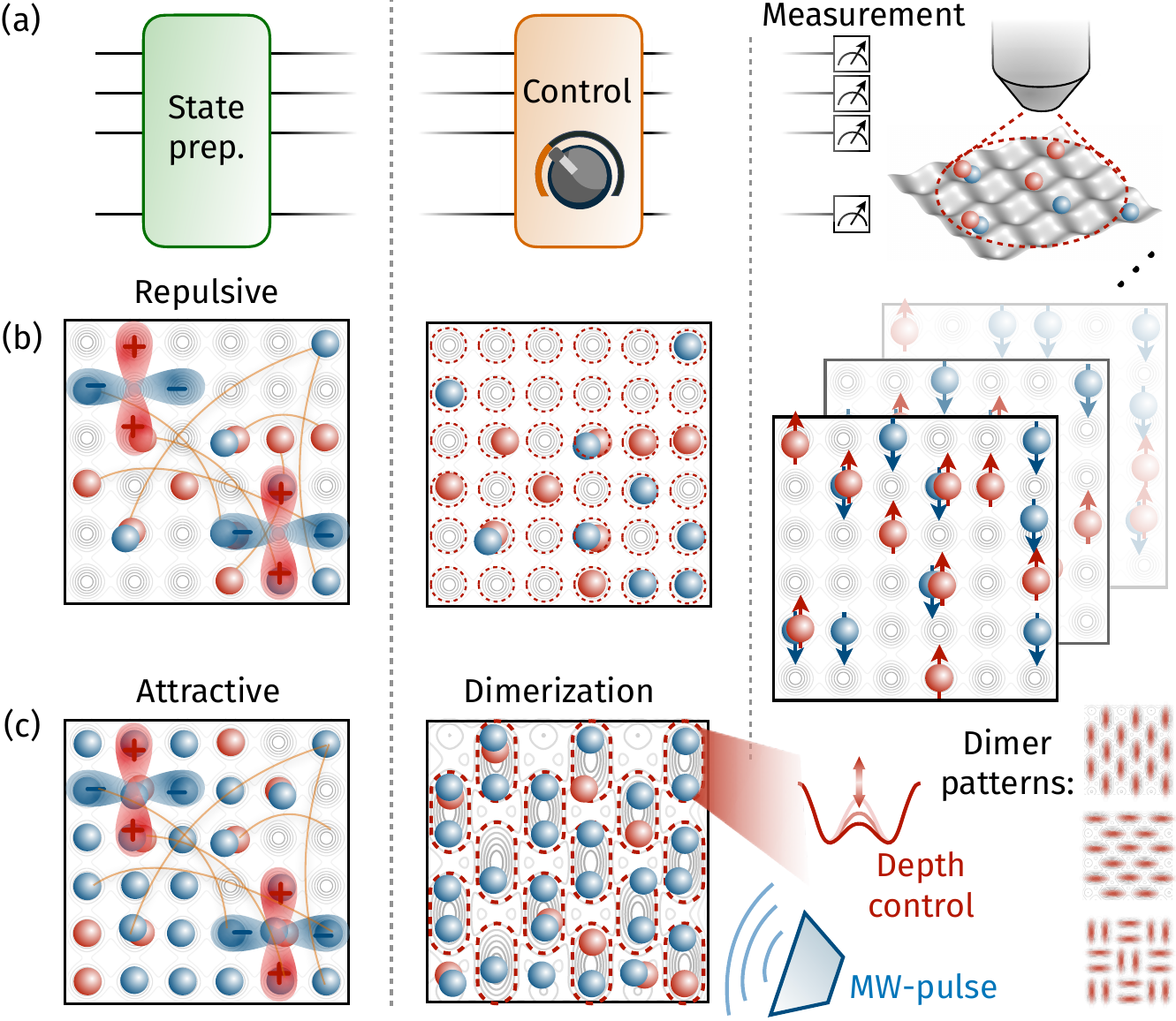}
    \caption{\textbf{Overview of protocol.} (a) By applying control pulses after state preparation, a broad class of observables can be measured.
    (b) Conventional approaches involve site-resolved particle number measurements on low-energy states of the repulsive Fermi-Hubbard model. These give charge- and spin-correlations but offer limited information for other quantities. (c) We propose preparing low-energy states of the \textit{attractive} Hubbard model 
    and applying global control pulses to efficiently measure observables such as $d$-wave pairing correlations. This only requires the ability to use optical superlattices to dimerize the lattice into isolated double wells, followed by globally tuning lattice depth and applying a global microwave (MW) pulse. Several dimer patterns are compatible with our scheme: parallel dimers reveal long-range pairing while alternating vertical and horizontal dimers also reveal $d$-wave angular dependence. 
    }
    \label{fig:overview}
\end{figure}

Quantum gas microscopes of ultracold fermionic particles in an optical lattice offer an approach to study the Fermi-Hubbard and related models~\cite{esslinger2010fermi,gross2017quantum,tarruell2019quantum,bohrdt2021exploration} [Fig.~\ref{fig:overview}(b)]. These experiments offer unprecedented microscopic control and detection of these strongly-correlated quantum systems~\cite{gross2021quantum}, and major experimental efforts are underway to explore the doped Fermi-Hubbard phase diagram and address outstanding questions regarding $d$-wave superconductivity and quantum magnetism~\cite{esslinger2010fermi,bohrdt2021exploration}.
Such experiments have observed precursors to superconductivity~\cite{hart2015observation,mazurenko2017cold,chiu2019string,brown2019bad,ji2021coupling,shao2024antiferromagnetic} and made consistent progress towards achieving cold effective temperatures where superconductivity is expected~\cite{onofrio2016physics,mazurenko2017cold,spar2022realization}. 

However, challenges remain, including the natural question of how to verify the presence of superconductivity in a quantum gas microscope.  One possibility is to indirectly measure superconductivity via transport measurements, similar to traditional solid-state experiments. However, this approach involves additional overhead, since such transport measurements themselves need to be simulated with cold atom techniques, such as with two-terminal~\cite{chien2015transport,krinner2017transport} experiments.

In this work, we propose a method to directly measure a broad class of observables, including $d$-wave pairing correlators.
Our method requires minimal experimental requirements (all of which have been demonstrated) and is applicable in all parameter regimes, while existing proposals hold in certain limits, including where interactions are dominant (the so-called ``$t-J$ limit")~\cite{schlomer2024localcontrolmixeddimensions} or effectively zero~\cite{kessler2014single,impertro2023localreadoutcontrolcurrent,impertro2024realizationstronglyinteractingmeissnerphases}.

\begin{figure*}[ptb!]
    \centering
    \includegraphics[width=\linewidth]{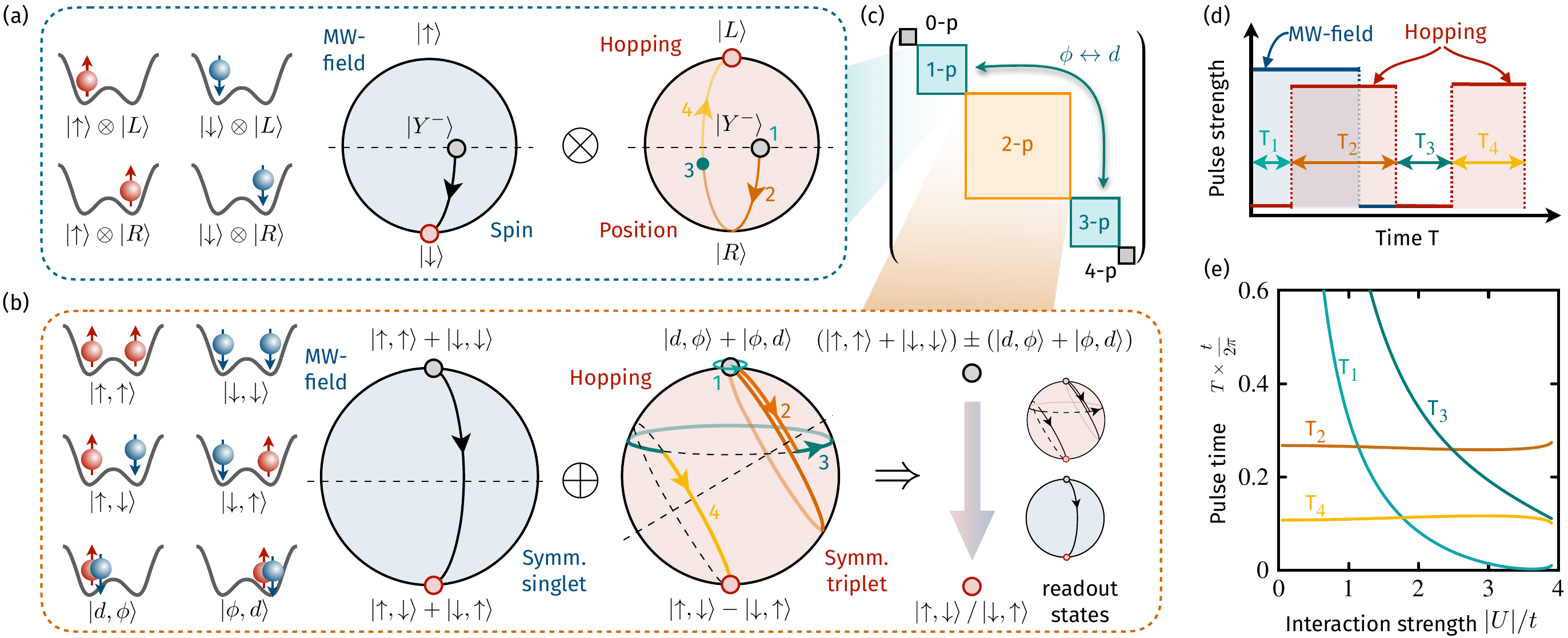}
    \caption{\textbf{Details of protocol.} 
    The protocol leverages the Hilbert space structure of spin-1/2 fermions on two sites.
    (a) The 1-particle (1-p) sector consists of the tensor product of spin and positional two-level systems. Microwave (MW) and hopping control perform $X$-axis rotations on each Bloch sphere. These are our desired mappings sending $\ket{\uparrow} \pm i \ket{\downarrow}$ and $\ket{L} \pm i \ket{R}$ into readout basis states; we require a $3\pi/2$ over-rotation on the positional Bloch sphere (right). (b) The 2-p sector is spanned by six states which we organize by spin quantum number and inversion symmetry. A MW pulse effects an $X$-axis rotation on the two-level system spanned by the inversion symmetric, spin-triplet states. Meanwhile, hopping and interactions rotate the two-level system of inversion symmetric, spin-singlet states about a tilted axis (right). MW and hopping pulses achieve the desired north-to-south-pole mappings.
    (c) To measure long-range correlations, we seek solutions that are simultaneously correct on all particle-number sectors, transforming all eigenstates of the pairing observable $\tilde{O}_{i,i'}$ into readout states. Since the 0-/4-p sectors are one-dimensional and the 1-/3-p sectors are isomorphic, the problem reduces to finding a pulse sequence that is correct on both 1-p and 2-p sectors. (d) Our solution consists of a MW pulse (blue) in parallel with alternating hopping and idling steps (red). (e) We plot the duration of each pulse step across a range of interaction strengths $|U|/t$; this simplest \textit{four-pulse} sequence is valid in the range $|U|/t \in [0,3.9071]$ (see SM~\cite{SM}).}
    \label{fig:details}
\end{figure*}
We use two key concepts: (i) mapping from the repulsive to \emph{attractive} Hubbard model~\cite{ho2009quantum,mitra2018quantum,chan2020pair,gall2020simulating,hartke2020doublon,hartke2023direct} and (ii) efficient control of the quantum state of fermions on two lattice sites to map previously inaccessible quantities into easily measured ones. We only require the ability to perform the following global operations: partition the lattice into ``dimers" of two sites each~\cite{atala2013direct,kessler2014single,impertro2023localreadoutcontrolcurrent,impertro2024realizationstronglyinteractingmeissnerphases}, tune lattice depth after dimerization, perform spin-rotations, and finally make spin-and charge-resolved measurements of each site~\cite{bakr2009quantum,boll2016spin,koepsell2020robust} [Fig.~\ref{fig:overview}(c)]. The protocol is sample-efficient and we use optimal control techniques to optimize its robustness to experimental imperfections. 

While $d$-wave pairing is our motivating example, with only five types of global control, the protocol generalizes to any two-site observable that is particle-number preserving. We prove this universality, construct pulse sequences for observables of interest, such as energy density, and discuss their potential applications.

\textit{Attractive Hubbard model and pairing correlator.---} 
We use the relationship between the attractive and repulsive Hubbard model to map the pairing correlations into a measurable quantity. In the repulsive model, the superconducting order parameter is of the form $\Delta^\dagger_{i,i'} \Delta_{j,j'} \equiv c^\dagger_{i,\uparrow} c^\dagger_{i’,\downarrow} c_{j,\uparrow}^{} c_{j’,\downarrow}^{}$: one creates a Cooper pair at sites $i$ and $i’$, removes a Cooper pair at potentially faraway sites $j$ and $j’$, and verifies that the system has remained globally phase coherent. Measuring these correlations requires the coherent transport of particles over large distances, a challenging task. Working with the attractive Hubbard model circumvents this issue. The attractive Hubbard model has negative Hubbard interaction $U$ and is related to its repulsive counterpart by a particle-hole transformation on the spin-down particles~\cite{ho2009quantum,SM}. Crucially, this transforms the above order parameter into $\tilde{\Delta}^\dagger_{i,i'} \tilde{\Delta}_{j,j'} \equiv c^\dagger_{i,\uparrow} c_{i’,\downarrow}^{} c^\dagger_{j,\downarrow} c_{j’,\uparrow}^{}$. Unlike its repulsive version, this quantity conserves particle number separately on $(i,i')$ and $(j,j')$; it can be measured through correlations of local spin quantities.
Measuring $\tilde{\Delta}_{i,i'}^\dagger\tilde{\Delta}_{j,j'}^{}$ on a low-energy state of the attractive Hubbard model is equivalent to measuring $\Delta_{i,i'}^\dagger\Delta_{j,j'}^{}$ on the corresponding low-energy state of the repulsive Hubbard model. Recent work has utilized this property of the repulsive-to-attractive mapping~\cite{schlomer2024localcontrolmixeddimensions}. Here, we develop a simple protocol to measure such spin correlations with minimal experimental requirements, that applies in any parameter regime. 

\textit{Measurement protocol.---}
Our strategy is to transform the desired order parameter into a readily-measured quantity, diagonal in the particle-number basis, via global control. After dimerizing the lattice, we apply a global unitary $V = \mathcal{T} \exp\left[-\int iH(T)dT\right]$ generated by the time-dependent Hamiltonian
\begin{align}
    H(T) = t(T) H_\text{Hop} + B_X(T) H_\text{MW} + U H_\text{Int},
\end{align}
where $H_\text{Hop}=-\sum_{\langle i, i'\rangle, \sigma} (c^\dagger_{i,\sigma}c^{}_{i',\sigma} + \text{h.c.})$ is the hopping term, restricted to hopping within dimers, $H_\text{MW}=\sum_{i} (c^\dagger_{i,\uparrow}c^{}_{i,\downarrow} + \text{h.c.})$  is the microwave (MW) spin-flip term, whose time-dependent strengths $t(T)$ and $B_X(T)$ are controllable, and $H_\text{Int}$ is the Hubbard interaction term which we assume has fixed strength $U$.  We design a pulse sequence $t(T)$ and $B_X(T)$ such that the resultant unitary $V$ transforms the desired order parameter into a directly measurable observable. That is, the rotated observable 
\begin{equation}
V \tilde{O}_{i,i'}  V^\dagger = \text{diag}(\{\lambda_a\}) \label{eq:mapping_to_diagonal_obs}
\end{equation}
is diagonal in the measurement basis, where $\tilde{O}_{i,i'}$ is a spin- and Hermitian-symmetrized variant of $\tilde{\Delta}_{i,i'}$ (with eigenvalues $\lambda_a$): $\tilde{O}_{i,i'} \equiv c^\dagger_{i,\uparrow} c_{i’,\downarrow}^{} - c^\dagger_{i',\uparrow} c_{i,\downarrow}^{} + \text{h.c.}$
The measured correlations between $V \tilde{O}_{i,i'}  V^\dagger$ and $V \tilde{O}_{j,j'}  V^\dagger$ give the correlators  $\langle \tilde{O}_{i,i'} \tilde{O}_{j,j'} \rangle$, in turn equal to the desired pairing correlations $\langle \tilde{\Delta}_{i,i'}^\dagger \tilde{\Delta}_{j,j'}^{} \rangle$ up to a factor of 8, see Supplemental Material (SM)~\cite{SM}.

Only a restricted set of controls are available, which constrains the possible unitary transformations. In particular, coherent local control of a quantum gas microscope must preserve particle number, and the resulting unitary cannot couple between particle number sectors of the dimer Hilbert space. In order to measure the desired long-range correlations, one must achieve the transformation Eq.~\eqref{eq:mapping_to_diagonal_obs} on each sector separately and simultaneously.

The number sectors are remarkably simple. The 16-dimensional Hilbert space of spin-1/2 fermions on a dimer separates into sectors of 0 to 4 particles, with dimensions 1,4,6,4 and 1 respectively. The 0- and 4-particle sectors are trivial, and the 1- and 3-particle sectors are related by exchanging empty and doubly-occupied sites: a solution on the 1-particle sector is automatically a solution on the 3-particle sector. Therefore, the task simplifies to designing one that simultaneously performs desired transformations on the 1- and 2-particle sectors  [Fig.~\ref{fig:details}(a-c)].

The Hamiltonians $H(T)$ in the 1- and 2-particle sectors are:
\begin{align}
    &H^{\text{(1-p)}} = \begin{pmatrix}
        0 & B_X & -t & 0\\
        B_X & 0 & 0 & -t \\
        -t & 0  & 0 & B_X\\
        0 & -t & B_X & 0
    \end{pmatrix}
    \begin{matrix}
        {\scriptstyle \ket{\uparrow, \phi}}\\
        {\scriptstyle \ket{\downarrow, \phi}}\\
        {\scriptstyle \ket{\phi, \uparrow}}\\
        {\scriptstyle \ket{\phi, \downarrow}}
    \end{matrix}~,
    \label{eq:Ham_1p}\\
    &H^{\text{(2-p)}} = 
    \begin{pmatrix}
        U & -2t & 0 &  &  & 
        \\
        -2t & 0 & 0 &   \multicolumn{3}{c}{\Large\bm{0}} \\
        0 & 0 & U &  &  & \\
         &  &  & 0 & 0 & 0\\
         & {\Large\bm{0}} &  & 0 & 0 & 2B_X\\
         &  &  & 0 & 2 B_X &0 
    \end{pmatrix} 
    \begin{matrix}
        {\scriptstyle \ket{d, \phi} + \ket{\phi, d}}\\
        {\scriptstyle \ket{\uparrow, \downarrow} - \ket{\downarrow, \uparrow}}\\
        {\scriptstyle \ket{d, \phi} - \ket{\phi, d}}\\
        {\scriptstyle \ket{\uparrow, \uparrow} - \ket{\downarrow, \downarrow}}\\
        {\scriptstyle \ket{\uparrow, \downarrow} + \ket{\downarrow, \uparrow}}\\
        {\scriptstyle \ket{\uparrow, \uparrow} + \ket{\downarrow, \downarrow}}
    \end{matrix}~,
    \label{eq:Ham_2p}
\end{align}
with $t$ and $B_X$ the strengths of the hopping and MW controls (suppressing their time-dependence), and $U$ the Hubbard interaction strength. We have denoted our basis choice on each sector; we use a non-standard basis on the 2-particle sector.
$\phi$ and $d$ denote empty and doubly-occupied sites, $\uparrow,\downarrow$ denote singly-occupied sites, and we have omitted state normalization factors.

These Hamiltonians are highly structured.
First, the 1- and 3-particle Hilbert spaces naturally factorize into positional ($|L/R\rangle$) and spin ($\ket{\uparrow\!/\!\downarrow}$) two-level systems, respectively manipulated by hopping and MW controls [Fig.~\ref{fig:details}(a)].
Meanwhile, the 2-particle Hamiltonian is block-diagonal on the spin-triplet $\{\ket{\uparrow, \uparrow}, \ket{\uparrow, \downarrow} + \ket{\downarrow, \uparrow}, \ket{\downarrow, \downarrow}\}$ and spin-singlet $\{\ket{d, \phi}, \ket{\uparrow, \downarrow} - \ket{\downarrow, \uparrow}, \ket{\phi, d}\}$ sectors. The triplet sector is spatially symmetric. Therefore, it is invariant under hopping control $t(T)$, but transforms under MW control $B_X(T)$. Conversely, the singlet sector transforms under $t(T)$ and is invariant under $B_X(T)$. Further symmetries mean that each control only couples two symmetric states within each spin multiplet [Eq.~\eqref{eq:Ham_2p}, Fig.~\ref{fig:details}(b)]. 

\begin{figure*}[tbp!]
    \centering
    \includegraphics[width=1\linewidth]{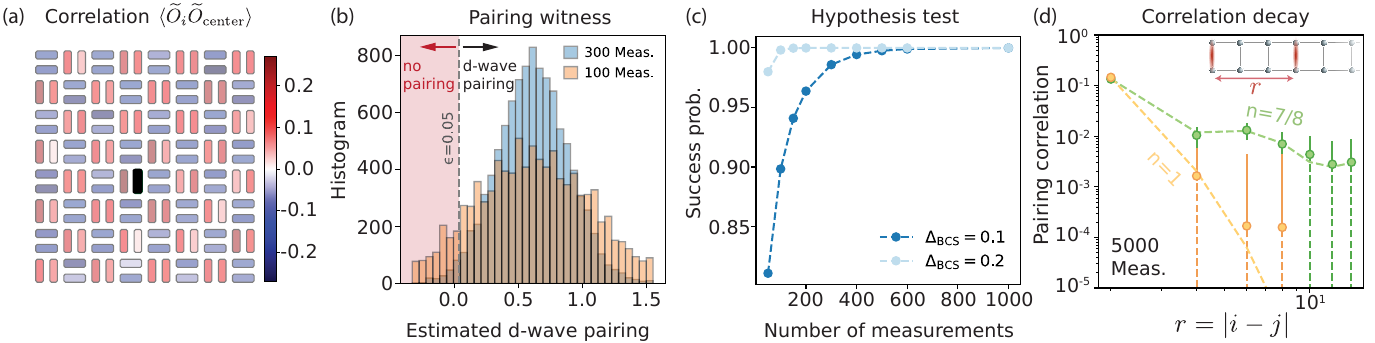}
    \vspace{-0.8cm}\caption{\textbf{Sample efficiency.} (a) Dimer-dimer $d-$wave pairing correlations of a BCS wavefunction. The central vertical dimer (shaded black) is fixed, and the color of each surrounding dimer indicates its estimated correlation with this central dimer. 
    Just 500 global measurements allow for the determination of pairing correlations with minimal statistical fluctuations (color variations). (b) Histogram of the estimated $d$-wave pairing strength.
    As the number of measurements increases, the distribution becomes more concentrated around its true value; with 300 measurements, we can reliably distinguish $d$-wave pairing from the lack thereof. (c) We perform a hypothesis test for $d$-wave pairing with a threshold of $\epsilon = 0.05$ [(b)]. We plot the correctness probability of our test as a function of the number of measurements: with 200 measurements, we misclassify with $\leq5\%$ probability. (d) We simulate the ground state of a two-leg Fermi-Hubbard ladder with  $U/t = 8$  at various electron fillings. At an average electron density of  $n = 7/8$  ($1/8$ hole doping), pairing correlations decay as a power law (green). In contrast, at  $n = 1$  (half-filling), the correlations decay exponentially (orange). 5000 total measurements allow correlations up to a distance of 8 to be accurately estimated. 
    }
    \label{fig:numerics}
\end{figure*}

Therefore, our problem can be understood as using hopping and MW control to simultaneously control two qubit degrees of freedom each, in close analogy with the design of the Levine-Pichler entangling gate in neutral-atom quantum computers~\cite{Levine2019Parallel}. Our strategy to achieve Eq.~\eqref{eq:mapping_to_diagonal_obs} is to design a unitary that maps the eigenstates of $\tilde{O}$ onto readout basis states. On the 1-particle sector, the eigenstates of $\tilde{O}$ are $\ket*{Y^\pm}_\text{sp}\otimes \ket*{Y^{\pm'}}_\text{pos}$: the product of Pauli-$Y$ eigenstates on the spin and positional Bloch spheres. Meanwhile, on the 2-particle sector the nontrivial eigenstates of $\tilde{O}$ are $(\ket{d,\phi} + \ket{\phi,d}) \pm (\ket{\uparrow, \uparrow}+\ket{\downarrow, \downarrow})$, supported entirely on states which are controlled by hopping and MW fields.

As depicted in Fig.~\ref{fig:details}(a,b), our solution performs the following transformations with MW control:
\begin{align}
    \ket{Y^\pm}_\text{sp} \equiv \ket{\uparrow} \pm i \ket{\downarrow} &\mapsto \ket{Z^{\pm}}_\text{sp} \equiv \ket{\uparrow}/\ket{\downarrow}, \label{eq:sp_map_1p}\\
    \ket{\uparrow,\uparrow}+\ket{\downarrow,\downarrow} &\mapsto \ket{\uparrow,\downarrow} + \ket{\downarrow, \uparrow}, \label{eq:sp_map_2p}
\end{align}
Meanwhile, we use hopping control to map
\begin{align}
    |Y^\pm\rangle_\text{pos} \equiv |L\rangle \mp i |R\rangle &\mapsto |Z^{\mp}\rangle_\text{pos} \equiv |R\rangle/|L\rangle, \label{eq:hop_map_1p}\\
    |d,\phi\rangle+|\phi,d\rangle &\mapsto \ket{\uparrow,\downarrow} - \ket{\downarrow, \uparrow}. \label{eq:hop_U_1_map_2p}
\end{align}
Together, Eq.~(\ref{eq:sp_map_1p}-\ref{eq:hop_U_1_map_2p}) map the eigenstates of $\tilde{O}$ onto the 1-particle readout basis and the 2-particle states $\ket{\uparrow,\downarrow}/\ket{\downarrow,\uparrow}$.  While hopping and MW control do not provide universal control on their respective two-level systems and hence would not be expected to suffice for general eigenstates, in this case they do give the desired transformations. MW control acts as $X$-axis rotations on its respective 1- and 2-particle Bloch spheres; from Eq.~(\ref{eq:Ham_1p},\ref{eq:Ham_2p}), a pulse of duration $(\pi/2)/(2B_X)$ achieves both maps Eq.~(\ref{eq:sp_map_1p},\ref{eq:sp_map_2p}). Likewise, hopping acts as an $X$-rotation on its 1-particle Bloch sphere. However, the presence of interactions $U$ means that hopping control is no longer a simple $X$-axis rotation on its 2-particle Bloch sphere [Eq.~\eqref{eq:Ham_2p}, Fig.~\ref{fig:details}(b)]. Yet, the combination of hopping with a fixed strength $t$ and idling with no hopping [$t(T)=0$] gives universal control on this particular Bloch sphere, allowing us to engineer the map Eq.~\eqref{eq:hop_U_1_map_2p}. As discussed in the End Matter (EM), we design a family of trajectories on the Bloch sphere with a tunable parameter, which enables us to find a pulse sequence that performs both maps Eq.~(\ref{eq:hop_map_1p},\ref{eq:hop_U_1_map_2p}) simultaneously. Our simplest class of solutions with two idling-hopping steps hold for $0\leq |U|/t \leq 3.9071$ [Fig.~\ref{fig:details}(d,e)]. Larger values of $|U|/t$ require sequences with more idling-hopping steps: in the SM we provide solutions up to $|U|/t \approx 12$~\cite{SM}, see also reference code available online~\cite{github_code}.  

\textit{Sample efficiency.---}
Quantum gas microscopes are constrained by the number of measurements that they can feasibly perform; each experimental run is relatively slow, typically requiring a gas to be cooled to quantum degeneracy.
Therefore, it is important to extract useful information with as few measurements as possible. This protocol is provably the most sample-efficient to measure the desired correlations, see SM~\cite{SM}.
Here, we show numerically that a modest number of measurements (thousands or fewer) can resolve signatures of $d$-wave pairing in realistic settings.
We simulate two candidate superconducting states: a BCS state with explicitly introduced $d$-wave superconducting order, and the ground state of interacting fermions on a two-legged ladder, obtained through density matrix renormalization group (DMRG) methods (details in SM~\cite{SM}), and then simulate the estimation of their pairing correlations with the protocol.
In Fig.~\ref{fig:numerics}(a), we plot the correlations $\langle\tilde{O}_{i} \tilde{O}_\text{center}\rangle$ between a central vertical dimer and all other dimers. With only 500 measurement shots, we clearly observe the $d$-wave angular dependence---correlations are positive for vertical-vertical and negative for vertical-horizontal dimer pairs---with minimal statistical noise. 
Instead of resolving individual correlations, we can also appropriately average over them to quantify $d$-wave order. In Fig.~\ref{fig:numerics}(b), we plot the histogram of the estimated $d$-wave order parameter $\langle \tilde{D}_{i} \tilde{D}_{i+L/2}\rangle$, for 100 and 300 measurements, where $\tilde{D}_{i}\equiv\widetilde{O}_{i,i+\hat{x}}+\widetilde{O}_{i+\hat{y}-\hat{x},i+\hat{y}-2\hat{x}}-\widetilde{O}_{i+\hat{y},i+2\hat{y}}-\widetilde{O}_{i-\hat{x},i-\hat{x}-\hat{y}}$ encodes the expected angular dependence. For a fixed number of samples, the estimator is a random variable, peaked around its true, non-zero value indicating $d$-wave superconductivity. With small probability, the estimate will be near zero or negative, and one would falsely conclude that $d$-wave superconductivity is absent. 
Fig.~\ref{fig:numerics}(c) demonstrates that only a few hundred measurement shots suffice to achieve $\geq 95\%$ success probability for such classification, with a chosen threshold of $\epsilon=0.05$. Finally, we study $d$-wave pairing in a more realistic setting: the ground state of a two-leg ladder. 
Unlike the BCS state, here the dimer-dimer correlations decay with their distance. In the superconducting phase, the correlator $\langle \widetilde{O}_i \widetilde{O}_{i+r}\rangle$ exhibits a power-law decay over distance $r$, while the Mott-insulating phase exhibits exponentially decaying correlations instead~\cite{PhysRevB.92.195139}. In Fig.~\ref{fig:numerics}(d), we demonstrate that one can observe this phenomenon with 5000 measurements: we accurately reconstruct the power-law/exponential decay of pairing correlations in the ground states of Fermi-Hubbard ladders with average electron density $n=7/8$ and $n=1$, respectively.

\textit{Optimization and robustness.---} 
\begin{figure}[t!]
    \centering
    \includegraphics[width=0.99\linewidth]{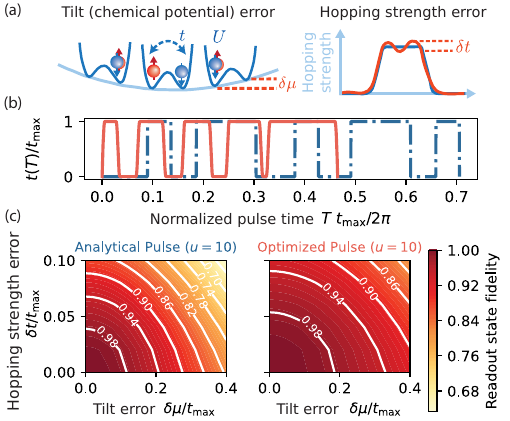}
    \vspace{-0.6cm}
    \caption{\textbf{Robust pulse control}. (a) We optimize our pulse sequence for robustness against two common error sources: unbalanced chemical potential, i.e.~a tilt $\delta \mu\equiv\mu_{i}-\mu_{i'}$ and hopping strength error $\delta t$. (b)  Using optimal control techniques, we find a robust pulse $t(T)$ (red solid line), which is more robust and faster than our analytical pulse (blue, dot-dashed) in the parameter regime $|U|/t\approx 10$. (c) Contour plots of the readout state fidelity with respect to $\delta \mu$ and $\delta t$. The optimized pulse (right) is significantly more robust than the analytical pulse (left), with $\geq94\%$ fidelity under realistic values of error strength.}
    \label{fig:RobustControl}
\end{figure}
Our pulse sequence is exact under ideal conditions, mapping the eigenstates of the pairing correlator $\tilde{O}$ onto Fock states. Here, we analyze the robustness of our pulse sequence to experimental imperfections and develop optimized sequences using optimal control techniques.
In particular, we consider two error sources: dimer tilt and hopping strength error. Our analytical solution assumes a uniform chemical potential on the dimer. The presence of a tilt $\delta\mu \equiv \mu_i - \mu_{i'}$ admixes the two states $|d,\phi\rangle \pm |\phi,d\rangle$, introducing infidelity. Such tilts may arise from various sources including the harmonic trap used to confine the optical lattice [Fig.~\ref{fig:RobustControl}(a)] or laser speckle~\cite{meyrath2005bose,gaunt2013bose,navon2021quantum}. 
Meanwhile, inhomogeneous hopping strengths lead to different amounts of rotation on each dimer.

We use the optimal control technique of direct collocation~\cite{directcollocation,PhysRevApplied.17.014036} to optimize for robustness against these errors. We fix the interaction strength $U$ and optimize the hopping pulse $t(T)$ between values $0$ and a maximum strength $t_\text{max}$. We maximize the fidelity between the target readout states, and initial states (eigenstates of $\widetilde{O}$) evolved in the presence of error, see EM for details. In particular, we target the regime $|U|/t_\text{max} \approx 10$, where the analytic pulse sequences are more sensitive to error. Starting from an analytic solution with five hopping and idling steps interleaved [blue curve in Fig.~\ref{fig:RobustControl}(b)], our optimization yields a smooth pulse $t(T)$ (red curve). In Fig.~\ref{fig:RobustControl}(c), we show contour plots of the fidelity over a wide range of error values of tilt $\delta \mu$ and hopping strength $\delta t$: the optimized pulse fidelity is considerably higher than the analytical pulse fidelity, at $\geq94\%$ for experimentally realistic error strengths~\cite{lebrat2024nagaoka}.

\textit{Universal control and tomographic completeness.---}
Finally, this protocol generalizes to measuring a large class of observables. In fact, \textit{any} number-conserving observable on a dimer can be measured with five types of global control: (i) hopping, (ii) global lattice tilt, (iii) MW drive, (iv) $Z$-magnetic field (or optical detuning), and (v) a global gradient in the $Z$-magnetic field. Remarkably, each control only couples specific pairs of states (see EM), which allows for the easy design of efficient pulse sequences to measure various quantities, including the kinetic energy and the total energy density across a dimer (SM and reference code~\cite{SM,github_code}). We further explore possible applications of this protocol, namely using global pulse sequences to: (i) perform thermometry in a quantum gas microscope, in which the current approach of comparing against classical simulations is expected to fail at low temperatures, and (ii) enhance the parity readout capabilities commonly seen in quantum gas microscopes~\cite{parsons2016site} to effectively full spin- and charge- resolution~\cite{hartke2023direct,koepsell2020robust}, see SM~\cite{SM}. 

\textit{Outlook.---}
Using hitherto-unnoticed structure in the Hilbert space of spin-1/2 fermions on two sites, we develop an experimental protocol that allows a large class of quantities to be measured, including our motivating example of $d$-wave pairing, only using global controls. We anticipate that this protocol will enable further applications such as variational state preparation~\cite{stanisic2022observing}, and more generally serve as a robust quantum control toolbox for fermionic quantum simulators.

\textit{Acknowledgements---} 
We thank the following for insightful discussions: Anant Kale, Aaron Young, Martin Lebrat, Botond Oreg, Carter Turnbaugh, Minh C. Tran, Shang Liu, Zhaoyu Han, Aaron Trowbridge, Andy Goldschmidt, Yi-Zhuang You, Adrian Kantian, Peter Zoller, John Preskill, Sarang Gopalakrishnan, and David Huse.
We acknowledge financial support through the Center for Ultracold Atoms, an NSF Physics Frontiers Center (PHY-2317134), the NSF QLCI Award OMA-2016245, the NSF CAREER award 2237244, the Heising-Simons Foundation (grant \#2024-4851), and the DOE QUACQ program (DE-SC0025572). C.K. acknowledges support from the NSF through a grant for the ITAMP at Harvard University.

\newpage
\let\oldaddcontentsline\addcontentsline
\renewcommand{\addcontentsline}[3]{}
\bibliography{main}

\clearpage
\newpage
\section*{End matter}
\textit{Pulse design.---}We briefly summarize how to determine our pulse sequence parameters: the pulse durations $T_{1,2,3,4}$ plotted in Fig.~\ref{fig:details}(d, e). As discussed in the main text, we have a spin-flip microwave (MW) pulse of strength $B_X$ and duration $(\pi/2)/(2 B_X)$. This performs a $\pi/2-$pulse on the 1-particle (1-p) spin states and a $\pi-$pulse on the two levels of the 2-particle (2-p) inversion symmetric, triplet states, mapping $\ket{\uparrow} \pm i \ket{\downarrow} \mapsto \ket{\downarrow}/\ket{\uparrow}$ and $\ket{\uparrow, \uparrow} + \ket{\downarrow, \downarrow} \mapsto \ket{\uparrow, \downarrow} + \ket{\downarrow, \uparrow}$. We also require a non-trivial pulse sequence of hopping strength over time. This performs the mappings on the 1- and 2-p sectors:
\begin{align}
    |Y^\pm\rangle_\text{pos} \equiv |L\rangle \pm i |R\rangle &\mapsto |Z^{\pm'}\rangle_\text{pos} \equiv |L\rangle/|R\rangle, \label{eq:EM_hop_U_1_map_1p}\\
    |d,\phi\rangle+|\phi,d\rangle &\mapsto \ket{\uparrow,\downarrow} - \ket{\downarrow, \uparrow}, \label{eq:EM_hop_U_1_map_2p}
\end{align}
where on the 1-p sector, $\pm'$ indicates that $|Y^+\rangle_\text{pos}$ can be mapped to either $|L\rangle$ or $|R\rangle$. Taken together, the field and hopping pulses map the eigenstates of $\tilde{O}$ onto readout states: $|Y^\pm\rangle_\text{sp} \otimes |Y^{\pm'}\rangle_\text{pos} \mapsto |Z^{\pm''}\rangle_\text{sp} \otimes |Z^{\pm'''}\rangle_\text{pos}$ and $(\ket{\uparrow, \uparrow} + \ket{\downarrow, \downarrow}) \pm (\ket{d,\phi}+\ket{\phi,d}) \mapsto \ket{\uparrow, \downarrow}/\ket{\downarrow, \uparrow}$.

As discussed in the main text, the effect of hopping and idling in both 1-/2-p sectors can be visualized on four Bloch spheres [Fig.~\ref{fig:details}(b,c)]. Our goal is to perform a $\pi/2$ $X-$rotation on the 1-p positional Bloch sphere, while also mapping the north to south pole on the inversion-symmetric, spin-singlet 2-p states. The challenge lies in the fact that in the presence of interactions, hopping results in a rotation on this 2-p Bloch sphere about a \textit{tilted} axis. To remedy this, we add an idling step, corresponding to a rotation about the vertical axis, to transport between tilted arcs, as illustrated in Fig.~\ref{fig:details}(c). 

We must choose the correct pulse parameters in order to achieve Eq.~(\ref{eq:EM_hop_U_1_map_1p},\ref{eq:EM_hop_U_1_map_2p}) simultaneously. We parametrize the hopping sequence in terms of the angles $\phi_\text{hop}$ and $\phi_\text{idle}$ that the trajectory traces on the 2-p Bloch sphere; these can be converted to the duration of each step by 
$\phi_\text{idle}/2 = (U/2) T_\text{idle}$ and $
 \phi_\text{hop}/2 = \sqrt{4t^2 + U^2/4} T_\text{hop}$. Starting from the north pole, we use the Rodrigues formula~\cite{cheng1989historical} to determine the endpoint $\vec{r}^{(+)}$ of the first hopping step in terms of $\phi^{(+)}_\text{hop}$:
\begin{equation}
\vec{r}^{(+)}  = \begin{pmatrix}
\cos \theta \sin \theta \big(1- \cos \phi^{(+)}_\text{hop}\big) \\
-\sin\theta \sin \phi^{(+)}_\text{hop}\\
\cos^2\theta + \sin^2\theta \cos \phi^{(+)}_\text{hop}
\end{pmatrix},
\end{equation}
where $\cot \theta \equiv |U|/4t \equiv u/4$. Similarly, backward-rotating the south pole gives the starting point $\vec{r}^{(-)}$ of the second hopping step:  
\begin{equation}
\vec{r}^{(-)}  = \begin{pmatrix}
-\cos \theta \sin \theta \big(1- \cos \phi^{(-)}_\text{hop}\big) \\
-\sin\theta \sin \phi^{(-)}_\text{hop}\\
- \cos^2\theta + \sin^2\theta \cos \phi^{(-)}_\text{hop}
\end{pmatrix}.
\end{equation}
$\vec{r}^{(+)}$ and $\vec{r}^{(-)}$ can be connected by an idling step if the $z$-coordinates of $\vec{r}^{(\pm)}$ are equal. This relates $\phi^{(+)}_\text{hop}$ and $\phi^{(-)}_\text{hop}$ by the equation:
\begin{equation}
    \frac{u^2}{16} = - \cos \left(\frac{\phi^{(+)}_\text{hop}+\phi^{(-)}_\text{hop}}{2}\right) \cos \left(\frac{\phi^{(+)}_\text{hop}-\phi^{(-)}_\text{hop}}{2}\right).
    \label{eq:EM_matching_z_ccord}
\end{equation}
Expressed in this form, it is convenient to incorporate the constraint from the 1-p sector: the total hopping time must equal $m(\pi/2)/(2t)$ for some integer $m$, which, in terms of the angles $\phi^{(\pm)}_\text{hop}$ is equivalent to 
\begin{equation}
\phi^{(+)}_\text{hop}+\phi^{(-)}_\text{hop} = 2 \sqrt{1+ \frac{u^2}{16}} \frac{m \pi}{2}  \label{eq:EM_1p_hopping_time}
\end{equation}
Combining Eqs.~(\ref{eq:EM_matching_z_ccord},\ref{eq:EM_1p_hopping_time}) gives a nonlinear equation for $\phi^{(+)}_\text{hop}-\phi^{(-)}_\text{hop}$ as a function of $m$ and $u$. As mentioned in the main text, choosing $m=3$ gives solutions in the region $0\leq u \leq 3.90709$. The angles $\phi^{(\pm)}_\text{hop}$  [proportional to $T_{2,4}$ in Fig.~\ref{fig:details}(d,e)] immediately determine the rest of the pulse sequence. The idling angle $\phi^{(+\rightarrow -)}_\text{idle}$ ($\propto T_3$) to connect the two tilted arcs can be obtained from $\vec{r}^{(\pm)}$ as
\begin{equation}
     \tan \phi^{(+\rightarrow -)}_\text{idle} = \frac{r^{(+)}_xr^{(-)}_y - r^{(-)}_x r^{(+)}_y}{r^{(+)}_xr^{(-)}_x + r^{(+)}_yr^{(-)}_y}.
\end{equation}
Finally, we require an initial idling step, whose angle $\phi^\text{(init)}_\text{hop}$ ($\propto T_1$) is simply the difference between the phases accumulated in the triplet vs.~the singlet sector. This ensures that our pulse sequence ends in the desired superpositions $\ket{\uparrow, \downarrow} \pm \ket{\downarrow, \uparrow}$. These four angles fully specify our pulse sequence, and their corresponding pulse times are plotted as a function of $u$ in Fig.~\ref{fig:details}(e). In the SM, we discuss pulse sequences valid for larger values of $u$, either with larger values of $m$ or pulse sequences with more steps. 

\textit{Pulse optimization.---}We use the method of \textit{direct collocation}~\cite{directcollocation,PhysRevApplied.17.014036} to optimize our pulse sequences for robustness against two sources of error: tilt error $\delta \mu \equiv \mu_i - \mu_{i'}$, in which the two sites on each dimer have unequal chemical potential, and amplitude error $\delta t$, where the hopping strength $t$, or equivalently interaction strength $U$ in a dimer differs from its expected value. We quantify the robustness of our pulse through the fidelity
\begin{equation}
f_{m,s}\equiv \big|\langle\psi_{s}^{(f)}|V[t(T);\delta t^{(m)}, \delta \mu^{(m)}]|\psi_s^{(i)}\rangle \big|^2
\end{equation}
between the target readout state $|\psi_{s}^{(f)}\rangle$, and the initial state $|\psi_{s}^{(i)}\rangle$ (i.e.~nontrivial eigenstates of the observable $O$), evolved with an imperfect unitary $V[t(T);\delta t^{(m)}, \delta \mu^{(m)}]$~\cite{RobustNMR,RobustLearning}. We then average the fidelity over the 1-p and 2-p eigenstates, indexed by $s \in \{1,2,…6\}$, and uniformly over a range of error strengths, indexed by $m$. There are multiple ways one could perform this average, but we have verified that this choice does not significantly affect the optimization. A large fidelity would certify that the expectation values measured with the protocol are close to their true value, even in the presence of errors.

Within the framework of directed collocation, we minimize the average infidelity
\begin{equation}
\mathrm{min}~\ell(V[t(T)])  = 1-\frac{1}{MS}\sum_{m=1}^{M}\sum_{s=1}^{S}f_{m,s},
\label{eq:optimization_loss_function}
\end{equation}
subject to the constraints that
\begin{align}
    V_{k+1}-\mathrm{exp}\left[-i H(t_k)\Delta T\right]V_k&=0, \label{eq:Scrh_eq_1}\\
V_1&=I_{16}, \label{eq:Scrh_eq_2}\\
0 \leq t_k &\leq t_{\mathrm{max}},\\
\left|(t_{k+1}-t_{k})/\Delta T\right| &\leq C_1,\\
\text{and } \left|(t_{k+1}-2t_{k}+t_{k-1})/\Delta T^2\right|&\leq C_2,
\label{eq:optimization_constraint_4}
\end{align}
where we have discretized the unitary $V[t(T)]$ into a sequence of unitaries $V_k$ at time-points $T_k$ separated by an interval $\Delta T$, with hopping strengths $t_k$. These unitaries are constrained to satisfy the Schr\"odinger equation [Eq.~\eqref{eq:Scrh_eq_1}], and the initial condition that $V_1$ is the $16\times 16$ identity $I_{16}$ [Eq.~\eqref{eq:Scrh_eq_2}]. We have also constrained the pulse to be smooth by requiring its first and second derivatives to have magnitudes smaller than $C_1$ and $C_2$ respectively. Our optimization task (\ref{eq:optimization_loss_function}-\ref{eq:optimization_constraint_4}) is in the form of a typical nonlinear programming (NLP) problem, and we use an interior-point optimizer (IPOPT) \cite{nocedal2009adaptive,doi:10.1137/S1052623403426544,doi:10.1137/S1052623497325107} to find a numerical solution. This method has been used in recent quantum optimal control literature \cite{directcollocation} and is available online as \verb|QuantumCollocation.jl|~\cite{QuantumCollocationJL}. 

Compared to gradient-based pulse optimization methods, such as gradient ascent pulse engineering (GRAPE), the directed collocation approach has several advantages: (i) enforcing the Schr\"odinger dynamics as a constraint enables the solver to explore \emph{unphysical} regions before converging to an optimal physical trajectory, (ii) it is easy to add additional constraints on the pulse shape (e.g.~maximum value and derivatives), and (iii) this approach is compatible with widely available and highly optimized NLP solvers.

Our analytic pulses show a modest degree of robustness to experimental error (Fig.~\ref{app_fig:robustness_analytical} in SM). Unsurprisingly, the more complicated 10-pulse sequences, necessary in the regime $|U|/t = 10$, are significantly less robust than the simpler 4-pulse sequences valid for small $|U|/t$. As illustrated in Fig.~\ref{fig:RobustControl}, in this regime our optimization scheme shows significant gains, with no additional experimental complexity, achieving over $95\%$ fidelity under realistic values of $\delta \mu$ and $\delta t$.

\textit{Universal experimental control set.---} We find that five types of global experimental controls allow for universal manipulations of the quantum states on each particle-number sector. These controls are:
\begin{enumerate}[label=(\roman*)]
    \item Hopping,
    \item Global lattice tilt,
    \item Spin-flip MW drive,
    \item $Z$-magnetic field (or optical detuning), and a
    \item Global gradient in the $Z$-magnetic field.
\end{enumerate}  
This can be seen by directly computing the action of each control on each particle-number sector. For example, in the 2-p sector, we have
\begin{align}
       H^\text{(2-p)} &= \begin{pmatrix}
        U  & -2t & \delta\mu & 0 & 0 & 0\\
        -2t & 0 & 0 & 0 & \delta B_Z & 0\\
        \delta\mu & 0 & U & 0 & 0 & 0\\
        0 & 0 & 0 & 0 & 0 & 2\bar{B}_Z\\
        0 & \delta B_Z & 0 & 0 & 0 & 2\bar{B}_X\\
        0 & 0 & 0  & 2 \bar{B}_Z & 2\bar{B}_X & 0\\
    \end{pmatrix}  
    \begin{matrix}
        {\scriptstyle \ket{d, \phi} + \ket{\phi, d}}\\
        {\scriptstyle \ket{\uparrow, \downarrow} - \ket{\downarrow, \uparrow}}\\
        {\scriptstyle \ket{d, \phi} - \ket{\phi, d}}\\
        {\scriptstyle \ket{\uparrow, \uparrow} - \ket{\downarrow, \downarrow}}\\
        {\scriptstyle \ket{\uparrow, \downarrow} + \ket{\downarrow, \uparrow}}\\
        {\scriptstyle \ket{\uparrow, \uparrow} + \ket{\downarrow, \downarrow}}
    \end{matrix}~,
    \nonumber
\end{align}
where $t$ is the hopping strength, $\delta \mu \equiv \mu_i - \mu_{i'}$ is the chemical potential difference (``tilt") between the sites $i$ and $i'$ of the dimer, $\bar{B}_X \equiv (B_{X,i}+B_{X,i'})/2$ and $\bar{B}_Z \equiv (B_{Z,i}+B_{Z,i'})/2$ are the average spin-flip and Zeeman fields experienced by the dimer, and $\delta B_Z \equiv B_{Z,i}-B_{Z,i'}$ is the difference in Zeeman fields across the dimer. Here, we have used the same basis as in Eq.~\eqref{eq:Ham_2p}. Remarkably, each control selectively couples different states in the 2-p Hilbert space, as is also the case in the 1-/3-p Hilbert space (see SM Fig.~\ref{fig:universal_control_app}). 
Using this selectivity, we explicitly show how \textit{any} number-preserving observable can be measured using these five controls. 
We also design efficient pulses to measure specific observables of interest. In all examples we study, we do not need all five controls, as summarized in Table~\ref{tab:observable_pulses}.

Finally, we mention a subtlety: we propose to realize the $\delta B_Z$ term with a \textit{global} field gradient. This not only introduces the $\delta B_Z$ term, but also shifts the value of $\bar{B}_Z$ from dimer to dimer. Generically, this leads to spatially-varying rotations between the $\ket{\uparrow,\uparrow} \pm \ket{\downarrow,\downarrow}$ states. However, if we only use the global field gradient to perform $\pi$-pulses between $\ket{\uparrow,\downarrow} - \ket{\downarrow,\uparrow}$ and $\ket{\uparrow,\downarrow} + \ket{\downarrow,\uparrow}$, we show in the SM that the action on the $\{\ket{\uparrow,\uparrow} \pm \ket{\downarrow,\downarrow}\}$ subspace is trivial. Despite this restriction, this suffices for our universality argument, and we can also utilize general rotations when the action on the $\{\ket{\uparrow,\uparrow} \pm \ket{\downarrow,\downarrow}\}$ states does not affect the protocol outcome.

\begin{table}[bp!]
    \centering
    \begin{tabular}{c c c c c c c}
    \toprule
    Obs./Gate & Hop. & Tilt & $\bar{B}_X$ & $\bar{B}_Z$ & $\delta B_Z$ & Ref.\\
    \midrule
    $d$-wave pairing     & $\checkmark$ & & $\checkmark$& & &   \\
    Kinetic energy & $\checkmark$& $\checkmark$ & & & & SM Fig.~\ref{fig:pulse_sequences}\\
    Energy density &$\checkmark$ & $\checkmark$ & & & $\checkmark$& \\
    \midrule
    \textbf{pSWAP} & $\checkmark$ & & & & &\\
    \textbf{odd-SWAP} & $\checkmark$ & & & & & SM Fig.~\ref{fig:more_pulse_sequences}\\
    Spin-resolved current & $\checkmark$ & $\checkmark$ & & & $\checkmark$& \\
    \bottomrule
    \end{tabular}
    \caption{Summary of pulse sequences designed in this work and their required experimental controls. 
    While all five controls are needed to universally manipulate the states on a dimer, for the pulse sequences developed here, we most frequently utilize hopping, tilt and field-gradient. We also list gates developed for thermometry: the \textbf{pSWAP} gate applies a SWAP between the two sites of the dimer along with parity-dependent phase, and the \textbf{odd-SWAP} gate applies a SWAP only when the parity is odd, see SM~\cite{SM}.
    }
    \label{tab:observable_pulses}
\end{table}

\let\addcontentsline\oldaddcontentsline
\clearpage
\newpage
\onecolumngrid

\appendix
\tableofcontents

\setcounter{figure}{0}
\renewcommand{\figurename}{Fig.}
\renewcommand{\thefigure}{S\arabic{figure}}
\setcounter{table}{0}
\renewcommand{\tablename}{Table}
\renewcommand{\thetable}{S\arabic{table}}

\section{Superconducting pairing observables and Bardeen–Cooper–Schrieffer (BCS) theory}
In this work, we present a protocol to measure the long-distance correlations of the Cooper pairing creation observable, after a spin-singlet and Hermitian symmetrization and a particle-hole transformation mapping the repulsive to attractive Fermi-Hubbard model. Here, we describe this symmetrization, the repulsive-to-attractive mapping, and describe our numerical methods investigating Bardeen–Cooper–Schrieffer (BCS) states and the ground states of Fermi-Hubbard model on a two-leg ladder.

\subsection{Superconducting pairing observables}
A signature of superconductivity is the formation of a condensate of fermion pairs. 
In textbook treatments of superconductivity, a superconducting state is distinguished by a non-zero expectation value of the pairing creation operator $\Delta^{\dagger}_{k}=\cd{k,\uparrow}\cd{-k,\downarrow}$~\cite{Ashcroft76,Girvin_Yang_2019}. Such a state cannot have fixed particle number, as would be expected in quantum gas microscopes. Instead, superconductivity would manifest in \textit{off-diagonal long-range order}, i.e. in the correlators $\langle c^\dagger_{k,\uparrow} c^\dagger_{-k,\downarrow} c^{}_{k',\uparrow} c^{}_{-k',\downarrow}\rangle$ taking non-zero value. These correlators can be obtained by the Fourier transform of real-space four point correlators $\langle c^\dagger_{i,\uparrow} c^\dagger_{i',\downarrow} c^{}_{j,\uparrow} c^{}_{j',\downarrow}\rangle$. 

Instead of the vanilla pairing operator~$\cd{i,\uparrow}\cd{i',\downarrow}$, we symmetrize the pairing operator as follows:
\eqs{
O_{i,i'}\equiv S^{\dagger}_{i,i'}+\text{h.c.}~,
\label{eq:pairing_correlator}
}
where $S^{\dagger}_{i,i'}\equiv \cd{i,\uparrow}\cd{i',\downarrow}-\cd{i,\downarrow}\cd{i',\uparrow}$ creates a singlet on $(i,i')$. This has two forms of symmetrization: spin-singlet symmetrization, and Hermitian symmetrization. First of all, since $s$- and $d$-wave superconductivity involve spatially symmetric and hence spin-antisymmetric Cooper pairs, only the singlet pairing function $S^{\dagger}_{i,i'}$ \cite{doi:10.1073/pnas.2112806119,PhysRevLett.127.097003} is expected to be non-zero, while the triplet pairing observable $\cd{i,\uparrow}\cd{i',\downarrow}+\cd{i,\downarrow}\cd{i',\uparrow}$ will have zero expectation value. The long-range spin-singlet pairing correlator is related to the vanilla pairing correlation by $\langle S^{\dagger}_{i,i'}S_{j,j'}^{}\rangle=4\langle c_{i,\uparrow}^{\dagger}c_{i',\downarrow}^{\dagger}c_{j,\uparrow}^{} c_{j',\downarrow}^{}\rangle$. 

This protocol enables the measurement of dimer-dimer correlations of Hermitian observables. This contains the terms
\eqs{
\langle O_{i,i'} O_{j,j'}\rangle
=\langle S^{\dagger}_{i,i'}S_{j,j'}^{}\rangle+\langle S^{\dagger}_{j,j'}S_{i,i'}^{}\rangle+\langle S^{\dagger}_{i,i'}S^{\dagger}_{j,j'}\rangle+\langle S_{i,i'}^{}S_{j,j'}^{}\rangle~.
\label{eq:OO_correlation}
}
We show that $\langle O_{i,i'} O_{j,j'}\rangle=2\langle S^{\dagger}_{i,i'}S_{j,j'}^{}\rangle$ for any thermal state of the Fermi-Hubbard model with fixed particle number. Due to the conservation of particle number, the last two terms in Eq.~\eqref{eq:OO_correlation} are always zero. Furthermore, the fact that the Fermi-Hubbard model is time-reversal symmetric implies that $\langle S^{\dagger}_{i,i'}S_{j,j'}\rangle=\langle S^{\dagger}_{j,j'}S_{i,i'}\rangle$. Therefore, the correlations $\langle O_{i,i'} O_{j,j'}\rangle$ measured in this protocol are equal to $8\langle c_{i,\uparrow}^{\dagger}c_{i',\downarrow}^{\dagger}c_{j,\uparrow}^{} c_{j',\downarrow}^{}\rangle$.

We remark that for BCS states (discussed below), which are time-reversal symmetric but do not have fixed particle number, one can show with Wick's theorem that $\langle O_{i,i'} O_{j,j'}\rangle=16\langle c_{i,\uparrow}^{\dagger}c_{i',\downarrow}^{\dagger}c_{j,\uparrow}^{} c_{j',\downarrow}^{}\rangle$ instead, since $\langle S^{\dagger}_{i,i'}S^\dagger_{j,j'}\rangle$ is non-zero. Therefore, particle-number conservation is a key assumption if one wants to quantitatively estimate $\langle S^{\dagger}_{i,i'}S_{j,j'}^{}\rangle$. Furthermore, upon the repulsive-to-attractive mapping, particle-number conservation is mapped to magnetization conservation, and the total magnetization may not necessarily be fixed in an experimental state. Fortunately, this assumption is not necessary, after a slight modification of the protocol in this work: by also measuring the correlators $\langle O^{(a)}_{i,i'} O^{(a)}_{j,j'}\rangle$ where $O^{(a)}_{i,i'}\equiv i(S^{\dagger}_{i,i'}-\text{h.c.})$, we can independently estimate $\langle S^{\dagger}_{i,i'}S_{j,j'}^{}\rangle + \langle S^{\dagger}_{j,j'}S_{i,i'}^{}\rangle$ and $\langle S^{\dagger}_{i,i'}S^{\dagger}_{j,j'}\rangle + \langle S_{i,i'}^{}S_{j,j'}^{}\rangle$.

We next turn to determining pairing order. $s$-wave and $d$-wave superconductivity can be distinguished by their real-space pairing symmetries. For example, in $s$-wave superconductivity, the pairing correlation does not depend on the relative orientation of $i,i',j$ and $j'$, while in $d-$wave superconductivity, the correlator changes sign when the orientation of $j$ to $j'$ is rotated by $\pi/2$. Therefore, to distinguish between $s$- and $d$-wave superconductivity, the following spatial symmetrizations can be used:
\eqs{
D_{s}(i)&\equiv O_{i,i+\hat{y}}+O_{i,i+\hat{x}}+O_{i,i-\hat{y}}+O_{i,i-\hat{x}},\\
D_{d}(i)&\equiv -O_{i,i+\hat{y}}+O_{i,i+\hat{x}}-O_{i,i-\hat{y}}+O_{i,i-\hat{x}},
}
and the long-range correlations $\langle D_{s}(i)D_{s}(j)\rangle$ and $\langle D_{d}(i)D_{d}(j)\rangle$ would distinguish pairing order. This will be the observable we aim to measure, upon a repulsive-to-attractive mapping which we discuss below.

\subsection{Repulsive-to-attractive particle-hole transformation}
The mapping between repulsive and attractive Fermi-Hubbard models is accomplished by the following particle-hole transformation, which only acts non-trivially on the spin-down particles~\cite{ho2009quantum}:
\eqs{
\begin{array}{ll}
    c_{i,\uparrow}\mapsto c_{i,\uparrow},\\
c^{}_{i,\downarrow}\mapsto(-1)^{i}c_{i,\downarrow}^{\dagger}.
\end{array}
 \label{eq:PHtransform}}
This is a unitary transformation, and the $(-1)^i$ phase factor indicates that this transformation has a $-1$ sign on one sublattice of a bipartite lattice, here the square lattice. This transformation leaves the operators of the spin-up fermions unchanged, but maps the spin-down occupation operator $n_{i,\downarrow}\rightarrow 1-n_{i,\downarrow}$. With this phase factor, the interaction term changes sign, $U\rightarrow -U$, while the hopping terms are unchanged. This also exchanges the roles of total particle number $\sum_i n_{i,\uparrow} + n_{i,\downarrow}$ with the total magnetization $\sum_i n_{i,\uparrow} - n_{i,\downarrow}$: a low-energy state at a certain filling and zero magnetization on the repulsive Hubbard model would be related by this particle-hole transformation to a low-energy state at half-filling and corresponding magnetization on the attractive Hubbard model.

In momentum space, the mapping can be written as
\eqs{
c_{k,\downarrow}\mapsto c^{\dagger}_{-k+Q,\downarrow},~ c_{k,\uparrow}\mapsto c_{k,\uparrow},
\label{eq:transform}}
where $Q=(\pi/a,\pi/a)$ with $a$ as the lattice spacing. This can be used to obtain different order parameters and corresponding phases transformation between the repulsive ($U>0$) and attractive ($U<0$) Fermi-Hubbard models. In our case of interest, $d$-wave superconductivity in the repulsive Hubbard model, with order parameter $\Delta^{(d)}\equiv \sum_{k}(\cos k_x-\cos k_y)\langle c_{-k,\downarrow}c_{k,\uparrow} \rangle$ gets mapped to $d$-wave anti-ferromagnetic order, with order parameter $M^{(d)}_{Q} \equiv\sum_{k}(\cos k_x-\cos k_y)\langle c^{\dagger}_{k+Q,\downarrow}c^{}_{k,\uparrow}\rangle$, see Ref.~\cite{ho2009quantum} for more details.

\subsection{BCS theory and efficient simulation of BCS states}
We describe our numerical simulations performed on Bardeen–Cooper–Schrieffer (BCS) states. These states are toy models for superconducting states, based on BCS theory. The fermionic wavefunction can be written as~\cite{Ashcroft76,Girvin_Yang_2019}

\eqs{
\ket{\psi}=\left[\prod_{\vec{k}}\left(u_{\vec{k}}+v_{\vec{k}}\cd{\vec{k},\uparrow}\cd{-\vec{k},\downarrow}\right)\right]\ket{\text{vac}},
}
where
\begin{align}
    u_{\vec{k}}\equiv\frac{1}{\sqrt{1+|a(\vec{k})|^2}}, ~v_{\vec{k}}\equiv\frac{a(\vec{k})}{\sqrt{1+|a(\vec{k})|^2}},
\end{align}
and the momentum space pairing function $a(\vec{k})$ is defined as $a(\vec{k})=\Delta(\vec{k})/\left[\xi_{\vec{k}}+\sqrt{\xi_{\vec{k}}^2+\Delta(\vec{k})^2}\right]$, where the dispersion and gap functions are given respectively by
\eqs{
\xi_{\vec{k}}\equiv-2\left[\cos(k_x)+\cos(k_y)\right]-\mu
}
\eqs{
\Delta(\vec{k}) \equiv \left\{
\begin{array}{lll}
    \Delta & \text{if $s$-wave}\\
    \Delta[\cos(k_x)-\cos(k_y)] & \text{if $d$-wave}
\end{array}
\right.
}
In our numerical simulations, we use the parameter values $\Delta = 0.1\sim 0.3$ and $\mu = 0.5$, firmly in the BCS regime of the BCS-BEC (Bose-Einstein condensate) crossover.

Since the BCS wavefunction is a fermionic Gaussian state, its properties are fully determined by its two-fermion correlations. In our numerical simulation, we consider a $L\times L$ square lattice with periodic boundary conditions. We use the following useful correlations:
\eqs{
\langle c^\dagger_{\vec{p},\uparrow}c^\dagger_{-\vec{q},\downarrow}\rangle&=\delta_{\vec{p},\vec{q}}v_{\vec{p}}^*u_{\vec{p}},~~~~\langle c^\dagger_{\vec{p},\uparrow}c^{}_{\vec{q},\uparrow} \rangle=\delta_{\vec{p},\vec{q}}|v_{\vec{p}}|^2,~~~~\langle c^\dagger_{\vec{p},\downarrow}c^{}_{\vec{q},\downarrow} \rangle=\delta_{\vec{p},\vec{q}}|v_{-\vec{p}}|^2.
}
Since there is no spin-triplet pairing, $\langle c^\dagger_{\vec{p},\uparrow}c^\dagger_{\vec{q},\uparrow}\rangle=0$. Furthermore, the state has magnetization 0, and hence $\langle c^\dagger_{\vec{p},\uparrow}c^{}_{\vec{q},\downarrow}\rangle=\langle c^\dagger_{\vec{p},\downarrow}c^{}_{\vec{q},\uparrow}\rangle=0$. Finally, using Wick's theorem, we calculate the four-fermion correlators:
\eqs{
\langle(c^\dagger_{\vec{p},\uparrow}c^\dagger_{-\vec{q},\downarrow})( c^{}_{-\vec{q}',\downarrow} c^{}_{\vec{p}',\uparrow})\rangle&= \langle c^\dagger_{\vec{p},\uparrow} c^\dagger_{-\vec{q},\downarrow}\rangle\langle c^{}_{-\vec{q}',\downarrow}\cc{\vec{p}',\uparrow}\rangle-\langle c^\dagger_{\vec{p},\uparrow} c^{}_{-\vec{q}',\downarrow}\rangle\langle c^\dagger_{-\vec{q},\downarrow}  c^{}_{\vec{p}',\uparrow}\rangle+\langle c^\dagger_{\vec{p},\uparrow} c^{}_{\vec{p}',\uparrow}\rangle\langle c^\dagger_{-\vec{q},\downarrow} c^{}_{-\vec{q}',\downarrow}\rangle\\
&=\delta_{\vec{p},\vec{q}}\delta_{\vec{p}',\vec{q}'} v^*_{\vec{p}} u_{\vec{p}}^{} u^*_{\vec{p}'} v_{\vec{p}'}^{}+\delta_{\vec{p},\vec{p}'}\delta_{\vec{q},\vec{q}'}|v_{\vec{p}}^{}|^2|v_{-\vec{q}}^{}|^2
}

\subsubsection{Reduced density matrix of a fermionic bilinear state}
In our numerical simulations, we evaluate the spatial correlations $\langle c^\dagger_i c^\dagger_{i'} c^{}_j c^{}_{j'}\rangle$ on the BCS state by computing its reduced density matrix supported on the neighborhood of $i,i',j,$ and $j'$.

Given a fermionic bilinear wave function
\eqs{
|\psi\rangle=C \exp\left(\frac{1}{2}\sum_{i,j}G_{ij}\cd{i}\cd{j}\right)|\text{vac}\rangle \label{eq:bilinear},
}
where matrix $G$ denotes the fermion pairing in the real space, one can efficiently calculate its reduced density matrix on a subsystem $A$. We rearrange the pairing matrix $G$ into four sub-matrices: $G_{AA}$, $G_{A\bar{A}}$, $G_{\bar{A}A}$ and $G_{\bar{A}\bar{A}}$ depending on whether the index $i,j$ are in the region $A$ or its complement $\bar{A}$. Then the reduced density matrix can be written as~\cite{PhysRevB.64.064412}
\eqs{
\rho_A = |C|^2\exp\left(\sum_{i,j \in A}\alpha_{ij}\cd{i}\cd{j}\right)\exp\left(\sum_{i,j \in A}(\log \beta)_{ij}\cd{i}\cc{j}\right)\exp\left(\sum_{i,j \in A}-\alpha_{ij}\cc{i}\cc{j}\right),
\label{eq:RDM}
}
with
\eqs{\left\{
\begin{array}{ll}
v=G_{A\bar{A}}(\mathrm{1}-G_{\bar{A}\bar{A}})^{-1}\\
    \alpha=\frac{1}{2}(G_{AA}-v G_{\bar{A}\bar{A}}v^{T}) \\
    \beta=vv^{T}\\
\end{array}
\right.\label{eq:alpha}
}
where $v^T$ is the matrix transpose of $v$. This can be shown by calculating the overlap of the reduced density matrix with the fermionic coherent states  $\ket{\xi_1,\xi_2,\dots,\xi_L}=\exp\left(-\sum_i\xi_i c^{\dagger}_i\right)|\text{vac}\rangle$ as
\eqs{
&\langle \xi_{A_1},\xi_{A_2},\dots,\xi_{A_M}|\rho_{A}|\xi_{A_1}',\xi_{A_2}',\dots,\xi_{A_M}'\rangle\\
&=|C|^2\int \Pi_{j\in \bar{A}}e^{-\xi_{\bar{A}_j}^{*}\xi_{\bar{A}_j}}\langle \xi_{A_1},\dots,\xi_{A_M},-\xi_{\bar{A}_1},\dots,-\xi_{\bar{A}_L}|\rho_0| \xi_{A_1}',\dots,\xi_{A_M}',\xi_{\bar{A}_1},\dots,\xi_{\bar{A}_L}\rangle,
}
with
\eqs{
&c_i|\xi_1,\dots,\xi_L\rangle=\xi_i|\xi_1,\dots,\xi_L\rangle\\
&\rho_0=\exp(\frac{1}{2}\sum_{i,j}G_{ij}\cd{i}\cd{j})|\text{vac}\rangle\langle \text{vac}|\exp(-\frac{1}{2}\sum_{i,j}G_{ij}\cc{i}\cc{j})
}
This results in
\eqs{
\langle \xi_{A_1},\dots,\xi_{A_M}|\rho_{A}|\xi_{A_1}',\dots,\xi_{A_M}'\rangle=|C|^2\exp\left(\sum_{i,j \in A}\alpha_{ij}\xi_{i}^{*}\xi_{j}^{*}\right)\exp\left(\sum_{i,j \in A}\beta_{ij}\xi_i^{*}\xi_j^{'}\right)\exp\left(\sum_{i,j \in A}-\alpha_{ij}\xi_{i}^{'}\xi_{j}^{'}\right).
}
A detailed derivation can be found in Ref.~\cite{PhysRevB.64.064412}, and one should notice a sign difference between the definition in \eqnref{eq:alpha} for $\alpha$ and the one in Ref.~\cite{PhysRevB.64.064412}.

In our numerical simulations of a lattice system with the periodic boundary condition, we transform the BCS wavefunction into real space with bilinear form Eq.~\eqref{eq:bilinear} by using two spinless fermions to represent one spinful fermion. We then apply Eq.~\eqref{eq:RDM} to get the reduced density matrix. A point of note is that when using Eq.~\eqref{eq:RDM}, the definition of $G$ should be the same as Eq.~\eqref{eq:bilinear}, and the extra coefficient $1/2$ cannot be absorbed into the normalization constant $|C|$. In order to simulate measurements on a state with the attractive Fermi-Hubbard model, we use the particle-hole transformation Eq.~\eqref{eq:PHtransform}.  
One can show that one can take the partial trace before or after the particle-hole transformation. In our case, we first obtain the reduced density matrix from Eq.~\eqref{eq:RDM} before applying the particle-hole transformation. 

In all the simulations with the BCS state, we choose the system to be a $50 \times 50$ square lattice, comparable in size to typical experiments.

\subsection{Ground state of Fermi-Hubbard ladders}

In this work, we numerically study the ground state of the Fermi-Hubbard model on a two-leg ladder with length $N=40$ (obtained with density matrix renormalization group (DMRG) methods~\cite{PhysRevLett.69.2863}), as a more realistic state than BCS states. On the repulsive Hubbard model, we observe (i) power-law decay of pairing correlations at average electron density $n=7/8$ ($7N/16$ spin-up electrons and $7N/16$ spin-down electrons) with $U=8$, and (ii) exponentially-decaying correlations at half filling $n=1$ ($N/2$ spin-up electrons and $N/2$ spin-down electrons) with $U=8$. On the attractive Fermi-Hubbard model, this corresponds to (i) $7N/16$ spin-up electrons and $9N/16$ spin-down electrons with $U=-8$ and (ii) $N/2$ spin-up electrons and $N/2$ spin-down electrons with $U=-8$. We set the maximum bond dimensions to be 1800 for the hole-doped state and 900 for the half-filled state, implemented with the \verb|ITensor.jl| package~\cite{itensor}. Similar parameters are also used in Ref.~\cite{PhysRevB.92.195139}. 

\section{Parameters for hopping pulse sequences}
\label{app:dwave_hopping_pulse}
Here, we describe our analytic solution to the hopping pulse sequence to measure $d-$wave pairing correlations. To restate our approach outlined in the main text, we develop pulse sequences that map the eigenstates of the observable of interest to the readout basis. The $d$-wave pairing observable, Eq.~\eqref{eq:pairing_correlator}, has nontrivial eigenstates in the 1-p, 2-p, and 3-p sectors:
\begin{center}
\begin{tabular}{ c c  c }
\toprule
Eigenvalue  &  Eigenstate & Readout state\\
\midrule
$+1$  & $|Y^+\rangle_\text{sp}|Y^+\rangle_\text{pos}$ & $|\!\downarrow, \phi\rangle$\\
$+1$ & $|Y^-\rangle_\text{sp}|Y^-\rangle_\text{pos}$ & $|\phi, \uparrow\rangle$\\
$-1$ & $|Y^+\rangle_\text{sp}|Y^-\rangle_\text{pos}$ & $|\phi, \downarrow\rangle$\\
$-1$ & $|Y^-\rangle_\text{sp}|Y^+\rangle_\text{pos}$ & $|\!\uparrow,\phi\rangle$\\
\midrule
$+2$ & $|d,\phi\rangle + |\phi,d\rangle - |\!\uparrow,\uparrow\rangle - |\!\downarrow,\downarrow\rangle$ & $|\!\downarrow,\uparrow\rangle$\\
$-2$ & $|d,\phi\rangle + |\phi,d\rangle + |\!\uparrow,\uparrow\rangle + |\!\downarrow,\downarrow\rangle$ & $|\!\uparrow,\downarrow\rangle$\\
\midrule
$+1$  & $|Y^-\rangle_\text{sp,3p}|Y^+\rangle_\text{pos}$ & $|\!\downarrow, d\rangle$\\
$+1$ & $|Y^+\rangle_\text{sp,3p}|Y^-\rangle_\text{pos}$ & $|d, \uparrow\rangle$\\
$-1$ & $|Y^-\rangle_\text{sp,3p}|Y^-\rangle_\text{pos}$ & $|d, \downarrow\rangle$\\
$-1$ & $|Y^+\rangle_\text{sp,3p}|Y^+\rangle_\text{pos}$ & $|\!\uparrow,d\rangle$\\
\bottomrule
\end{tabular} 
\end{center}

As stated in the main text, the MW pulse is simple: a pulse of duration $(\pi/2)/(2 B_X)$ effects a $\pi/2$-rotation on the spin Bloch sphere in the 1-p and 3-p sectors, and performs a $\pi$-rotation on the triplet, inversion-symmetric 2-p Bloch sphere, sending the state $|\!\uparrow,\uparrow\rangle + |\!\downarrow,\downarrow\rangle$ to $|\!\uparrow,\downarrow\rangle + |\!\downarrow,\uparrow\rangle$. It remains to design a pulse sequence of hopping strength over time, that effects a $\pi/2$-rotation on the positional Bloch spheres in the 1-p and 3-p sectors, simultaneously with a $\pi$-rotation on the singlet, inversion-symmetric 2-p Bloch sphere.

The pulse sequence that achieves this goal, which we dub \textbf{Hop+U$^1$}, will turn out to be useful not just to measure $d-$wave pairing. In Appendix~\ref{app:further_pulse_sequences}, we develop pulse sequences to measure other quantities of interest, including the kinetic energy and energy density. These sequences make use of \textbf{Hop+U$^1$} extensively, in conjunction with the other types of experimental control discussed in Appendix~\ref{app:universal_control_set}. 

This pulse sequence achieves the mappings on the 1-p and 2-p sectors
\begin{align}
    |Y^\pm\rangle_\text{pos} &\mapsto |Z^{\pm'}\rangle_\text{pos},\\
    |d,\phi\rangle+|\phi,d\rangle &\mapsto |\!\uparrow,\downarrow\rangle - |\!\downarrow, \uparrow\rangle, \label{eq:SM_hop_U_1_map_2p}
\end{align}
where on the 1-p sector, $\pm'$ denotes that the $|Y^+\rangle_\text{pos}$ state can be mapped to either $|Z^+\rangle_\text{pos}$ or $|Z^-\rangle_\text{pos}$.
For simplicity, we assume that we only have the ability to toggle the hopping strength off or on, i.e.~between the values $0$ and $t$. Meanwhile, we assume that the interaction strength is always fixed to a value of $U$. 

Under idling and hopping, the dynamics on the 2-p sector is only nontrivial on the two-level system formed by the states $\{|d,\phi\rangle + |\phi,d\rangle, |\!\uparrow, \downarrow\rangle - |\!\downarrow, \uparrow\rangle\}$, and designing the mapping Eq.~\eqref{eq:SM_hop_U_1_map_2p} is equivalent to designing a trajectory on the equivalent Bloch sphere. Our pulse sequences will depend on the ratio of $u \equiv |U|/t$. In particular, our simplest pulse sequence has four steps: idling, hopping, idling and hopping for different amounts of time. On the Bloch sphere, idling gives rise to trajectories on horizontal arcs, while hopping results in trajectories on tilted arcs. The simple four-pulse sequences correspond to connecting two tilted circles that intersect the north and south poles with a single horizontal circle [Fig.~\ref{fig:n_pulse_solutions}(a)]. Geometrically, this is only possible when $u \leq 4$.

Even within this feasible range $u \leq 4$, the parameters of the \textit{four-pulse sequence}, i.e.~the duration of each step, depend on the value of $u$. In Appendix~\ref{app:four_pulse_sequence} we describe how to determine these parameters. 
Our solutions for the four-pulse sequences depend on an integer $m$ describing the amount of over-rotation on the 1-p positional Bloch sphere. Solutions exist for the simplest case $m=3$ when $u \leq 3.919$. In the rest of the feasible range $3.919 \leq u \leq 4$, we determine an infinite sequence of integers $m$ for which solutions exist. We then seek to develop pulse sequences in the experimentally relevant regime $u\approx 10$. In Appendix~\ref{app:n_pulse_sequences}, we study more complex \textit{6-,8- and 10-pulse sequences} [Fig.~\ref{fig:n_pulse_solutions}(b-d)] that provide solutions for larger values of $u$, providing solutions for their pulse parameters.

Our pulse sequence solutions hold regardless of the sign of $U$ or $t$: such sign changes result in axes of rotation that are either reflected about the $X-Y$ plane of the Bloch sphere, or with reversed rotation directions. The resulting trajectories in the Bloch sphere are simply reflected versions of the trajectory with positive $U$ and $t$. Therefore, without loss of generality, we take $U$ and $t$ to be positive. 

Finally, we remark on our assumption that the interaction strength $U$ is fixed. In reality, it can change when hopping is toggled on and off, since a deeper lattice may have more narrow motional eigenstates and hence doublons may experience stronger interaction when hopping is turned off. In this case, a simple modification suffices: we rescale the idling times appropriately by the modified interaction strength $U'$.

\subsection{General strategy}
\label{app:solution_strategy}
\begin{figure}[t!]
    \centering
    \includegraphics[width=0.95\linewidth]{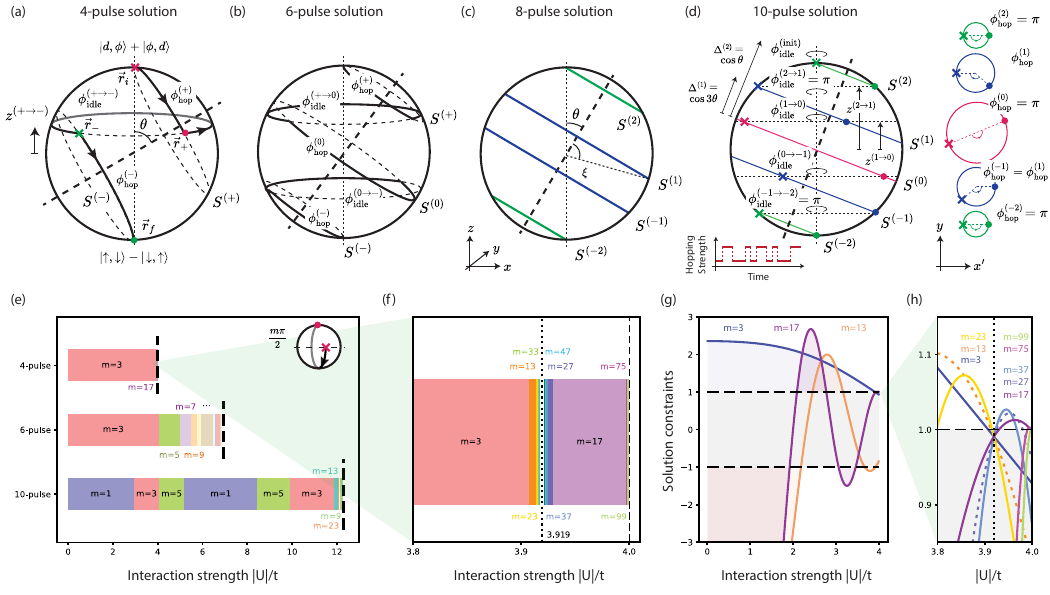}
    \caption{\textbf{Details of hopping pulse sequence solutions.} For larger values of $u\equiv|U|/t$, we require sequences with a larger number of pulses to connect the north pole of the Bloch sphere with the south pole. (a-d) We explore several classes of solutions, 4-, 6-, 8- and 10-pulse solutions. These solutions are parametrized by angles $\phi_\text{idle}$ and $\phi_\text{hop}$ of each arc. We provide solutions as a function of $u$ for the 4-pulse sequence in App.~\ref{app:four_pulse_sequence}, the 6-pulse sequence in App.~\ref{app:6_pulse} and the 8- and 10-pulse sequences in App.~\ref{app:10_pulse}. We specify relevant geometric quantities for each pulse sequence. (d) The particular class of 10-pulse solutions that we pursue fixes certain angles to simple values, which we indicate on the Bloch sphere, as well as on the right column, which shows cross sectors of the tilted arcs, as viewed from along the tilted axis. (e) Each class of solutions has a free parameter $m$ which is an odd integer, describing the amount of over-rotation (inset). Across the range of $u$, we plot the minimum $m$ for which a solution exists. (f) For the 4-pulse solutions, $m=3$ spans the range $u \in [0,3.90709]$. In order to reach the entire feasible range $[0,4]$, larger values of $m$ are required. These values form three intricate infinite series and cover all points except $u = 8\sqrt{6}/5 = 3.919$ and $u = 4$. Similar infinite series are observed for the 6- and 10-pulse sequences, as illustrated in (e). (g) The range of validity of each solution class, and value $m$ is given by regions in which an inequality is satisfied. We illustrate this for the four pulse sequence: the inequality Eq.~\eqref{eq:4-pulse_inequality} is satisfied when the function $\text{cos}\Big(\frac{m\pi}{2}\sqrt{1+\frac{u^2}{16}}\Big)/\Big(\frac{u^2}{16}\Big)$ lies above $+1$ or below $-1$, i.e.~outside of the shaded gray region. We plot this function $m=3,13,17$, which provide the three simplest solutions. (h) We plot these curves in the region $u \in [3.8,4]$ to illustrate the infinite series. For compact presentation, we plot the absolute values of the function and compare it to the value 1 (negative values denoted by dashed curves). The curves in both series $m=3,13,23,...$ and $m=17,27,37,...$ all intersect precisely at the point $u=8\sqrt{6}/5$ and take value below 1, hence there is no solution precisely at this point. We also indicate the curves $m=75,99$, which provide solutions approaching $u\rightarrow 4$.
    }
    \label{fig:n_pulse_solutions}
\end{figure}

For all $n$-pulse solutions, our analysis follows three steps. The pulse sequences are specified by the positions $\Delta_j$ of the arcs $S_j$ along the tilted axis, as well as the heights $z_{j,j'}$ where $S_j$ and $S_{j'}$ are connected vertically [Fig.~\ref{fig:n_pulse_solutions}(a-d)]. The analysis steps are:
\begin{enumerate}
    \item Given heights $z_{j,j'}$ and positions $\Delta_j$, determine the required angles $\phi^{(j)}_\text{hop}$ along the arcs $S_j$ using the Rodrigues formula~\cite{cheng1989historical}. We then match the required evolution time in the 2-particle sector (LHS) with that of the 1-particle sector (RHS). 
    \begin{equation}
       \sum_{j}\phi^{(j)}_\text{hop} =  2\sqrt{1+\frac{u^2}{16}} \frac{m \pi}{2} ,
       \label{eq:hopping_phase_constraint}
    \end{equation}
    where $m$ is an odd integer describing the overrotation in the 1-particle sector. This is the only nontrivial step, at which there may not be a solution. For $n$-pulse sequences with large $n$, there may be many degrees of freedom, and we can arbitrarily fix some of their values (equivalently, fix the angles $\phi^{(j)}$), as we do in App.~\ref{app:n_pulse_sequences}.
    \item Once $\{\phi^{(j)}_\text{hop}\}$ is determined, the entire trajectory is specified and we can immediately determine the idling times $\phi^{(j \rightarrow j')}_\text{idle}$ by computing the angle between the position vectors $\vec{r}^{(j)}_{f}$ and $\vec{r}^{(j')}_{i}$, the final and initial points on the Bloch sphere of the arcs $S_j$ and $S_{j'}$ respectively. 
    \item Finally, we fix the accumulated phase to match the phase accumulated during the magnetic field pulse with an additional idling time $\phi^\text{(init)}_\text{idle}$ at the start of the sequence. This is the only step in which quantum mechanical calculations on $\mathbb{C}^2$ (as opposed to on the sphere $S^2$) is required.
\end{enumerate}

\subsection{Four-pulse sequence}
\label{app:four_pulse_sequence}
Here, we outline in detail the analysis to solve for the pulse parameters in the measurement sequence for $d$-wave pairing. The essential steps have been condensed in the End Matter (EM), and here we give an extended presentation. For fixed values of interaction strength $U$ and tunable hopping strength $t$, the protocol involves four steps: (i) idling for time $T_\text{idle}^{\text{(init)}}$, (ii) hopping for time $T_\text{hop}^{(+)}$, (iii) idling for time $T_\text{idle}^{(+\rightarrow-)}$, and finally (iv) hopping for time $T_\text{hop}^{(-)}$, which we label as $T_{1,2,3,4}$ in the main text. We derive and provide analytic expressions for these times, for $u \equiv |U|/t \leq 4$, and in Appendix~\ref{app:n_pulse_sequences} we discuss necessary modifications to the protocol for larger values of $u$. The pulse duration of each step is determined by the angle of its arcs on the Bloch sphere. We first solve for the angles of the hopping arcs $\phi^{(\pm)}_\text{hop}$, which yields the idling angle $\phi^{(+\rightarrow-)}_\text{idle}$. Finally, the angle $\phi^\text{(init)}_\text{idle}$ of the first idling step is computed to offset the total dynamical phase accumulated. The physical durations of each pulse are related to their respective angles by

\begin{align}
 \phi_\text{idle}/2 = (U/2) T_\text{idle}\\
 \phi_\text{hop}/2 = \sqrt{4t^2 + U^2/4} T_\text{hop}
\end{align}
where the factor of two is due to the spin-1/2 nature of the Bloch sphere.

To summarize our results for this simplest protocol, we have, as plotted in Fig.~\ref{fig:details}(e),
\begin{align}
    \sqrt{16t^2 + U^2} T^{(+)}_\text{hop} &= \frac{3\pi}{2} \sqrt{1+ \frac{u^2}{16} } + \arccos\left[-\frac{u^2}{16} \Big/\cos \left(\frac{n \pi}{2} \sqrt{1+ \frac{u^2}{16} }\right) \right] \label{eq:hopping_time_1}\\
    \sqrt{16t^2 + U^2} T^{(-)}_\text{hop} &= \frac{3\pi}{2} \sqrt{1+ \frac{u^2}{16} } - \arccos \left[-\frac{u^2}{16} \Big/ \cos \left(\frac{n \pi}{2} \sqrt{1+ \frac{u^2}{16}}\right) \right] \label{eq:hopping_time_2}\\
    U T^{(+\rightarrow -)}_\text{idle} &= \arctan\frac{r^{(+)}_xr^{(-)}_y - r^{(+)}_yr^{(-)}_x}{r^{(+)}_xr^{(-)}_x + r^{(+)}_xr^{(-)}_x} \label{eq:idling_time_2}\\
    U T^\text{(init)}_\text{idle} &= \text{arg}\bra{\text{sing}}
    V_{\theta}\left(T^{(-)}_\text{hop}\right)V_{0}\left(T^{(+\rightarrow-)}_\text{idle}\right) V_{\theta}\left(T^{(+)}_\text{hop}\right)
    \ket{d\phi^+} - \frac{3 \pi}{2} \label{eq:idling_time_1}
\end{align}
valid for $0 \leq u \leq 3.90709$. These parameters are respectively obtained from Eqs.~(\ref{eq:matching_hopping_times}, \ref{eq:hopping_times_angles}, \ref{eq:idling_phase_2}, \ref{eq:idling_phase_1}), where $\vec{r}^{(\pm)}$ are defined in Eqs.~(\ref{eq:v1_m}, \ref{eq:v2_m}), and $\ket{\text{sing}}$, $\ket{d\phi^+}$ and the unitaries $V_\theta(T)$ are defined in Eqs.~(\ref{eq:rotation_unitary}, \ref{eq:idling_phase_1}).
\subsubsection{Hopping times}
We solve for the hopping angles $\phi^{(\pm)}_\text{hop}$ by utilizing the Rodriguez rotation formula~\cite{cheng1989historical}. It states that in three dimensions, rotating a vector $\vec{r} \in \mathbb{R}^3$ by an angle $\phi$ about an axis $\hat{w} \in \mathbb{R}^3$ gives the new vector
\begin{equation}
\vec{r}_\text{rot} = \vec{r} \cos \phi + (\hat{w} \times \vec{r}) \sin \phi + \hat{w} (\hat{w} \cdot \vec{r}) (1-\cos \phi)~.
\label{eq:rodrigues_formula}
\end{equation}

The values of $U$ and $t$ fix the rotation axis $\hat{w} = (\sin \theta,  0, \cos \theta)^T = (1,0,U/(4t))^T/\sqrt{1 + U^2/(16t^2)}$. Given this, we parameterize the trajectory on the Bloch sphere in terms of rotation angles $\phi^{(+)}_\text{hop}$ and $\phi^{(-)}_\text{hop}$ about $\hat{w}$. On the Bloch sphere, we have initial and final points $\vec{r}_i = (0,0,1)^T$ and $\vec{r}_f= (0,0,-1)^T$. The rotation by the first hopping step gives
\begin{equation}
\vec{r}^{(+)}  = \begin{pmatrix}
\cos \theta \sin \theta (1- \cos \phi^{(+)}_\text{hop}) \\
-\sin\theta \sin \phi^{(+)}_\text{hop}\\
\cos^2\theta + \sin^2\theta \cos \phi^{(+)}_\text{hop}
\end{pmatrix} = 
\begin{pmatrix}
\frac{u/4}{1 + u^2/16} (1- \cos \phi^{(+)}_\text{hop}) \\
-\frac{1}{\sqrt{1 + u^2/16}} \sin \phi^{(+)}_\text{hop}\\
\frac{u^2/16 +  \cos \phi^{(+)}_\text{hop}}{1 + u^2/16}  
\end{pmatrix}
\label{eq:v1_m}
\end{equation}
This equation is equivalent to the parametric equation of the tilted arc $S^{+}$ (with position $\Delta$, here equal to $\cos \theta$):
\begin{equation}
    \vec{r}(\phi)= \Delta \begin{pmatrix}
        \sin \theta \\ 0 \\ \cos\theta
    \end{pmatrix} + \sqrt{1-\Delta^2} \begin{pmatrix}
        - \cos \phi \cos \theta \\ -\sin \phi  \\ \cos \phi \sin\theta
    \end{pmatrix}
    \label{eq:equation_of_arc}
\end{equation}

Likewise, rotating $\vec{r}_f$ by $-\phi^{(-)}_\text{hop}$ gives
\begin{equation}
\vec{r}^{(-)}  = \begin{pmatrix}
-\cos \theta \sin \theta (1- \cos \phi^{(-)}_\text{hop}) \\
-\sin\theta \sin \phi^{(-)}_\text{hop}\\
- \cos^2\theta + \sin^2\theta \cos \phi^{(-)}_\text{hop}
\end{pmatrix} = 
\begin{pmatrix}
-\frac{u/4}{1 + u^2/4} (1- \cos \phi^{(-)}_\text{hop}) \\
-\frac{1}{\sqrt{1 + u^2/16}} \sin \phi^{(-)}_\text{hop}\\
-\frac{u^2/16 - \cos \phi^{(-)}_\text{hop}}{1 + u^2/16}  
\end{pmatrix}
\label{eq:v2_m}
\end{equation}
Matching the $z$-coordinates of $\vec{r}^{(+)} $ and $\vec{r}^{(-)}$ [the third entries of Eq.~\eqref{eq:v1_m} and \eqref{eq:v2_m}] gives a relation between  $\phi^{(+)}_\text{hop}$ and $ \phi^{(-)}_\text{hop}$, which can be written as $\frac{u^2}{16} = - \cos \left(\frac{\phi^{(+)}_\text{hop}+\phi^{(-)}_\text{hop}}{2}\right) \cos \left(\frac{\phi^{(+)}_\text{hop}-\phi^{(-)}_\text{hop}}{2}\right)$. Finally, the dynamics in the 1-particle sector constrains the total hopping angle 
\begin{equation}
\phi^{(+)}_\text{hop}+\phi^{(-)}_\text{hop} = 2 \sqrt{1+ \frac{u^2}{16}} \frac{m \pi}{2}  \label{eq:matching_hopping_times}
\end{equation}
where $m$ is a free parameter taking odd-integer values, representing the amount of overrotation on the 1-p Bloch  sphere. This gives the equation
\begin{equation}
\frac{u^2}{16} = - \cos \left(\frac{m \pi}{2} \sqrt{1+\frac{u^2}{16} }\right) \cos \left(\frac{\phi^{(+)}_\text{hop}-\phi^{(-)}_\text{hop}}{2}\right) 
\label{eq:hopping_times_angles}
\end{equation}
which allows us to solve for $\phi^{(\pm)}_\text{hop}$ as a function of $u$. For each $m$, Eq.~\eqref{eq:hopping_times_angles} has solutions as long as the following inequality is satisfied
\begin{equation}
    \left|\cos \left(\frac{m \pi}{2} \sqrt{1+\frac{u^2}{16} }\right) \bigg/ \frac{u^2}{16}\right| \geq  1
\label{eq:4-pulse_inequality}
\end{equation}
As mentioned in the main text, $m=1$ does not have any solutions. However, setting $m=3$ yields a solution in the range $0 \leq u \leq 3.90709$, illustrated in Fig.~\ref{fig:n_pulse_solutions}(e,g). 

As a technical point, we may achieve solutions for \textit{almost any} $0\leq u \leq 4$ with different values of $m$ [Fig.~\ref{fig:n_pulse_solutions}(f)]. To illustrate this, we consider the point $u = 4$. There is a valid solution of $\phi^{(+)}_\text{hop}-\phi^{(-)}_\text{hop}$ to Eq.~\eqref{eq:hopping_times_angles} as long as $m(\pi/2) \sqrt{2} \approx n \pi$, where $n$ is an integer. In other words, we seek a rational approximation $m/n\approx \sqrt{2}$, with odd numerator $m$. The sequence of best rational approximations to $\sqrt{2}$ is well studied, and is given by~\cite{oeis_sqrt2_numerators,oeis_sqrt2_denominators}
\begin{equation}
    \frac{1}{1}, \frac{3}{2}, \frac{7}{5},\frac{17}{12}, \frac{41}{29},\frac{99}{70},\frac{239}{169} \cdots \rightarrow \sqrt{2}
\end{equation}
which guides our search for over-rotation integers $m$. For example, while the cases $m=7$ and $m=41$ do not yield new solutions, while $m=17$ extends the range of validity to $3.92979 \leq u \leq 3.99725$, and $m=99$ covers the range $3.99794 \leq u \leq 3.99992$ [Fig.~\ref{fig:n_pulse_solutions}(f)]. While there is no solution at exactly $u = 4$, since we can achieve arbitrarily precise rational approximations of $\sqrt{2}$ (or any other number), we can achieve a solution for almost any value of $u < 4$. 

We say \textit{almost any} value because there are isolated points for which there is no solution. If $\sqrt{1+u^2/16}$ is an irrational number and $u < 4$, $\cos (\frac{m \pi}{2} \sqrt{1+ u^2/16 })$ will densely fill the interval $[-1,1]$ over all odd $m$, and hence there are values of $m$ that satisfy the inequality Eq.~\eqref{eq:hopping_times_angles}. However, there are certain values of $u$ for which there is no solution to Eq.~\eqref{eq:hopping_times_angles}. For example, precisely at the point $u = 4 \sqrt{(7/5)^2 - 1} = \frac{8 \sqrt{6}}{5}\approx  3.919$, the quantity $\cos(\frac{m \pi}{2} \sqrt{1+ u^2/16 })/(u^2/16)$ takes on five distinct values, $\{0, \pm 0.612276, \pm 0.990684\}$, all of which have magnitude smaller than $1$. Around this special point however, the sequences $n=13,23,33,...$ and $n=17,27,37,...$ provides solutions [Fig.~\ref{fig:n_pulse_solutions}(f)]. One can also verify that  $u = 4 \sqrt{(41/29)^2 - 1} = 3.99762$ and $u = 4 \sqrt{(239/169)^2 - 1} = 3.99993$ are similar exceptional points. Finally, we emphasize that these higher over-rotation sequences are likely only of theoretical interest; in practice, they are superseded by simpler protocols with more pulse steps, discussed below in Appendix~\ref{app:n_pulse_sequences}.

\subsubsection{Idling times}
Using our solutions of Eq.~\eqref{eq:hopping_times_angles}, we can immediately solve for the idling angle $\phi^{(+\rightarrow -)}_\text{idle}$ from the vectors $\vec{r}^{(+)}$ and $\vec{r}^{(-)}$ [Eqs.~(\ref{eq:v1_m}, \ref{eq:v2_m})], projected onto the $X-Y$ plane, and using both the dot and cross product to fully resolve the angle $\phi^{(+\rightarrow -)}_\text{idle}$:
\begin{equation}
 \tan \phi^{(+\rightarrow -)}_\text{idle} = \frac{r^{(+)}_xr^{(-)}_y - r^{(-)}_x r^{(+)}_y}{r^{(+)}_xr^{(-)}_x + r^{(+)}_yr^{(-)}_y} 
 \label{eq:idling_phase_2}
\end{equation}

Finally, $\phi^\text{(init)}_\text{idle}$ is obtained by equating the dynamical phases accumulated in the singlet Bloch sphere (by the lattice depth pulse) and in the triplet Bloch sphere (by the MW pulse). These phases are most easily computed by multiplying the unitaries from each step to obtain the total phase accumulated. With $H = \begin{pmatrix}
    U & 2t \\
    2t & 0\\
\end{pmatrix}$, we have eigenvalues $U/2 \pm \lambda$, with $\lambda \equiv \sqrt{U^2/4 + 4 t^2}$, and eigenstates
\begin{equation}
    \ket{E_+} = \begin{pmatrix}
        \cos \frac{\theta}{2}\\
        \sin \frac{\theta}{2}
    \end{pmatrix}~,~\ket{E_-} = \begin{pmatrix}
        -\sin \frac{\theta}{2}\\
        \cos \frac{\theta}{2}
    \end{pmatrix}
\end{equation}
with $\theta \equiv \cot^{-1}(u/4)$ the angle of the rotation axis from the vertical. Exponentiating $H$, we obtain the unitary
\begin{equation}
V_\theta( T) = e^{- i \frac{UT}{2}} \begin{pmatrix}
    \cos(\lambda T/2) - i \sin(\lambda T/2) \cos \theta & - i \sin(\lambda T/2) \sin \theta \\
    - i \sin(\lambda T/2) \sin \theta & \cos (\lambda T/2) + i \sin(\lambda T/2) \cos \theta
\end{pmatrix}
\label{eq:rotation_unitary}
\end{equation}
The phase accumulated during the idling step is exactly $-UT_\text{idle}^\text{(init)}$, therefore
\begin{equation}
    UT_\text{idle}^\text{(init)} = \text{arg}\bra{\text{sing}}V_{\theta}\left(T^{(-)}_\text{hop}\right)V_{0}\left(T^{(+\rightarrow -)}_\text{idle}\right) V_{\theta}\left(T^{(+)}_\text{hop}\right)
    \ket{d\phi^+} - \frac{3 \pi}{2}
    \label{eq:idling_phase_1}
\end{equation}
where we represent the states $\ket{\text{sing}} \equiv (0,1)^T$ and $\ket{d\phi^+} \equiv (1,0)^T$ on the Bloch sphere, and $3\pi/2$ is the phase accumulated by the field step [as can be verified by setting $\theta = \lambda T = \pi/2$ in Eq.~\eqref{eq:rotation_unitary}].

\subsection{$n$-pulse sequences}
\label{app:n_pulse_sequences}
Here, we describe pulse sequences with larger number of steps, which allow for the measurement of the superconducting order parameter at larger values of $u \equiv |U|/t$. As established above, our pulse sequence in its simplest incarnation is valid for $0 \leq u \leq 3.90709$ and may be extended up to $u < 4$. Beyond $u=4$, the protocol should be modified, by introducing $n$ tilted arcs connected by $n-1$ vertical arcs. Below, we examine the cases $n=3,4,5$, which we refer to as 6-, 8- and 10-pulse sequences. These pulse sequences have more degrees of freedom and we therefore restrict ourselves to solutions with a high degree of symmetry [Fig.~\ref{fig:n_pulse_solutions}(b,c,d)], which simplifies our analysis below.

\subsubsection{6-pulse sequence}
\label{app:6_pulse}
We first discuss a 6-pulse sequence: comprising 3 tilted arcs connected by 2 vertical arcs [Fig.~\ref{fig:n_pulse_solutions}(b)]. Such sequences are potentially valid up to $u = 4\sqrt{3} \approx 6.928$, above which the three tilted arcs cannot be made to vertically overlap. Our simplest two solutions are valid for $u \leq 4.039$ and $u \leq 5.016$ respectively, and increasing the amount of over-rotation again yields solutions for the entire interval $0 \leq u < 4\sqrt{3}$. Although the space of possible sequences is larger, here we restrict ourselves to a class of symmetric pulses which are simpler to analyze, surprisingly even simpler than the four-pulse sequence discussed in the main text. 

We consider three tilted arcs. The first and third arcs, $S^{(+)}$ and $S^{(-)}$, are fixed to intersect the north and south poles of the Bloch sphere respectively, but we have freedom to place the middle arc $S^{(0)}$ anywhere in between. For simplicity, we choose to place $S^{(0)}$ on the `tilted' equator [Fig.~\ref{fig:n_pulse_solutions}(b)].

As in the four-pulse sequence, our solution consists of angles $\phi^{(\pm/0)}_\text{hop}$ on the tilted arcs, which can be used to determine the idling angles $\phi^{(+\rightarrow 0/ 0 \rightarrow -)}_\text{idle}$. We make the following crucial simplifying assumptions: we choose a symmetric solution $\phi^{(+)}_\text{hop} = \phi^{(-)}_\text{hop}$, and we set $\phi^{(0)}_\text{hop} = \pi$ [Fig.~\ref{fig:n_pulse_solutions}(b)]. In analogy to Eq.~\eqref{eq:matching_hopping_times}, matching the required hopping times between 1- and 2-particle sectors gives the equation
\begin{equation}
    2 \phi^{(+)}_\text{hop} + \pi = m \pi \sqrt{1+ \frac{u^2}{16}} 
    \label{eq:matching_times_six_pulse}
\end{equation}
for odd integer $m$, which directly allows us to solve for $\phi^{(+)}_\text{hop}$. In order to check the validity of the solution, we also note that for a given $u$, only a range of angles $\phi^{(+)}_\text{hop}$ are valid, in the sense that they vertically overlap with the central tilted arc $S^{(0)}$. The value $\phi^{(+)}_\text{hop}$ satisfies
\begin{equation}
    \cos \phi^{(+)}_\text{hop} \leq \sqrt{1+ \frac{u^2}{16 }} - \frac{u^2}{16} 
\end{equation}

As with Eq.~\eqref{eq:hopping_times_angles}, there is no solution for $m=1$. The case $m=3$ admits solutions for $0 \leq u \leq 4.039$, and $m=5$ includes the interval $1.04737 \leq u \leq 5.016$. As with the four-pulse sequences, solutions across the entire range $0 \leq u \leq 4 \sqrt{3}$ exist for sufficiently large $m$. In this case, however, every larger value of $m$ monotonically increases the feasible range, up to the point $u = 4 \sqrt{3}$ [Fig.~\ref{fig:n_pulse_solutions}(e)]. Precisely at this point, there is no solution, but unlike the four-pulse case, there are no other exceptional points within the interval $u \in [0,4\sqrt{3})$ for which no solutions exist.

\subsubsection{8- and 10-pulse sequences}
\label{app:10_pulse}
We next move on to discuss higher-pulse sequences, with the aim of providing a concrete sequence in the regime $u\equiv |U|/t \approx 10$. To do so, we examine the feasibility regions of 8- and 10-pulse sequences. We first examine its `feasibility regions', i.e.~the maximum $u$ for which these sequences are possible. As above, we restrict ourselves to solutions where the tilted arcs are symmetric about the equator. In the 8- and 10-pulse cases respectively, we consider symmetrically-placed arcs $S^{(\pm 2)}$ and $S^{(\pm 1)}$ [Fig.~\ref{fig:n_pulse_solutions}(c)]. In the 10-pulse case, there is an additional middle arc $S^{(0)}$ which we fix to be on the tilted equator [Fig.~\ref{fig:n_pulse_solutions}(d)].

In both case, the locations of arcs $S^{(\pm 1)}$ and $S^{(\pm 2)}$ are specified by two numbers: $\Delta^{(1)}$ and $\Delta^{(2)}$, i.e.~their heights along the tilted axis [Fig.~\ref{fig:n_pulse_solutions}(d)]. $\Delta^{(2)}$ is fixed by demanding that it intersects with the north pole of the Bloch sphere, this gives $\Delta^{(2)} = \cos\theta$. We have two constraints on $\Delta^{(1)}$: (i) that $S^{(2)}$ and $S^{(1)}$ overlap vertically at height $z^{(2\rightarrow 1)}$, and that (ii) $S^{(1)}$ and $S^{(-1)}$ overlap at height $z^{(1\rightarrow -1)}= 0$, or $S^{(1)}$ and $S^{(0)}$ overlap vertically at height $z^{(1\rightarrow 0)}$ in the 8- and 10-pulse cases respectively. With $\theta \equiv \cot^{-1}(u/4)$, constraint (i) can be written as
\begin{equation}
    \cos 2\theta \leq z^{(2\rightarrow 1)} \leq \cos(\xi - \theta) = \Delta^{(1)} \cos \theta + \sqrt{1-(\Delta^{(1)})^2} \sin \theta \Rightarrow \Delta^{(1)} \geq \cos 3\theta 
\end{equation}
where $\xi = \cos^{-1} \Delta^{(1)}$ is the conical angle of $S^{(1)}$ with the tilt axis [Fig.~\ref{fig:n_pulse_solutions}(c)]. Meanwhile, constraint (ii) can be written as
\begin{align}
    \cos(\xi + \theta) \leq z^{(1\rightarrow -1)} = 0 \Rightarrow \cos 3\theta 
 \leq \Delta^{(1)} \leq \sin \theta & \text{ (8-pulse)}\\
    \cos(\xi + \theta) \leq z^{(1\rightarrow 0)} \leq \sin \theta \Rightarrow \cos 3\theta \leq \Delta^{(1)} \leq \sin 2\theta  & \text{ (10-pulse)} \label{eq:10_pulse_delta1_constraint}
\end{align}

Combining these constraints, we find that 8- and 10-pulse solutions could exist for $u \leq 4(1+\sqrt{2}) \approx 9.657$ and $u \leq 4 \cot \frac{\pi}{10} \approx 12.311$ respectively. Since we seek solutions for the regime $u\approx 10$, we focus on the 10-pulse sequence in the rest of this section, following the steps outlined in App.~\ref{app:solution_strategy}. We have three free parameters in the 10-pulse solution: the vertical heights $z^{(2\rightarrow 1)}$ and $z^{(1\rightarrow 0)}$ where the arcs intersect, and the position of $S^{(1)}$ along the tilt axis, $\Delta^{(1)}$. For simplicity, we fix $\Delta^{(1)} = \cos~3\theta$, i.e.~at the lower end of the range specified by Eq.~\eqref{eq:10_pulse_delta1_constraint}. This fixes $\phi^{(+2)}_\text{hop} = \pi$ and $\phi^{(2\rightarrow 1)}_\text{idle} = \pi$, since it connects the lowest point of $S^{(2)}$ with the highest point of $S^{(1)}$. Additionally, we design the path symmetrically such that the paths on $S^{(\pm 1)}$ are the same, as are those on $S^{(\pm 2)}$, and that $\phi^{(0)}_\text{hop} = \pi$, illustrated in Fig.~\ref{fig:n_pulse_solutions}(d). This leaves $\phi^{(+1)}_\text{hop}$ as the only free parameter, which is in turn fixed by matching the 1- and 2-particle sectors:
\begin{equation}
    2 \phi^{(+1)}_\text{hop} + 3\pi = m \pi \sqrt{1+ \frac{u^2}{16} } 
    \label{eq:matching_times_ten_pulse}
\end{equation}
for $m$ an odd integer. There is an additional constraint that there be a path from the endpoint on $S^{(1)}$ to $S^{(0)}$, which is equivalent to
\begin{equation}
    -\sin \theta  \leq \Delta^{(1)} \cos \theta + \sqrt{1-(\Delta^{(1)})^2} \sin \theta \cos \phi^{(+1)}_\text{hop} \leq \sin \theta 
    \label{eq:ten_pulse_ineq}
\end{equation}
with $\Delta^{(1)} = \cos 3\theta$, $\cot \theta = u/4$, and $\phi^{(+1)}_\text{hop}$ given by Eq.~\eqref{eq:matching_times_ten_pulse}. For $m=1$, Eq.~\eqref{eq:ten_pulse_ineq} is satisfied for $u \in [0,2.94602] \cup [5.19039,  8.44756]$, for $m=3$, $u \in [0, 4.05348] \cup [6.08537, 7.73347] \cup [9.89222, 11.8799]$, and for $m=5$, $u \in [0, 5.39352] \cup [6.35307, 7.48752] \cup [8.42478, 10.3614]$, as illustrated in Fig.~\ref{fig:n_pulse_solutions}(e). As with the four- and six-pulse cases, with arbitrarily large $m$ one can cover (almost every point in) the entire feasible range $0 \leq u < 4 \cot \frac{\pi}{10}$.

\section{Optimal sample complexity for estimating observable expectation values}
Here, we prove that to estimate the expectation value $\langle O \rangle \equiv \text{tr}(\rho O)$ of an observable $O$, the optimal measurement one can perform is the approach we take in this work: to measure in its eigenbasis $\{|o_j\rangle\langle o_j|\}$.
While this appears to be a fairly basic fact in quantum estimation theory, we have not found it elsewhere in the literature, hence we provide a self-contained proof here.

We consider the following setting: given a Hermitian observable $O$, consider the positive operator-valued measure (POVM) $\{M_a\}$ consisting of elements $M_a$. A finite number of measurements with this POVM on the state $\rho$ yields a noisy estimate $\langle \tilde{O} \rangle \approx \langle O \rangle \equiv \text{tr}(\rho O)$. We ask about the sample complexity of estimating $\langle O \rangle$, i.e.~ the variance in the estimates $\text{var}(\langle \tilde{O}\rangle)$ as a function of the number of samples $n$. We show that the POVM consisting of projectors onto the eigenbasis of $O$, $\{|o_j\rangle\langle o_j|\}$ yields an unbiased estimator of $\langle O\rangle$ with the smallest sample complexity over all POVMs $\{M_a\}$ which can estimate $\langle O\rangle$.

We analyze the sample complexity of estimating $\langle O\rangle$ with an arbitrary POVM $\{M_a\}$, i.e.~where $M_a \succeq 0$ are positive semi-definite Hermitian operators that sum to the identity, $\sum_a M_a = \mathbb{I}$. From $n$ total measurements with this POVM, we obtain counts $\hat{n}_a$, i.e.~the number of times each element $M_a$ is observed, satisfying $\sum_a\hat{n}_a = n$. The counts $\{\hat{n}_a\}$ are weakly correlated random integers following the \textit{multinomial distribution}~\cite{bickel2015mathematical}
\begin{equation}
    \{\hat{n}_a\} \sim \text{Multinomial}(\{\text{tr}(\rho M_1), \text{tr}(\rho M_2), \dots\};n).
\end{equation}
The expectation value $\tr(O \rho)$ can be estimated from our POVM for all $\rho$, if and only if $O$ can be expressed as a linear combination of the elements $M_a$. That is, there are real coefficients $c_a$ such that $O = \sum_a c_a M_a$. Then, the estimate of $\langle O \rangle$ is
\begin{equation}
    \langle \tilde{O} \rangle = \sum_a c_a \frac{\hat{n}_a}{n},
    \label{eq:O_estimator_POVM}
\end{equation}
As an example, for the POVM $\{|o_j\rangle \langle o_j|\}$ (here, we reserve the index $j$ to label the eigenbasis), we have $c_j = \lambda_j$, the eigenvalue $O|o_j\rangle = \lambda_j |o_j\rangle$. Here, as expected, our estimator is $\langle \tilde{O} \rangle = \sum_j \lambda_j \frac{\hat{n}_j}{n}$.

Using the fact that for multinomial distributions $\mathbb{E}[\hat{n}_a] = n\text{tr}(\rho M_a)$, we can verify that Eq.~\eqref{eq:O_estimator_POVM} is an unbiased estimator of $\langle O\rangle$. Using the covariances of the multinomial distribution $\text{cov}(\hat{n}_a,\hat{n}_b) = n\left[\text{tr}(\rho M_a)\delta_{a,b} - \text{tr}(\rho M_a)\text{tr}(\rho M_b)\right]$, the uncertainty in our estimate $\langle\tilde{O}\rangle$ is given by:
\begin{align}
    \text{var}(\langle\tilde{O}\rangle) = \mathbb{E}[\langle \tilde{O}\rangle^2] -\mathbb{E}[\langle \tilde{O}\rangle]^2 &= \sum_{a,b} c_a c_b \frac{\text{cov}(\hat{n}_a,\hat{n}_b)}{n^2}\\
    & = \frac{1}{n}\left(\sum_a c_a^2 \text{tr}(\rho M_a)  - \langle O \rangle^2\right) \label{eq:sample_complexity_POVM}
\end{align}

We can immediately see that the POVM $\{|o_j\rangle\langle o_j|\}$ has a sample complexity equal to the variance of $O$ on the state $\rho$: 
\begin{equation}
\text{var}_\text{eigenbasis}(\langle \tilde{O}\rangle) = \frac{1}{n}\left[\text{tr}(\rho O^2) - \langle O\rangle^2\right] \equiv \frac{1}{n}\left(\langle O^2 \rangle - \langle O \rangle^2\right). \label{eq:eigenbasis_sample_complexity}   
\end{equation}
We now prove that Eq.~\eqref{eq:eigenbasis_sample_complexity}  is a lower bound for Eq.~\eqref{eq:sample_complexity_POVM}, i.e. projectors on the eigenbasis have the optimal sample complexity over \textit{all} POVMs that can estimate $\langle O\rangle$. The positive semi-definiteness of $M_a$ allows us to write $M_a = K^\dagger_a K_a$ in terms of possible rectangular matrices $K_a$. It suffices to show that for every pure state $\rho = |\psi\rangle\langle \psi|$, 
\begin{align}
    \sum_a c_a^2 \langle \psi| M_a |\psi\rangle \geq \langle \psi|O^2 |\psi \rangle~.
    \label{eq:sample_complexity_inequality}
\end{align}
To do so, we use the Cauchy-Schwarz inequality on the left-hand-side, noting that for \textit{any} normalized state $|\Phi\rangle$, 
\begin{align}
    \sum_a c_a^2 \langle \psi| M_a |\psi\rangle = \Big(\sum_b \Vert K_b |\Phi\rangle\Vert^2 \Big)\Big(\sum_a c_a^2 \Vert K_a |\psi\rangle \Vert^2\Big) \geq \Big\vert \sum_a \langle \Phi | K^\dagger_a c_a K^{}_a |\psi\rangle \Big|^2 = \big|\langle \Phi | O |\psi \rangle\big|^2, 
    \label{eq:AppC_Cauchy_Schwarz}
\end{align}
where since $\Vert|\Phi\rangle\Vert = 1$, we have $\sum_b \Vert K_b |\Phi\rangle\Vert^2 = 1$. Finally, setting $|\Phi\rangle = O|\psi\rangle/\sqrt{\langle \psi|O^2|\psi\rangle}$ gives $\langle \psi |O^2 |\psi \rangle$ for the right hand side of Eq.~\eqref{eq:AppC_Cauchy_Schwarz}, equal to the desired inequality Eq.~\eqref{eq:sample_complexity_inequality}.

\section{Optimization for pulse robustness with direct collocation}

\begin{figure}[t]    \includegraphics[width=0.99\linewidth]{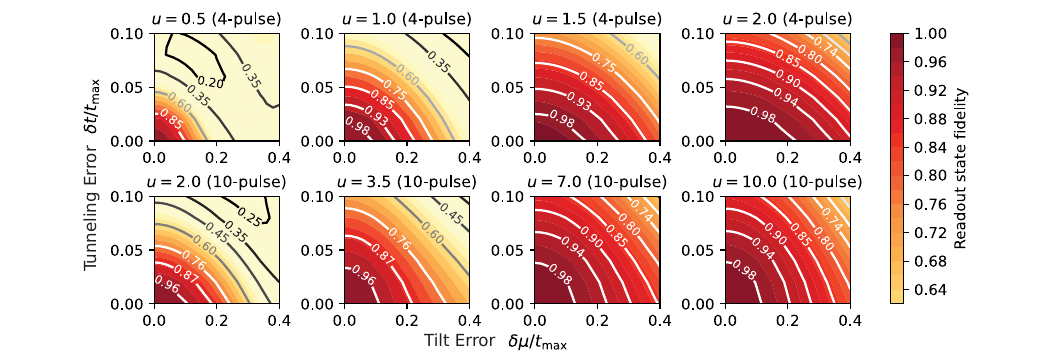}
    \caption{\textbf{Robustness of analytical pulses}. Contour plots of the state fidelity with respect to tilt error $\delta \mu$ and hopping error $\delta t$ for the analytical pulses at different $u\equiv|U|/t$. The first row shows the robustness of analytical four-pulse solutions, while the second row shows the robustness of analytical ten-pulse pulse solutions (note different values of $u$ between rows).}
    \label{app_fig:robustness_analytical}
\end{figure}
In this section, we describe our robust pulse optimization procedure using the
quantum optimal control method of direct collocation. We first study the robustness of our analytical pulses under the tilt error $\delta \mu$ and hopping error $\delta t$, for different $u\equiv |U|/t$ values. In Appendix \ref{app:dwave_hopping_pulse}, we discussed $n$-pulse analytical sequences valid for different values of $|U|/t$. In Fig.~\ref{app_fig:robustness_analytical}, we show the contour plots of the state fidelity for 4-pulse and 10-pulse solutions at different values of $|U|/t_{\mathrm{max}}$. As one might expect, sequences with fewer pulses are more robust at the same value of $|U|/t_{\mathrm{max}}$. 
In Fig.~\ref{fig:RobustControl}, we chose to work in the regime $|U|/t\approx 10$, relevant for typical experiments,
and we demonstrated how the analytical 10-pulse solution can be significantly optimized.
Here, we describe how we perform this optimization with quantum optimal control techniques.

The quantum control problem can be formulated as an optimization problem in the time-dependent state trajectories subject to the Schr{\"o}dinger equation. The unitary propagator $V$ evolves as
\eqs{
\frac{\partial V}{\partial T}=-iH[\boldsymbol{a}(T)]V, 
}
where $\boldsymbol{a}(T)$ are trainable pulse parameters. Similar to most quantum optimal control approaches, we set the Hamiltonian to be $H[\boldsymbol{a}(T)]=H_0 +\sum_i a_i(T)H_i$,
where $H_0$ is the \emph{drift} term that cannot be tuned, and $H_i$ are controllable Hamiltonians whose strengths can be tuned via $a_i(T)$. In our setup, the drift term would be the Hubbard interaction term $H_\text{Int}$, and tunable terms are the hopping and magnetic field terms. Since the magnetic field commutes with the rest of the Hamiltonian, we do not need to analyze it in our problem, and we only focus on the control of the hopping Hamiltonian $H_\text{Hop}$ with strength $t(T)$
\eqs{
\dfrac{\partial V}{\partial T}=-i[H_\text{Int}+t(T)H_\text{Hop}]V.
}
We discretize the time interval $[0,T]$ into $N$ points separated by intervals $\Delta T$, and we use $V_k$ and $t_k$ to denote the unitary and hopping strength at the $k$-th time point. Then the quantum optimal control problem can be formulated as the following optimization problem \cite{
directcollocation,PhysRevApplied.17.014036}:
\eqs{
\underset{t_1:t_{N-1}}{\mathrm{min}}~& \ell[V_N(t_1:t_{N-1})], \\
\mathrm{with}~~&V_N(t_1:t_{N-1})=\prod_{k=1}^{N-1}\mathrm{exp}\left[-i H(t_k)\Delta T\right]
}
where $\ell(\cdot)$ is the loss function that depends on the goal of quantum controls. For our purposes, we use the state infidelity as our loss function: 
\begin{equation}
\ell[V_N(t_1:t_{N-1})]=1-\frac{1}{S}\sum_{s=1}^{S}|\langle \psi_s^{(f)}|V_N(t_1:t_{N-1})|\psi_s^{(i)}\rangle|^2,
\end{equation}
where $|\psi_s^{(i)}\rangle$ (initial states) are the nontrivial eigenstates of observables $\widetilde{O}$, and $|\psi_s^{(f)}\rangle$ (final states) are the readout states. Since the 1-particle and 3-particle sector are dual to each other, we only need to keep the 4 nontrivial eigenstates from the 1-particle sector and 2 nontrivial eigenstates from the 2-particle sector, i.e. $S=6$. Unlike gradient-based quantum optimal control methods, direct collocation treats the unitary $V_k$ and pulse parameters $t_k$ as separate variables, related by the Schr{\"o}dinger equation as a constraint. Therefore, one can rewrite the above optimization problem to 
\eqs{
\underset{t_1:t_{N-1}}{\mathrm{minimize}}~~\ell[V_N(t_1:t_{N-1}&)] ,\\
\text{subject to }V_{k+1}-\mathrm{exp}\left[-i H(t_k)\Delta T\right]V_k&=0,\\
V_1&=I_{16},\\
0\leq t_k&\leq t_{\mathrm{max}},\\
\left|\frac{t_{k+1}-t_{k}}{\Delta T}\right|&\leq C_1,\\
\text{and }\left|\frac{t_{k+1}-2t_k+t_{k-1}}{(\Delta T)^2}\right|&\leq C_2,
}
where the last two constraints are optional but enforce the smoothness of the pulse. This is in the form of a typical nonlinear programming (NLP) problem:
\eqs{
\underset{\boldsymbol{x}\in \mathbb{R}^n}{\mathrm{minimize}}~&f(x)
\\
\text{subject to }&g_L\leq g(x)\leq g_U\\
\text{and }&x_L\leq x\leq x_U.
}

There are different methods for solving such an NLP. Since most optimization solvers only support real numbers, one needs to first map the complex-valued quantum optimization into a real-valued variable optimization. This can be done by using the isomorphic representation of quantum states and Hamiltonians \cite{PhysRevA.95.042318}:
\eqs{
\widetilde{\psi}=\begin{pmatrix}
    \Re \psi\\
    \Im \psi
\end{pmatrix}\text{, }\widetilde{G}\equiv-i\widetilde{H}\equiv\begin{pmatrix}
    \Im H & \Re H\\
    -\Re H & \Im H
\end{pmatrix},
} 
so that $\partial \widetilde{\psi}/\partial t=\widetilde{G}\widetilde{\psi}$. In particular, we use the interior-point optimizer (IPOPT) \cite{nocedal2009adaptive,doi:10.1137/S1052623403426544,doi:10.1137/S1052623497325107} for solving such an NLP, which has been used in recent quantum optimal control literature \cite{directcollocation} and is available open-source as \verb|QuantumCollocation.jl|~\cite{QuantumCollocationJL}. 

Enforcing dynamical equations as constraints in the optimization procedure is a key idea in direct collocation; it enables the solver to explore \emph{unphysical} regions during the optimization before converging to an optimal physical trajectory. Compared to gradient-based pulse optimization methods such as gradient ascent pulse engineering (GRAPE), it has several advantages: (i) it has better convergence by utilizing unphysical solutions during optimization, (ii) it is easy to add different types of constraints on the pulse shape (e.g.~maximum value and derivatives), and (iii) it is compatible with NLP solvers that are commonly used and highly optimized.

\begin{figure}[t]
    \includegraphics[width=0.95\linewidth]{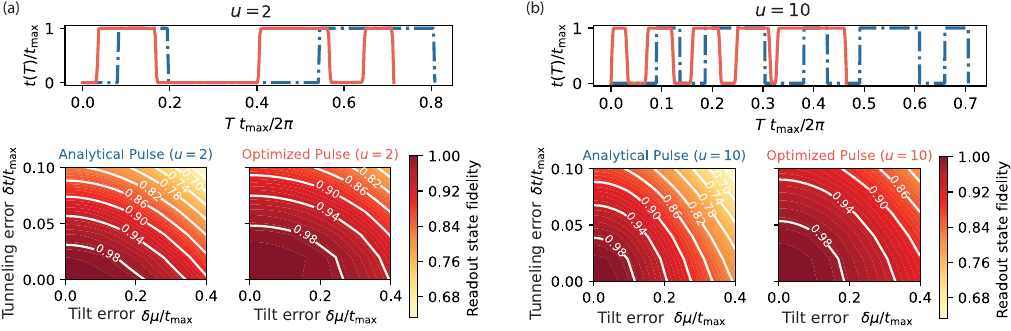}
    \caption{\textbf{Fidelity of analytic vs.~robust pulses}. (a) The state fidelity of the analytic pulse (blue dashed line) and optimized pulse (red line) at $|U|/t_{\mathrm{max}}=2$. (b) The state fidelity of the analytical pulse (blue dashed line) and optimized pulse (red line) at $|U|/t_{\mathrm{max}}=10$.}
    \label{app_fig:robustness_pulse}
\end{figure}

In our optimization, we fix the Hubbard interaction $U$ and treat the hopping strengths $t(T)$ as trainable parameters, where $T\in[0,T_{\mathrm{max}}]$ is time. We allow the hopping strength to take values between 0 and a maximum value $t_{\max}$. We want the optimal pulse $t(T)$ to be robust against tilt error $\delta \mu$ and hopping error $\delta t$. During the optimization, we use the analytical solution as a warm start, and we set the maximum $t_{\max}$ to be slightly larger than the height of the warm-start pulse. To find robust pulses, we use the idea of error sampling~\cite{RobustNMR,RobustLearning}. In this approach, the system with an error can be modeled as $H=H_\text{Int}+ [t(T)+\delta t] H_\text{Hop}+\delta \mu H_\text{Tilt}$, where $H_\text{Int}$ is the Hubbard interaction, $H_\text{Hop}$ is the hopping Hamiltonian, $H_\text{Tilt}$ is the tilt chemical potential, and $\delta \mu$, $\delta t$ are the error strengths. Even though the error strength can be stochastic, it has been shown that if the optimal pulse is robust to a range of error strengths, it is also robust to stochastic, potentially time-dependent, errors \cite{2022arXiv220814193K}. Therefore, we keep $\delta \mu$ and $\delta t$ constant during the unitary evolution in our optimization. We sample $\delta \mu^{(m)}$ and $\delta t^{(m)}$ uniformly from a range of values, $\delta \mu \in [-\delta \mu_{\max}, \delta \mu_{\max}]$ and $\delta t \in [-\delta t_{\max}, \delta t_{\max}]$, and use the average state infidelity as the loss function:
\eqs{
\ell(V[t(T)])  = 1-\frac{1}{MS}\sum_{m=1}^{M}\sum_{s=1}^{S}|\langle\psi_{s}^{(f)}|V[t(T);\delta t^{(m)}, \delta \mu^{(m)}]|\psi_s^{(i)}\rangle|^2,
}
where $m$ indexes the sampled error strengths (discretized to $M$ values). 
During our optimization, we first set the constraints on the first and second order derivatives of $t(T)$ to be large or infinity such that the solver converges to a robust pulse. Then we gradually enforce the first and second order derivatives constraints without jeopardizing the averaged state fidelity such that the optimal pulse is smoother but also robust. In Fig.~\ref{app_fig:robustness_pulse}, we plot two optimized pulses for $|U|/t_{\mathrm{max}}=2$ (starting from the four-pulse solution) and $|U|/t_{\mathrm{max}}=10$ (starting from the 10-pulse solution). In both cases, we see marked improvements in robustness compared to our analytical pulses; the optimized pulses we find are also faster and hence may be more robust to other types of error such as decoherence.

\section{Universal experimental control set}
\label{app:universal_control_set}
\begin{figure}[t]
    \centering
    \includegraphics[width=\linewidth]{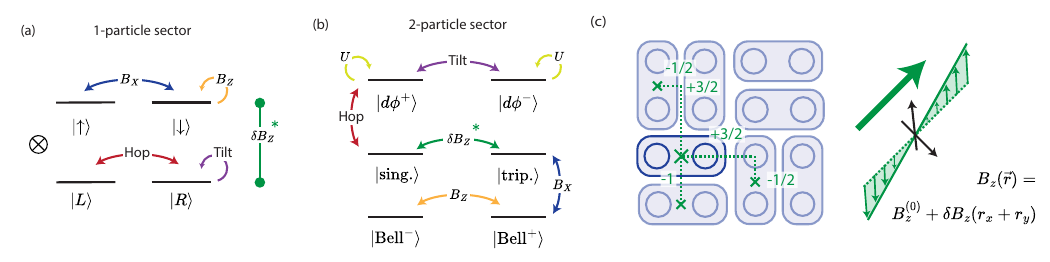}
    \caption{\textbf{Universal control set.} We identify five global controls that allow for universal control of the states in each particle-number sector: hopping, global lattice tilt, MW drive $B_X$, $Z$-magnetic field (or optical detuning) $B_Z$, and a global gradient in the $Z$-magnetic field $\delta B_Z$ (or, gradient in the optical detuning field). (a,b) We denote the action of each control on the 1-p (or 3-p) sector and the 2-p sector. Remarkably, our controls are selective: each form of control only couples certain pairs of states, yet together offer control over the entire Hilbert space of each number sector. (a) In the 1- or 3-p sector, the global $X-$ and $Z-$magnetic fields only act on the spin degrees of freedom, while hopping and tilt act on the positional degrees of freedom. Meanwhile, the magnetic field gradient serves as an entangling Ising-like interaction between spin and position (which we indicate schematically with a $CZ$ gate symbol). (b) In the 2-p sector, each control only couples two states and does not act on other states, in the basis defined in Eq.~(\ref{eq:control_basis_1}-\ref{eq:control_basis_3}). This enables our design of pulse sequences in Appendix~\ref{app:further_pulse_sequences} that measure a variety of quantities. An exception is the global magnetic field gradient, which we denote with an asterisk: a global field gradient also introduces a different value of average $Z$-magnetic field on each dimer and also couples $|\text{Bell}^\pm\rangle$. When the field gradient is used to effect a $\pi-$pulse between $|\text{sing}/\text{trip}\rangle$, this coupling vanishes. (c) This property of the field gradient holds when we choose the global magnetic field gradient to lie equally between the $X$- and $Y$-axes. As discussed in Appendix~\ref{app:B_field_gradient}, with this gradient, for any dimerization of the square lattice, all dimers will experience the same local gradient $\delta B_Z$, and their average $Z$-magnetic fields will shift by integer multiples of $\delta B_Z$, equal to the Manhattan distances between dimers: here we indicate the Manhattan distances between a central dimer (thick outline) and other dimers. 
    } \label{fig:universal_control_app}
\end{figure}

In Appendix~\ref{app:dwave_hopping_pulse}, we designed our first nontrivial pulse sequence \textbf{Hop+U$^1$} using a combination of hopping and idling. These sequences can be understood by their effects on the Hilbert space of each particle number sector. On the 1-p sector, hopping performs an $X-$axis rotation on the positional Bloch sphere and idling has no effect. Meanwhile, on the 2-p sector, hopping and idling enable arbitrary control on the $\{|d,\phi\rangle+|\phi,d\rangle, |\!\uparrow,\downarrow\rangle - |\!\downarrow,\uparrow\rangle\}$ two-level system. It is natural to ask whether the entire Hilbert space of each particle number sector can be manipulated with other types of global control. Here, we analyze a set of global experimental controls, acting on dimers, which together offer full control over the Hilbert space of each particle number sector. Surprisingly, only five experimental controls suffice, in conjunction with a Hubbard interaction $U$ that we assume is always fixed. Furthermore, each type of control selectively couples certain states in Hilbert space, illustrated in Fig.~\ref{fig:universal_control_app}(a,b), which allows us to design pulse sequences that effect desired transformations. In this section, we analyze this universal control set and prove its universality.

\subsection{Universal control set}
Our universal control set comprises of:
\begin{enumerate}
    \item Hopping strength $t$, $H_\text{hop} \equiv -t \sum_{\sigma\in \{\uparrow,\downarrow\}} \left(c^\dagger_{i,\sigma}c^{}_{i',\sigma} + \text{h.c.}\right)$
    \item Dimer tilt $\delta\mu$, $H_\text{tilt} \equiv \frac{\delta\mu}{2} \sum_{\sigma\in \{\uparrow,\downarrow\}} \left(\hat{n}_{i,\sigma} - \hat{n}_{i',\sigma}\right)$
    \item Global MW field $B_X$,  $H_{B_X} \equiv \bar{B}_X \sum_{r\in \{i,i'\}} \left(c^\dagger_{r,\uparrow}c^{}_{r,\downarrow} + \text{h.c.}\right)$
    \item Global $Z$-magnetic (or optical detuning) field $B_Z$, $H_{B_Z} \equiv \bar{B}_Z \sum_{r\in \{i,i'\}} \left(\hat{n}_{r,\uparrow} - \hat{n}_{r,\downarrow}\right)$
    \item Global $Z$-magnetic field gradient $\delta B_Z$, $H_{\delta B_Z} \equiv \frac{\delta B_Z}{2} \left(\hat{n}_{i,\uparrow} - \hat{n}_{i,\downarrow} -\hat{n}_{i',\downarrow} + \hat{n}_{i',\uparrow}\right)$
\end{enumerate}
where $\delta\mu \equiv \mu_i-\mu_{i'}$, $\delta B_Z \equiv B_{Z,i} - B_{Z,i'}$ are the differences in chemical potential and magnetic field respectively, and $\bar{B}_X \equiv (B_{X,i} +B_{X,i'})/2$, $\bar{B}_Z \equiv (B_{Z,i} +B_{Z,i'})/2$ are the average $X-$ and $Z-$fields experienced by the dimer, and we explicitly write the Hamiltonian of each control acting on the dimer sites $i$ and $i'$.

In Fig.~\ref{fig:universal_control_app}(a,b), we list the actions of these controls on the 1-p and 2-p Hilbert space. On the 1-p Hilbert space, the fields $\bar{B}_X$ and $\bar{B}_Z$ naturally act on the spin two-level system, while hopping and tilt are the analogues of $\bar{B}_X$ and $\bar{B}_Z$ on the positional two-level system. Meanwhile, the field gradient $\delta B_Z$ generates \textit{entangling} $ZZ$-type interactions between the spin and position degrees of freedom: a gradient of the MW field strength would also entangle spin and position: we do not list this here because it can be generated from 3., 4. and 5.

In detail, the Hamiltonian on the dimer, with all controls included, has the following matrix elements in the 1-p, 2-p, and 3-p sectors:
\begin{align}
    H^\text{(1-p)} &= \begin{pmatrix}
        \bar{B}_Z+\frac{1}{2}(\delta\mu+\delta B_Z) & \bar{B}_X & -t & 0\\
        \bar{B}_X & -\bar{B}_Z+\frac{1}{2}(\delta\mu-\delta B_Z) & 0 & -t\\
        -t & 0 & \bar{B}_Z +\frac{1}{2}(-\delta\mu-\delta B_Z) & \bar{B}_X\\
        0 & -t & \bar{B}_X & -\bar{B}_Z + \frac{1}{2}(-\delta\mu+\delta B_Z)\\
    \end{pmatrix}\\
   H^\text{(2-p)} &= \begin{pmatrix}
        U+\delta\mu & 0 & -t & t & 0 & 0\\
        0 & 2\bar{B}_Z & \bar{B}_X & \bar{B}_X & 0 & 0\\
        -t & \bar{B}_X & \delta B_Z & 0 & \bar{B}_X & -t\\
        t & \bar{B}_X & 0 & -\delta B_Z & \bar{B}_X & t\\
        0 & 0 & \bar{B}_X & \bar{B}_X & -2\bar{B}_Z & 0\\
        0 & 0 & -t & t & 0 & U-\delta\mu\\
    \end{pmatrix}\\  
    H^\text{(3-p)} &= \begin{pmatrix}
        U+\bar{B}_Z+\frac{1}{2}(-\delta\mu+\delta B_Z) & \bar{B}_X & t & 0\\
        \bar{B}_X & U-\bar{B}_Z+\frac{1}{2}(-\delta\mu-\delta B_Z) & 0 & t\\
        t & 0 & U+\bar{B}_Z +\frac{1}{2}(\delta\mu-\delta B_Z) & \bar{B}_X\\
        0 & t & \bar{B}_X & U-\bar{B}_Z + \frac{1}{2}(\delta\mu+\delta B_Z)\\
    \end{pmatrix}
\end{align}
where the 1-p basis is $\{|\!\uparrow,\phi\rangle, |\!\downarrow,\phi\rangle, |\phi, \uparrow\rangle, |\phi, \downarrow\rangle\}$, the 2-p basis is $\{|d,\phi\rangle, |\!\uparrow,\uparrow\rangle, |\!\uparrow,\downarrow\rangle, |\!\downarrow,\uparrow\rangle, |\!\downarrow,\downarrow\rangle, |\phi,d\rangle\}$, and the 3-p basis is $\{|\!\uparrow,d\rangle, |\!\downarrow,d\rangle, |d, \uparrow\rangle, |d, \downarrow\rangle\}$. We have also used the Fock basis convention where we order the fermion creation operators first by site, before ordering them by spin. 
To illustrate, some non-trivial readout basis elements are, in terms of fermion creation operators acting on the vacuum:
\begin{align}
|\!\downarrow,\uparrow\rangle_{ij} &\equiv c^\dagger_{i,\downarrow} c^\dagger_{j,\uparrow}|\text{vac.}\rangle \\
|\!\downarrow,d\rangle_{ij} &\equiv c^\dagger_{i,\downarrow} c^\dagger_{j,\uparrow} c^\dagger_{j,\downarrow}|\text{vac.}\rangle
\end{align}
This site-first convention respects locality and conforms with our intuitions (for example, under this convention, $|\!\uparrow,\downarrow\rangle + |\!\downarrow,\uparrow\rangle$ is part of the multiplet which transforms under magnetic field as a spin-1 triplet). 

The Hamiltonian in the 2-p sector is particularly simple in the \textit{control basis}
\begin{align}
    |d\phi^+\rangle \equiv \frac{|d,\phi\rangle + |\phi,d\rangle}{\sqrt{2}},~~ & |d\phi^-\rangle \equiv \frac{|d,\phi\rangle - |\phi,d\rangle}{\sqrt{2}}, \label{eq:control_basis_1}\\
    |\text{sing}\rangle \equiv \frac{|\!\uparrow,\downarrow\rangle - |\!\downarrow,\uparrow\rangle}{\sqrt{2}},~~ & |\text{trip}\rangle \equiv \frac{|\!\uparrow,\downarrow\rangle + |\!\downarrow,\uparrow\rangle}{\sqrt{2}},\label{eq:control_basis_2}\\
    |\text{Bell}^+\rangle \equiv \frac{|\!\uparrow,\uparrow\rangle + |\!\downarrow,\downarrow\rangle}{\sqrt{2}},~~ & |\text{Bell}^-\rangle \equiv \frac{|\!\uparrow,\uparrow\rangle - |\!\downarrow,\downarrow\rangle}{\sqrt{2}}.\label{eq:control_basis_3}
\end{align}
In this basis, the Hamiltonian $H^\text{(2-p)}$ is
\begin{align}
       H^\text{(2-p)}_\text{ctrl. basis} &= \begin{pmatrix}
        U & \delta\mu & -2t & 0 & 0 & 0\\
        \delta\mu & U & 0 & 0 & 0 & 0\\
        -2t & 0 & 0 & \delta B_Z & 0 & 0\\
        0 & 0 & \delta B_Z & 0 & 2\bar{B}_X & 0\\
        0 & 0 & 0 & 2\bar{B}_X & 0 & 2\bar{B}_Z\\
        0 & 0 & 0 & 0 & 2\bar{B}_Z & 0\\
    \end{pmatrix},  
    \label{eq:H_2p_control_basis}
\end{align}
as we graphically depict in Fig.~\ref{fig:universal_control_app}(a,b).
 Eq.~\eqref{eq:H_2p_control_basis} is remarkably simple: each control only couples a specific pair of states in the 2-p manifold. This offers selective control of the control basis states in the 2-p sector, as indicated in Fig.~\ref{fig:universal_control_app}(b). Applying pulses of the various control knobs, in a specific order, allows for the measurement of arbitrary observables in this sector. We prove this general statement in Appendix~\ref{app:universality_proof}, and proceed to design efficient pulse sequences to measure various quantities in  Appendix~\ref{app:further_pulse_sequences}.

Finally we remark on a symmetry in the Hilbert space of the 2-p states. The 2-p basis naturally decomposes into two sets of three states: spin-singlet states $\{|d\phi^+\rangle,|d\phi^-\rangle, |\text{sing}\rangle\}$, and spin-triplet states $\{|\text{Bell}^+\rangle,|\text{Bell}^-\rangle, |\text{trip}\rangle\}$. While the singlet and triplet symmetry sectors have different multiplicities in spin-1/2 systems, with fermions on a dimer they have three states each and exhibit a remarkable symmetry under positional and spin controls. The spin-singlet states are by definition insensitive to magnetic field, yet couple when the lattice is manipulated with hopping or tilt. Meanwhile, the spin-triplet states mix under magnetic field, but are permutation invariant under exchange of the dimer sites, and hence are invariant under lattice control. These sectors are coupled by the magnetic field gradient, which mixes the positional and spin degrees of freedom. Their symmetry is broken by the presence of the Hubbard $U$ interaction, which adds an additional term on $|d\phi^\pm\rangle$ that is not present in $|\text{Bell}^\pm\rangle$. However, a ferromagnetic interaction would restore the singlet-triplet symmetry: it would give an additional energy to the $|\text{Bell}^\pm\rangle$ states analogous to the Hubbard $U$ term on $|d\phi^\pm\rangle$. Ferromagnetic interactions could be realized with, for example, magnetic atoms in optical lattices~\cite{su2023dipolar} and may offer avenues for more efficient control, since we would have full two-level system control on both $\{|d\phi^+\rangle, |\text{sing}\rangle\}$ and $\{|\text{Bell}^+\rangle, |\text{trip}\rangle\}$. However, such systems are relatively uncommon and we emphasize that even without ferromagnetic interaction, our control set is already universal and allow for efficient pulse sequence for all observables that we study in this work. One possible exception could be enabling a \textbf{SWAP} gate in which all number sectors gain an equal phase, as opposed to \textbf{pSWAP}, where the number sectors get imprinted with non-trivial phases (see Appendix~\ref{app:same_spin_three_site_interactions}).

\subsubsection{Magnetic field gradient}
\label{app:B_field_gradient}
As mentioned in the End Matter there is a subtlety with using dimer tilt and magnetic-field gradient (or gradient in an optical detuning field). These gradients can be implemented either with a sawtooth profile of chemical potential and field, or with a global gradient. The latter approach is experimentally simpler, but brings with it the challenge that the average chemical potential and field changes from dimer to dimer. The varying chemical potential does not affect the dynamics, since all dynamics on each dimer is block-diagonal in particle number. However, the varying field poses a problem, since there are non-trivial couplings between the triplet states $|\text{Bell}^\pm\rangle$ on top of the desired singlet-triplet coupling. This would lead to varying amounts of spin rotation on each dimer across the lattice.

These varying spin rotations can be set to zero due to a fortunate coincidence on the square lattice. In the presence of a gradient, the field is of the form $B_Z(\vec{r}) = B^{(0)}_Z + \vec{\nabla} B_Z\cdot \vec{r}$, where  $\vec{\nabla} B_Z$ is the gradient vector and $B^{(0)}_Z$ is the field at a reference point. The midpoints of each dimer are of the form $\vec{r} = a_0 [(n_x+1/2) \hat{x} + n_y \hat{y}]$ (horizontal dimer) or $\vec{r} = a_0 [n_x \hat{x} + (n_y + 1/2) \hat{y}]$ (vertical dimer) where $n_x$ and $n_y$ are integers and $a_0$ is the lattice spacing. We apply the field gradient along the diagonal of the $x$ and $y$ plane. Then the field at the midpoint of each dimer (i.e. average field experienced) is
\begin{equation}
        \bar{B}_Z(\vec{r}) = B^{(0)}_Z + \delta B_Z \left(n+\frac{1}{2}\right),
\end{equation}
for some integer $n$. That is, the field strengths across all dimers are offset by integer multiples of $\delta B_Z \equiv |\vec{\nabla} B_Z|a_0$, for all pairs of dimers, even with different orientations, see Fig.~\ref{fig:universal_control_app}(c). Meanwhile, the local gradient at each dimer is also $\delta B_Z$ for both horizontal and vertical dimers. The key difference is that in Eq.~\eqref{eq:H_2p_control_basis}, the average field $\bar{B}_Z$ has a factor of two strength relative to the gradient $\delta B_Z$. Therefore we can use a global uniform gradient to effect a $\pi$-pulse between $|\text{sing}\rangle$ and $|\text{trip}\rangle$, while performing a trivial operation ($2\pi-$pulse) on $\{|\text{Bell}^+\rangle, |\text{Bell}^-\rangle\}$. Note however, that in the 1-p basis, this will result in a $Z-$field $\pi$-pulse being applied on \textit{half} the dimers. This can usually be taken into account during the data-processing step. 

Taken together, this establishes that a global field gradient allows for selective swaps between the $|\text{sing}\rangle$ and $|\text{trip}\rangle$ states to be performed. This will be sufficient for most of our purposes. A notable exception is our pulse sequence to measure the energy density (Appendix~\ref{app:full_energy_pulse}), where we require a partial singlet-triplet oscillation, i.e.~a $\pi/2-$pulse between $|\text{sing}\rangle$ and $|\text{trip}\rangle$. This does not pose further complications to us since in that protocol, the $|\text{Bell}^\pm\rangle$ states and not used and never interact with the other states.

\subsection{Proof of universality}
\label{app:universality_proof}
Having introduced our control toolset, we proceed to establish that it is \textit{universal}. Specifically, we establish that any 1-p, 2-p or 3-p observable on a dimer can be measured with our scheme. By this we mean that any 1-/2-/3-p state, which we take to be an eigenstate of an observable, can be mapped into a readout basis state.

This result is almost immediate from Fig.~\ref{fig:universal_control_app}(a,b). In the 1- and 3-p sectors, the position and spin two-level systems can be thought of as two qubits. The field gradient provides an entangling gate between them, while hopping/tilt and $X-$ and $Z-$fields provide single-qubit control of each two-level system. Taken together, these ingredients allows for arbitrary two-qubit control, in particular the ability to map any state into a readout basis (i.e.~computational basis) state.

In the 2-p sector, it is convenient to consider the inverse process, i.e.~the ability to prepare any state on the 6-dimensional 2-p Hilbert space from an initial readout basis state, specifically $|d,\phi\rangle$. Our approach will be as follows: (i) first we map  $|d,\phi\rangle$ into the control basis state $|d\phi^+\rangle$. (ii) Next, the combination of hopping and idling provides the ability to prepare any state on the two-level system spanned by $\{|d\phi^+\rangle, |\text{sing}\rangle\}$. We control this two level system and use $\pi$-pulses of the other controls to sequentially transfer amplitude into each control basis state.

Step (i) can be accomplished by the sequence:
\begin{equation}
    \sqrt{2}|d,\phi\rangle = |d\phi^+\rangle + |d\phi^-\rangle \overset{\text{Hop}}{\rightarrow} i |\text{sing}\rangle + |d\phi^-\rangle \overset{\text{Tilt}}{\rightarrow} i |\text{sing}\rangle + i |d\phi^+\rangle \overset{\text{Hop + Idle}}{\rightarrow} |d\phi^+\rangle
\end{equation}

Next, we describe step (ii). Let the desired 2-p state be $(a,b,c,d,e,f)^T$ in the control basis $\{|d\phi^+\rangle,|d\phi^-\rangle,|\text{sing}\rangle,|\text{trip}\rangle,|\text{Bell}^+\rangle,|\text{Bell}^-\rangle\}$ (without loss of generality, with $a$ real and positive). After step (i), the state is $(1,0,0,0,0,0)^T$. Using hopping and idling, we prepare the state $(\sqrt{1-f^2},0,i f,0,0,0)^T$. We then perform $\pi-$pulses with $\delta B_Z$, $B_X$, and $B_Z$ in order, which will result in the state $(\sqrt{1-f^2},0,0,0,0,f)^T$. Each $\pi-$pulse includes a factor of $i$, which was why we initially included a factor of $i$. We proceed iteratively, preparing the state $(\sqrt{1-e^2-f^2},0,-e,0,0,f)^T$, which we then map $(\sqrt{1-e^2-f^2},0,0,0,e,f)^T$. Repeating this process prepares the state $(\sqrt{a^2+b^2+c^2},0,0,d,e,f)^T$. We next prepare the amplitude on $|d\phi^-\rangle$ with $(e^{i\phi_c} b,0,e^{i\phi_c}\sqrt{a^2+c^2},d,e,f)^T$, where $e^{i\phi_c} \equiv c/|c|$. We then swap population to $(0, i e^{i\phi_c} b, e^{i\phi_c}\sqrt{a^2+c^2},d,e,f)^T$. Finally, a hopping pulse gives the state $(i e^{i\phi_c} a ,i e^{i\phi_c} b, c,d,e,f)^T$, and idling for an appropriate amount of time fixes the phases on $|d\phi^\pm\rangle$ and gives the desired state $(a, b, c, d, e, f)^T$.

The above result is sufficient to measure any observable expectation value on a dimer. However, it is not sufficient to measure dimer-dimer correlation functions, as we do in the rest of this work. While an extended universality argument can be made demonstrating that \textit{any pair} of 1-p and 2-p states can be mapped into readout states using the same pulse, but we omit its details since it is not the focus of this work.

Finally, we emphasize that this construction is fairly inefficient: the sequences described in this sub-section are likely sub-optimal. 
Furthermore, our procedure involves the estimation of the population of each eigenstate (or each pair of eigenstates when measuring dimer-dimer correlations) with a separate pulse, which adds additional overhead. Yet, this discussion provides a theoretical guarantee that there exists pulse sequences within our framework to measure any arbitrary number-conserving operator. Backed by this  guarantee, in the following section we develop efficient pulse sequences for measuring a variety of relevant observables. 

\section{Designing further pulse sequences}
\label{app:further_pulse_sequences}
Having demonstrated that an entire class of observables can be measured with this protocol, we proceed to design efficient pulses to measure different quantities of interest. Using the control set discussed in Appendix~\ref{app:universal_control_set}, we develop a pulse to measure the kinetic energy, which we expand upon to measure the energy density (including both kinetic and interaction energy terms). As a concrete application of these pulse sequences, we describe how they can be used to measure the temperature of a state, with minimal assumptions. While our energy density measurement pulse is the primary workhorse for our thermometry application, we will also require additional pulses for certain edge cases, where we must estimate three-body quantities instead of dimer-dimer correlations. Fortunately, our framework of pulse sequences on dimers can still be used to measure the three-body quantities, by splitting the three sites into two overlapping dimers. We develop the required pulse sequences in Appendix \ref{app:three_body_pulse_sequences}, among them sequences that perform conditional swap operations on the dimer. These sequences turn out to have broader application, and we describe how they may be used to improve the parity-projection measurements common to quantum gas microscopes into fully spin- and charge-resolved measurements.

\subsection{Kinetic energy pulse sequence}

\label{app:kinetic_energy_pulse}
\begin{figure}[t!]
    \centering
    \includegraphics[width=\linewidth]{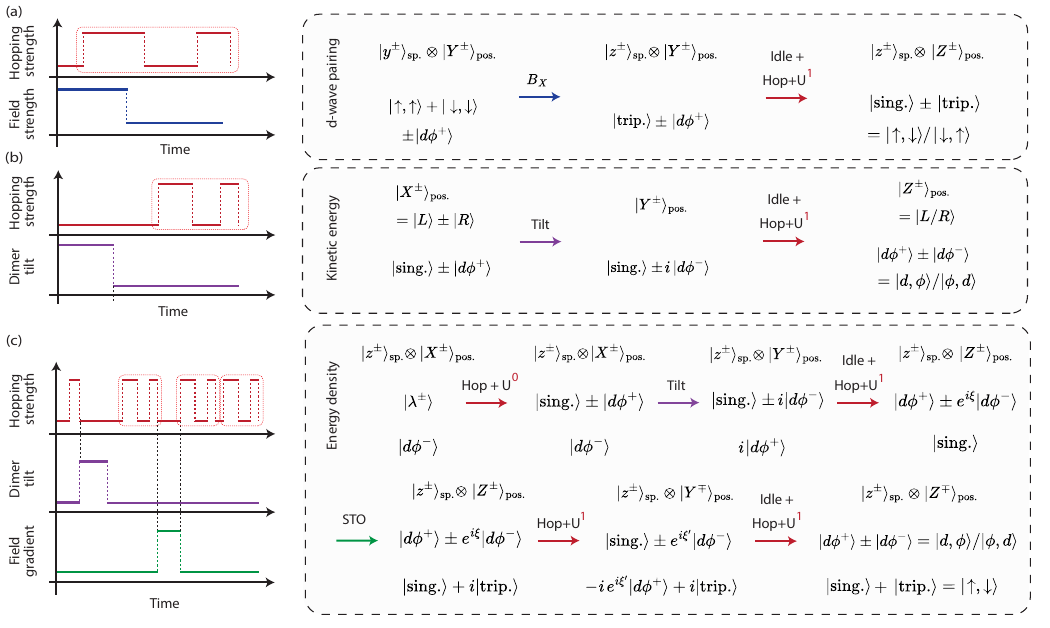}
    \caption{\textbf{Efficient pulse sequences.} We develop pulse sequences to measure (a) the $d$-wave pairing correlator, (b) kinetic energy, and (c) full energy densities (including both kinetic and interaction energy). In the left column, we depict the sequences of the required controls over time: The hopping control sequences are the most complex, but are simply repeated applications of the \textbf{Hop+U$^1$} pulse (dotted boxes). In the right column, we list how the relevant states in the 1-p and 2-p sectors transform under each step of the pulse sequence, beginning from the eigenstates of the observable in question and ending with states in the readout basis. The design of each sequence is explained in the main text and Appendix~\ref{app:further_pulse_sequences}. For brevity, we omit global phase and normalization factors when they can be safely ignored.
    } 
    \label{fig:pulse_sequences}
\end{figure}

We first develop a protocol to measure the kinetic energy across a dimer. As with the $d$-wave pairing function, we do so by finding a pulse sequence that rotates the eigenvalues of the observable in question (here, the kinetic energy operator) into the readout basis. Like all other observables accessible by this protocol, the kinetic energy operator preserves particle number, and hence can be analyzed in the different particle number sectors. As it turns out, the 1-p sector will pose the bigger challenge than the 2-p sector.

Across a pair of nearest-neighbor sites $\langle i,i'\rangle$, the kinetic energy operator on that bond is:
\begin{equation}
    O^\text{kin}_{i,i'} = \sum_{\sigma \in \{\uparrow,\downarrow\}}c^\dagger_{i,\sigma} c^{}_{i',\sigma} + \text{h.c.}
    \label{eq:kinetic_energy}
\end{equation}
Since this observable is spin-invariant, magnetic field controls are not required. Instead, in addition to particle-hopping control, we also use dimer tilt $\delta\mu$ introduced in Appendix~\ref{app:universal_control_set}. 
As illustrated in Fig.~\ref{fig:pulse_sequences}(b), our pulse sequence comprises of a $\pi/2$-rotation using dimer tilt, followed by the \textit{same} hopping pulse sequence \textbf{Hop+U$^1$} that we develop for the $d$-wave pairing measurement in Appendix~\ref{app:dwave_hopping_pulse}. 

We look for an appropriate sequence of control parameters that map the eigenstates of $O^\text{kin}_{i,i'}$ into readout basis states. In the 1-p and 2-p sectors, these eigenstates are:
\begin{gather}
    |X^\pm\rangle_\text{pos} \equiv |L\rangle \pm |R\rangle,\\
    |\text{sing}\rangle \pm |d\phi^+\rangle,
\end{gather}
where we have omitted the spin degree of freedom on the 1-p eigenstates. These 1-p eigenstates necessitate the use of dimer tilt control. Hopping control performs a rotation about the $X$-axis on the 1-p positional Bloch sphere, and hence the states $|X^\pm\rangle_\text{pos}$, as eigenstates of the hopping operator, are invariant. We must first use dimer tilt, which acts as a $Z$-axis rotation on the positional Bloch sphere. We perform a $\pi/2$ $Z-$rotation to map $|X^\pm\rangle_\text{pos}$ into $|Y^\pm\rangle_\text{pos}$, before performing a $\pi/2$ $X-$rotation to map these states to $|Z^\pm\rangle$.

However, introducing dimer tilt now complicates the 2-p sector. While hopping and interactions effectively act only on the two-level system spanned by $\{\ket{\text{sing}}, \ket{d\phi^+}\}$, dimer tilt couples the $\ket{d\phi^\pm}$ states. While this complicates our analysis, we use it to our advantage in several ways: (i) by allowing access to the $|d\phi^-\rangle$ state, we can prepare the superpositions $|d\phi^+\rangle \pm |d\phi^-\rangle$, which are the readout states $|d,\phi\rangle$ and $|\phi,d\rangle$, and (ii) by fully transferring population to the $|d\phi^-\rangle$ state, we can `shelve' information by making them invariant under hopping control. Meanwhile, (iii) the singlet state $|\text{sing}\rangle$ is invariant under dimer tilt. This greatly simplifies our analysis, simplifying the problem from controlling a generic three-level system to controlling two (coupled) two-level systems one after another.

The pulse sequence acts on the 2-p sector as indicated in the lower row of Fig.~\ref{fig:pulse_sequences}(b). Beginning from the eigenstates $|\text{sing}\rangle \pm |d\phi^+\rangle$, the dimer tilt coherently transfers population from $|d\phi^+\rangle$ to $|d\phi^-\rangle$. As with the $d$-wave pairing pulse, there is a factor of two difference in the coupling strengths in the
1-p and 2-p sectors, which coincides with the required $\pi/2$- vs.~$\pi$-pulses. The initial eigenstates are mapped to $|\text{sing}\rangle \pm |d\phi^-\rangle$. The hopping pulse then coherently maps  $|\text{sing}\rangle \mapsto |d\phi^+\rangle$, which prepares the desired states $|d\phi^+\rangle \pm |d\phi^-\rangle$. The hopping pulse is precisely the same \textbf{Hop+U$^1$} sequence as the $d$-wave pairing pulse: the parameters  $T_\text{hop}^{(\pm)}$ and  $T_\text{idle}^{(+\rightarrow -)}$ are identical [Eqs.~(\ref{eq:hopping_time_1}-\ref{eq:idling_time_2})], while the initial idling time $T_\text{idle}^\text{(init)}$ is slightly modified to match the phases in this pulse:
\begin{align}
    UT^{\text{(init)}}_\text{idle} = &-\text{arg}\bra{d\phi^+}
    V_{\theta}\left(T^{(-)}_\text{hop}\right)V_{0}\left(T^{(+\rightarrow-)}_\text{idle}\right) V_{\theta}\left(T^{(+)}_\text{hop}\right)V_\text{tilt}\Big(T_\text{tilt}\Big)
    \ket{\text{sing}} \label{eq:idling_time_1_KE}\\
    &+\text{arg}\bra{d\phi^-}
    V_{\theta}\left(T^{(-)}_\text{hop}\right)V_{0}\left(T^{(+\rightarrow-)}_\text{idle}\right) V_{\theta}\left(T^{(+)}_\text{hop}\right)V_\text{tilt}\Big(T_\text{tilt}\Big) 
    \ket{d\phi^+} \nonumber
\end{align}
where we have extended the unitary  $V_\theta(T)$ in Eq.~\eqref{eq:rotation_unitary} to act on the three-level system $\{|d\phi^+\rangle, |\text{sing}\rangle,|d\phi^-\rangle\}$.
\begin{align}
V_\theta(T) &= \exp \left[-i T \begin{pmatrix}
    U & 0 & - 2t\\
    0 & U & 0 \\
    -2t & 0 & 0 
\end{pmatrix} \right]
\label{eq:rotation_unitary_3LS}
\end{align}
on the basis $\{|d\phi^+\rangle,|d\phi^-\rangle,|\text{sing}\rangle\}$, and the unitary $V_\text{tilt}\left(T\right)$ is
\begin{align}
V_\text{tilt}\left(T\right) &\equiv \exp \left[-i T \begin{pmatrix}
    U & \delta \mu & 0\\
    \delta \mu & U & 0 \\
    0 & 0 & 0 
\end{pmatrix} \right]
\end{align}
with the dimer tilt of strength $\delta \mu \equiv \mu_i - \mu_{i'}$ applied for a time $T_\text{tilt} = \pi/(2 \delta\mu)$. This fully specifies the pulse sequence illustrated in Fig.~\ref{fig:pulse_sequences}(c).

\subsection{Energy density pulse sequence}
\label{app:full_energy_pulse}
Having designed a pulse sequence to measure the kinetic energy on a bond $\langle i,  i'\rangle$, in this section we present a protocol that measures, with a single pulse sequence, the \textit{total} energy density $h_{ii'}$, given by
\begin{equation}
    h_{ii'} = -t \sum_{\sigma = \{\uparrow,\downarrow\}} (c^\dagger_{i,\sigma} c_{i',\sigma}^{} + c^\dagger_{i',\sigma} c_{i,\sigma}^{}) + \frac{U}{4} (\hat{n}_{i,\uparrow}\hat{n}_{i,\downarrow} + \hat{n}_{i',\uparrow}\hat{n}_{i',\downarrow}), \label{eq:energy_density}
\end{equation}
where $\hat{n}_{i,\sigma} = c^\dagger_{i,\sigma} c^{}_{i,\sigma}$ and we have equally divided the interaction energy associated with site $i$ among its four adjacent nearest-neighbor bonds $\langle i,i' \rangle$ on the square lattice. While the energy density on a single dimer can be estimated by separately measuring the kinetic and interaction energies on the dimer, energy-energy correlations between two dimers cannot be estimated in this fashion, since there are kinetic-interaction cross terms which would be missed when we apply global pulses to measure the kinetic and interaction energies. Such energy-energy correlations may be relevant to hydrodynamics experiments, and in the following section we discuss how energy-energy correlations can be used to estimate the temperature of the experimental state.

The primary complicating factor for measuring the energy density is that we have \textit{three} non-trivial eigenstates of $h_{ii'}$ in the 2-p sector, which we want to map to three different readout states with a single pulse sequence while simultaneously achieving the desired $|X^\pm\rangle_\text{pos} \mapsto |Z^\pm\rangle_\text{pos}$ mapping in the 1-p sector. The eigenstates and their corresponding target states are:
\begin{align}
    |\sigma\rangle_\text{sp}\otimes |X^\pm\rangle_\text{pos} &\mapsto |\sigma \rangle_\text{sp}\otimes |Z^\pm\rangle_\text{pos}, \label{eq:fullE_1p}\\
    |\lambda^\pm\rangle \equiv |d\phi^+\rangle + \left( \frac{u}{16} \pm \sqrt{1+ \frac{u^2}{256}}\right)|\text{sing}\rangle &\mapsto |d,\phi\rangle/|\phi, d\rangle, \label{eq:fullE_2p_sym}\\
     |d\phi^-\rangle &\mapsto |\!\uparrow,\downarrow \rangle & \label{eq:fullE_2p_asym}
\end{align}
Note that Eq.~(\ref{eq:fullE_1p}) is the same in this case as in the kinetic energy measurement, since the interaction energy is only relevant in the 2-p sector. Meanwhile, the states $|\lambda^\pm\rangle$ can be mapped to the 2-p eigenstates of the kinetic energy $|\text{sing}\rangle \pm |d\phi^+\rangle$ by a simple pulse \textbf{Hop+U$^0$}. However, Eq.~(\ref{eq:fullE_2p_asym}) is entirely new, since the $|d\phi^-\rangle$ state has zero kinetic energy but carries interaction energy $U$.

The pulse sequence, illustrated in Fig.~\ref{fig:pulse_sequences}(c), relies on the kinetic energy pulse sequence as a backbone. Indeed, we first map the eigenstates $|\lambda^\pm\rangle$ into the kinetic energy eigenstates $|\text{sing}\rangle \pm |d\phi^+\rangle$, using a hopping pulse sequence which we dub \textbf{Hop+U$^{0}$} (details deferred to the end of this section)
\begin{gather}
|X^\pm\rangle_\text{pos} \overset{\textbf{Hop+U$^{0}$}}{\mapsto} |X^\pm\rangle_\text{pos}\\
   |\lambda^\pm\rangle \overset{\textbf{Hop+U$^{0}$}}{\mapsto} |\text{sing}\rangle \pm |d\phi^+\rangle
\end{gather}
We then apply the kinetic energy measurement sequence of tilt and hopping pulses discussed in App.~\ref{app:kinetic_energy_pulse}: 
\begin{gather}
   |X^\pm\rangle_\text{pos} \overset{\text{Tilt}}{\mapsto} |Y^\pm\rangle_\text{pos} \overset{\text{Idle+}\textbf{Hop+U$^{1}$}}{\mapsto} |Z^\pm\rangle_\text{pos}\\
   |\text{sing}\rangle \pm |d\phi^+\rangle \overset{\text{Tilt}}{\mapsto} |\text{sing}\rangle \pm i|d\phi^-\rangle  \overset{\text{Idle+}\textbf{Hop+U$^{1}$}}{\mapsto} |d\phi^+\rangle \pm e^{i \xi} |d\phi^-\rangle\\
   |d\phi^-\rangle \overset{\text{Tilt}}{\mapsto} |d\phi^+\rangle \overset{\text{Idle+}\textbf{Hop+U$^{1}$}}{\mapsto} |\text{sing}\rangle
\end{gather}
The kinetic energy pulse is slightly modified to imprint a phase difference $|d\phi^+\rangle \pm e^{i \xi} |d\phi^-\rangle$ in anticipation of future hopping and idling steps.

The final stage aims to map the singlet state $|\text{sing}\rangle$ to the readout state $|\!\uparrow,\downarrow\rangle \equiv |\text{sing}\rangle + |\text{trip}\rangle$. We do so with a magnetic field gradient $\delta B_Z$, which induces singlet-triplet oscillations~\cite{trotzky2010controlling,zhu2024quantumcircuitsbasedtopological}. This maps $|\text{sing}\rangle \mapsto |\text{sing}\rangle + i |\text{trip}\rangle$, and one can verify that the additional action on the states $\{|\text{Bell}^\pm\rangle\}$, which varies between dimers, is irrelevant, since none of our desired information is stored in these states. On the 1-p sector, the field gradient acts as a controlled $Z$-rotation, and the states $|z\rangle_\text{sp}\otimes |Z\rangle_\text{pos}$ simply acquire complex phases. 
\begin{align}
   |z\rangle_\text{sp}\otimes|Z^\pm\rangle_\text{pos} &\overset{\text{STO}}{\mapsto} \exp(iz Z^\pm \frac{\pi}{8}) |z\rangle_\text{sp}\otimes|Z^\pm\rangle_\text{pos} \\
   |\text{sing}\rangle &\overset{\text{STO}}{\mapsto} \frac{|\text{sing}\rangle+i|\text{trip}\rangle }{\sqrt{2}}
\end{align}

We now wish to remove the phase factor $i$ in the state $|\text{sing}\rangle + i |\text{trip}\rangle$. This cannot be achieved with field or field gradient controls, since they are realized by on-site fields and as such cannot change any entanglement, which we need to reach the product state $|\uparrow,\downarrow\rangle$. Indeed, hopping is our only control knob which is capable of generating entanglement. We use two applications of the familiar pulse \textbf{Hop+U$^1$} to map the $|\text{sing}\rangle$ part of the state to $|d\phi^+\rangle$, idle to remove the phase factor $i$, before mapping back to $|\text{sing}\rangle$ and giving the readout state $|\uparrow, \downarrow\rangle$. Meanwhile, this process also modifies the relative phase of the $|d\phi^+\rangle + e^{i \xi} |d\phi^-\rangle$ superposition. Appropriate choice of $\xi$ in the intermediate step ensures that we arrive at the readout states $|d,\phi\rangle, |\phi, d\rangle$.  In the 1-p sector, this sequence maps $|Z^\pm\rangle_\text{pos}$ to $|Y^\mp\rangle_\text{pos}$ and back into the readout basis $|Z^\mp\rangle_\text{pos}$. While we could use a pulse that for example, maps $ |Z^\pm\rangle_\text{pos} \mapsto |Z^\mp\rangle_\text{pos}$ at every stage, here we use \textbf{Hop+U$^1$} since it appears frequently in this work.
\begin{gather}
   |Z^\pm\rangle_\text{pos} \overset{\textbf{Hop+U$^{1}$}}{\mapsto} |Y^\pm\rangle_\text{pos} \overset{\text{Idle+}\textbf{Hop+U$^{1}$}}{\mapsto}|Z^\mp\rangle_\text{pos}\\
   |d\phi^+\rangle \pm e^{i \xi} |d\phi^-\rangle \overset{\textbf{Hop+U$^{1}$}}{\mapsto} |\text{sing}\rangle \pm e^{i \xi'} |d\phi^-\rangle \overset{\text{Idle+}\textbf{Hop+U$^{1}$}}{\mapsto} |d\phi^+\rangle \pm |d\phi^-\rangle \\
   |\text{sing}\rangle + i |\text{trip}\rangle \overset{\textbf{Hop+U$^{1}$}}{\mapsto} -i e^{i\xi'}|d\phi^+\rangle + i|\text{trip}\rangle \overset{\text{Idle+}\textbf{Hop+U$^{1}$}}{\mapsto} |\text{sing}\rangle + |\text{trip}\rangle 
\end{gather}
This completes the desired mapping (\ref{eq:fullE_1p}-\ref{eq:fullE_2p_asym}). 

Finally, we provide details on the pulse \textbf{Hop+U$^0$} that maps the eigenstates $|\lambda^\pm\rangle \mapsto |\text{sing}\rangle \pm |d\phi^+\rangle$. Since we achieve this only with hopping and idling, and the 1-p state remains in $|X^\pm\rangle_\text{pos}$, there are no non-trivial matching conditions on this pulse. Here, we design a pulse with the minimal number of steps [Fig.~\ref{fig:more_pulse_sequences}(c)]. The  idling step takes us to the appropriate tilted arc (this intersection always exists since $|\lambda^\pm\rangle$ lies between the tilted axis and the equator, for all values of $u$), and the hopping step brings us to the equator $|\text{sing}\rangle \pm |d\phi^+\rangle$. The relevant hopping parameters can be solved for by using Eq.~\eqref{eq:rodrigues_formula} and are numerically implemented in the reference code~\cite{github_code}.

The vertical arc containing the initial point $\vec{r}_i = (\sin \theta', 0 , \cos \theta')$, where $\theta' \equiv \cot^{-1}(u/16)$, has expression
\begin{equation}
    \vec{r}_\text{idle}(\phi_\text{idle}) = z \begin{pmatrix}
        0 \\ 0 \\ 1
    \end{pmatrix} + \sqrt{1-z^2} \begin{pmatrix}
        \cos \phi_\text{idle}  \\ \sin \phi_\text{idle} \\ 0
    \end{pmatrix},
    \label{eq:hopU0_vertical}
\end{equation}
where $z$ is the vertical coordinate $z \equiv \cos \theta'$. Meanwhile, backward-rotating from the final point $\vec{r}_f = (1, 0 , 0)^T$ gives the arc
\begin{equation}
    \vec{r}_\text{hop}(\phi_\text{hop}) = \Delta \begin{pmatrix}
        \sin \theta \\ 0 \\ \cos \theta
    \end{pmatrix} + \sqrt{1-\Delta^2} \begin{pmatrix}
        \cos\theta \cos \phi_\text{hop}  \\ -\sin \phi_\text{hop} \\ -\sin \theta \cos \phi_\text{hop}
    \end{pmatrix},
    \label{eq:hopU0_tilted}
\end{equation}
where $\Delta = \sin \theta$. Solving for the intersection point of Eqs.~(\ref{eq:hopU0_vertical}, \ref{eq:hopU0_tilted}) gives the values of $\phi_\text{hop}$ and $\phi_\text{idle}$.

\subsection{Application: thermometry}
In this section we discuss a concrete application of the energy-density measurement sequence. Indeed, as discussed above, while the energy density $\langle h_{ii'}\rangle$ can be estimated by summing the kinetic and interaction energies, estimating energy-energy correlations requires a protocol that directly measures the energy density. 

Here we discuss a practical application of such two-point energy correlations: estimating the temperature of a state. Despite its simplicity, estimating the temperature of a Gibbs state in a quantum many-body system remains an open problem. While it is straightforward to estimate the energy of a system, and the average energy is typically a monotonic function of the temperature, one does not know this function \textit{a priori}, which requires information or assumptions about the density of states (equivalently, the thermodynamic entropy). 

With experiments reaching sufficiently low temperatures where numerical simulations no longer serve as a reliable benchmark, developing methods to estimate the temperature of an experiment, in particular with as few assumptions as possible, remains an open problem. Approaches in the literature include coupling the system with a probe system, which suffers from certain fundamental uncertainties~\cite{stace2010quantum,hovhannisyan2018measuring,mehboudi2019thermometry}. Specific schemes have also been proposed for fermionic quantum gas microscopes~\cite{kohl2006tthermometry,hartke2020doublon}. Here, we propose a scheme that can be performed with the pulse sequences we have developed. 

We assume that there is an experimental tuning knob $\lambda$ which produces Gibbs states of varying temperatures 
\begin{equation}
 \rho_\beta = \frac{\exp[-\beta(\lambda)H]}{Z},
\end{equation}
where $Z$ is the partition function $Z\equiv \text{tr}(\exp[-\beta(\lambda)H])$. Our task is to determine the relation $\beta(\lambda)$. We do so by noting that the gradient $\partial_\lambda \beta(\lambda)$ can be estimated by measurements of the average energy and \textit{energy variance}. By further assuming that we can determine the temperature through other means at a given point (e.g.~at sufficiently high temperature through comparison with quantum Monte Carlo methods~\cite{varney2009quantum}), knowledge of the gradient allows reconstruction of the curve $\beta(\lambda)$.

The mean energy and energy variance are related to the temperature by thermodynamic relations. Specifically:
\begin{align}
    \langle E \rangle \equiv \text{tr}(\rho_\beta H) &= -\frac{\partial \log Z}{\partial \beta}\\
    \sigma_E^2 \equiv \text{tr}(\rho_\beta H^2 ) - \langle E\rangle^2 &= \frac{\partial^2 \log Z}{\partial \beta^2} = -\frac{\partial \langle E \rangle}{\partial \beta}
\end{align}
We now apply the chain rule on $\frac{\partial \langle E \rangle}{\partial \beta} = \frac{\partial \langle E \rangle}{\partial \lambda} \frac{d\lambda}{d\beta}$. We can independently estimate the derivative $\frac{\partial \langle E \rangle}{\partial \lambda}$ by measuring $\langle E\rangle$ at different values of $\lambda$. This gives the estimate of the gradient:
\begin{equation}
    \frac{d \beta}{d \lambda} \approx -\frac{1}{\sigma^2_E(\lambda)} \frac{\langle E \rangle (\lambda + \Delta \lambda) - \langle E \rangle (\lambda - \Delta \lambda)}{2 \Delta \lambda},
\end{equation}
which as discussed above, can be integrated to estimate the curve $\beta(\lambda)$. The energy variance is simply the sum over all two-point energy correlators between any two dimers $\langle i, i'\rangle, \langle j, j'\rangle$:
\begin{equation}
    \sigma_E^2 = \sum_{\langle i, i'\rangle} \sum_{\langle j, j'\rangle} \left[\text{tr}(\rho_\beta h_{i, i'} h_{j,j'})-\text{tr}(\rho_\beta h_{i, i'})\text{tr}(\rho_\beta h_{j,j'})\right]
\end{equation}
One can extract almost all of these two-point correlators immediately with the protocol in this work. Below, we comment on two subtleties: dimerization patterns which yield all pairs of bonds that do not overlap, and in the following subsection we sketch a protocol to measure the terms arising from overlapping bonds.

\subsubsection{Dimerization patterns}
\begin{figure}[t]
    \centering
    \includegraphics[width=\linewidth]{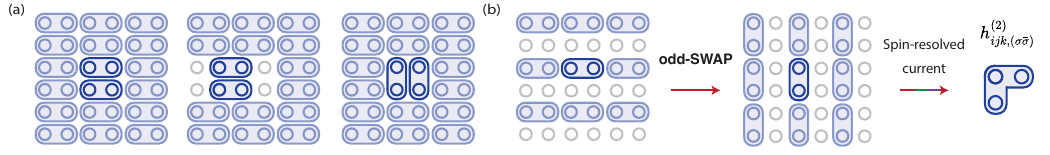}
    \caption{\textbf{Experimental details for thermometry.} Dimerization patterns to (a) measure dimer-dimer correlations at all distances and relative orientation, assuming translational and four-fold rotational invariance, and (b) to measure three-body observables which represent energy correlations between overlapping dimers (Appendix~\ref{app:three_body_pulse_sequences}). (a) Three dimerization patterns suffice to provide dimer-dimer pairs at all separations and orientations. We need two patterns for the horizontal-horizontal dimer pairs: a basic columnar dimerization (left) in addition to a pattern where a central pair of dimers are shifted by one lattice site (middle). The former captures dimers with even horizontal separation and the latter captures dimers with odd horizontal separation. We only need one pattern for the horizontal-vertical dimer pairs, where we rotate two dimers by $90^\circ$. Because we rotate two dimers, this pattern has pairs at both odd and even distances. Realizing the latter two patterns would require a limited degree of local control, which may be provided by a digital micromirror device or optical tweezers. (b) We measure the three-body correlations in question on an ``L"-shaped trimer by first forming horizontal dimers on \textit{every other} row, performing the \textbf{odd-SWAP} gate (\ref{eq:oddSWAP_1},~\ref{eq:oddSWAP_2}), then forming vertical dimers on every other column and measuring the spin-resolved current (Appendix~\ref{app:spin-resolved_current}).}
    \label{fig:dimerization}
\end{figure}

In order to measure all pairs of non-overlapping bonds, we cannot simply use the dimerization pattern in Fig.~\ref{fig:overview}(b). For example, this would only cover some of the possible horizontal-horizontal dimer pairs, even if we could assume translational and rotational invariance. One might guess that some small number of dimerization patterns could yield all dimer pairs. However, any combination of periodic dimerization patterns that could be formed by optical superlattices~\cite{kessler2014single,impertro2023localreadoutcontrolcurrent} would not capture all pairs of bonds, given by the least common multiple of the pattern periodicities. For simplicity, we consider a columnar dimerization (Fig.~\ref{fig:dimerization}(a), left) with periodicity two, and examine horizontal-horizontal dimer pairs on a straight line. This dimerization contains horizontal-horizontal dimer pairs with an even separation. Combining this with a columnar dimerization with periodicity three, for example, would now capture all dimer pairs that are separated by multiples of two or three lattice sites, but would miss every dimer pair with spacings $5 + 6 n$ for integer $n$. While this may be sufficient for the purpose of thermometry, where dimer-dimer correlations beyond a certain distance may be ignored, this approach is also experimentally challenging, since it is not straightforward to achieve superlattices of different periodicities.

Instead, we advocate for a different approach. Assuming that we have translational and rotational invariance, we only need three dimerization patterns, illustrated in Fig.~\ref{fig:dimerization}(a). The first pattern is simply the columnar dimerization. In order to capture the odd-spaced horizontal-horizontal dimer pairs, we then offset a central square by one site (middle). The final pattern involves \textit{rotating} this central square, which would give all vertical-horizontal dimer pair separations (right). With translational and rotational invariance, this gives us all possible separations between dimer pairs. While the overall dimerization pattern could be performed with an optical lattice, this offset and rotation of the central square would require a certain degree of local control, which could be provided by digital micromirror devices (DMDs) or optical tweezers, which have been recently demonstrated in optical lattice quantum gas microscopes~\cite{young2022tweezer,spar2022realization}.

\subsection{Pulse sequences for thermometry: measuring three-body quantities} 
\label{app:three_body_pulse_sequences}

\begin{figure}[t!]
    \centering
    \includegraphics[width=\linewidth]{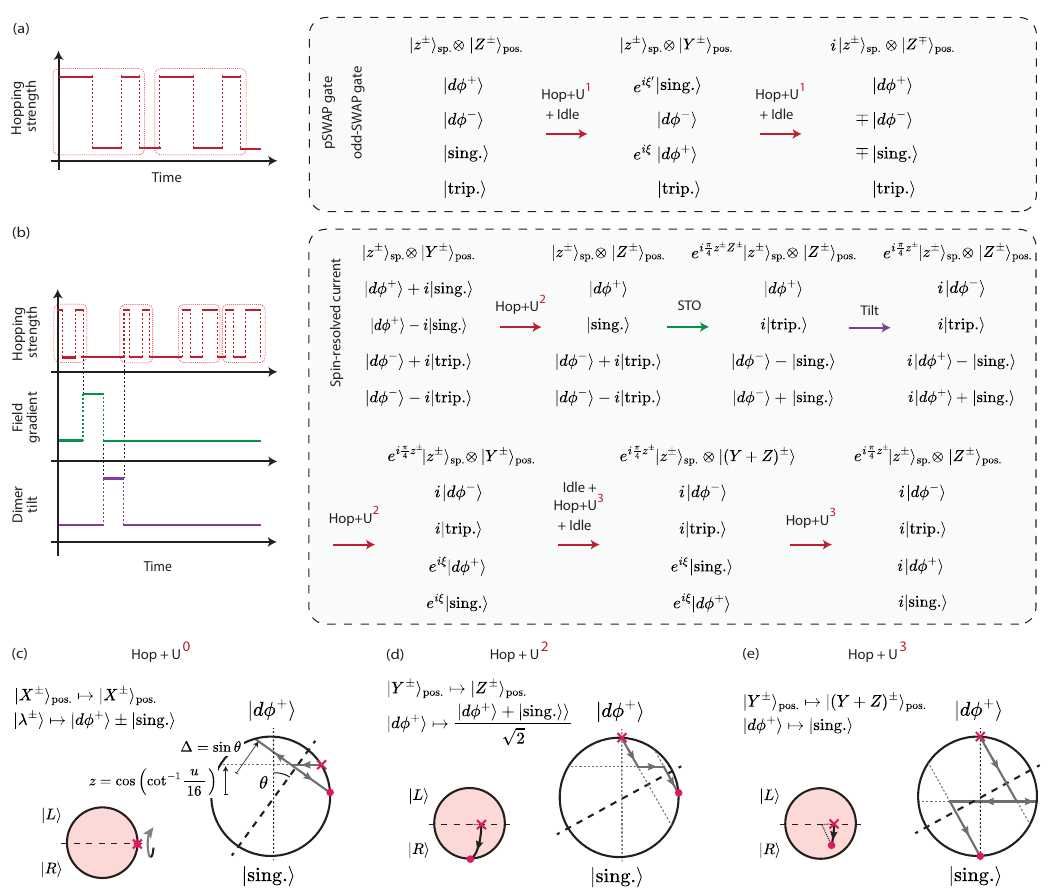}
    \caption{\textbf{Further pulse sequences.} Pulse sequences to (a) perform modified SWAPs [\textbf{pSWAP} (\ref{eq:pSWAP_1}-\ref{eq:pSWAP_5}) and \textbf{odd-SWAP} (\ref{eq:oddSWAP_1},\ref{eq:oddSWAP_2})] between sites $i$ and $i'$ of the dimer, and (b) measure the spin-resolved current $i c^\dagger_{i,\downarrow}c^{}_{i',\downarrow} + \text{h.c.}$, and (c-e) perform hopping and idling primitives \textbf{Hop+U$^0$}, \textbf{Hop+U$^2$}, and \textbf{Hop+U$^3$}, which we use in our energy-density pulse sequence [Fig.~\ref{fig:pulse_sequences}(c)] and to measure the spin-resolved current. (a,b) As in Fig.~\ref{fig:pulse_sequences}, we plot the time-sequence on the left, and list its actions on the relevant states of interest on the right, omitting phase and normalization factors when they can be ignored, but including them when they should be controlled by idling steps. For example, in (b), we require the final states $|d\phi^+\rangle$ and $|d\phi^-\rangle$ to have the same phase, since the eigenstates of the spin-resolved current are $|d,\phi\rangle \pm i |\!\uparrow, \downarrow\rangle$ and $|\phi,d\rangle \pm i |\!\downarrow, \uparrow\rangle$, which can be written as the superpositions $(|d\phi^+\rangle \pm i |\text{sing}\rangle) \pm' (|d\phi^-\rangle \pm i |\text{trip}\rangle)$. Note, however, that the displayed phase factors are in some sense schematic: for clarity of presentation, we have omitted some ``trivial" phase factors such as the phase continuously accumulated by the state $|d\phi^-\rangle$. All phases are explicitly computed in the reference code~\cite{github_code}. (c-e) We draw the trajectories on the relevant 1-p (shaded, smaller) and 2-p Bloch spheres (larger) for the \textbf{Hop+U$^0$}, \textbf{Hop+U$^2$}, and \textbf{Hop+U$^3$} pulses, state the mappings performed by each sequence, and in (c) also indicate relevant geometric quantities useful to solve for the pulse parameters in Eq.~(\ref{eq:hopU0_vertical},\ref{eq:hopU0_tilted}).}
    \label{fig:more_pulse_sequences}
\end{figure}

The only bond-bond correlations that cannot be measured with the scheme above are the bonds that overlap. These result in ``trimers" of three sites in a straight line or in an `L' shape. Our approach will be the same in either case: we will treat the trimer $\langle i,j,k\rangle$ as two dimers $\langle i,j\rangle$ and $\langle j, k\rangle$ [Fig.~\ref{fig:dimerization}(b)]. We first perform operations on $\langle i,j\rangle$, with site $k$ isolated, before isolating site $i$ and performing operations on dimer $\langle j,k\rangle$. This allows us to use our results on dimer control in this work. However, we note two subtleties: from an experimental point of view, sequentially forming two separate dimers in a coherent fashion may be more challenging to implement than the single-dimerization measurments. From a theoretical point of view, the operations on each dimer should be phase-coherent between multiple particle number sectors, since a number-definite state on the trimer may have indefinite particle number on a dimer. This merely requires us to keep track of the ``global phase" of each particle number sector, which we had previously ignored, and requires additional steps in the protocol.

We separate the energy density into kinetic and interaction energy terms. The overlapping terms consist of kinetic-kinetic terms, kinetic-interaction, and interaction-interaction terms. Here, we discuss the kinetic-kinetic terms; the other terms only require nontrivial operations on a single dimer. In turn, these kinetic-kinetic terms comprise products of hopping terms acting on the bonds $\langle i,j\rangle $ and $\langle j,k\rangle$ of a three-site system $\langle i,j,k\rangle$ and involving either (a) the same spin species or (b) opposite spin species. Our overall strategy will be to apply a pulse on dimer $\langle i,j \rangle$, followed by a pulse on dimer $\langle j,k\rangle$, then measuring all sites $i,j,k$. 

We develop several pulse sequences to measure the relevant terms. In summary, our proposed strategy is to estimate the kinetic-kinetic terms with: (a) for the same-spin terms, the \textbf{pSWAP} gate on dimer $\langle i,j \rangle$, followed by \textbf{Hop+U$^{1}$} on $\langle j,k \rangle$, (b) for opposite-spin terms, the \textbf{odd-SWAP} gate on dimer $\langle i,j \rangle$, followed by applying the spin-resolved current pulse on $\langle j,k \rangle$ [Fig.~\ref{fig:dimerization}(b)]. The kinetic-interaction terms can be estimated with either \textbf{Hop+U$^{1}$} on a dimer $\langle i,j\rangle$ along with projective readout of the site $k$, or a variant of the spin-resolved current pulse sequence. Finally, the interaction-interaction terms can be directly measured by projective readout. Below, we develop the necessary pulse sequences, which we also summarize in Fig.~\ref{fig:more_pulse_sequences}.

\subsubsection{Bond-bond correlations with same spin}
\label{app:same_spin_three_site_interactions}
The terms involving the same spin species are:
\begin{equation}
    h^{(2)}_{ijk,(\sigma\sigma)} \equiv (c^\dagger_{i,\sigma} c^{}_{j,\sigma}+c^\dagger_{j,\sigma} c^{}_{i,\sigma})(c^\dagger_{j,\sigma} c^{}_{k,\sigma}+c^\dagger_{k,\sigma} c^{}_{j,\sigma}) + \text{h.c.} = (c^\dagger_{i,\sigma} c^{}_{k,\sigma}+c^\dagger_{k,\sigma} c^{}_{i,\sigma}) (1-2\hat{n}_{j,\sigma})
\label{eq:same_spin_species_overlapping_KE}
\end{equation}
To measure this, we could perform a SWAP gate between sites $i$ and $j$, and then measure the kinetic energy on dimer $j,k$. Unfortunately, a SWAP gate, which directly permutes the sites $i$ and $j$, is difficult to achieve on all particle-number sectors at once without additional phase factors. Instead, we develop a ``\textbf{pSWAP}" gate, which performs the SWAP, along with a phase factor which only depends on the particle-number sector. Concretely, the 1-p, 2-p, and 3-p sectors gain phases of $i$, $1$, and $-i$ respectively. This will suffice for our purposes, and even be useful, since it is simpler to measure the current [i.e.~the eigenstates of $i(c^\dagger_j c^{}_k -c^\dagger_k c^{}_j)$] than the kinetic energy. Having performed the \textbf{pSWAP}, it is simply a matter of measuring the current on $j,k$ (using the \textbf{Hop+U}$^1$ pulse sequence) jointly with the particle density on $i$, and processing the measurements accordingly to obtain the eigenstates of Eq.~\eqref{eq:same_spin_species_overlapping_KE}.

The \textbf{pSWAP} gate performs the mapping:
\begin{align}
    |\sigma, \phi \rangle \mapsto i |\phi,\sigma\rangle,~&~|\phi,\sigma\rangle \mapsto i |\sigma,\phi\rangle, \label{eq:pSWAP_1}\\
    |d,\phi\rangle &\leftrightarrow |\phi,d\rangle,\\
    |\!\uparrow,\downarrow\rangle &\leftrightarrow|\!\downarrow,\uparrow\rangle,\\
    |\sigma,\sigma\rangle &\leftrightarrow|\sigma,\sigma\rangle,\\
    |\sigma, d \rangle \mapsto -i |d,\sigma\rangle,~&~|d,\sigma \rangle \mapsto -i |\sigma,d\rangle, \label{eq:pSWAP_5}
\end{align}
for $\sigma = \uparrow, \downarrow$. The mapping on the 2-p sector is equivalent to
\begin{align}
    |d\phi^+\rangle \mapsto |d\phi^+\rangle,~&~|d\phi^-\rangle \mapsto -|d\phi^-\rangle,\\
    |\text{trip}\rangle \mapsto |\text{trip}\rangle,~&~|\text{sing}\rangle \mapsto -|\text{sing}\rangle,\\
    |\text{Bell}^\pm\rangle &\mapsto |\text{Bell}^\pm\rangle,
\end{align}
a form which will be helpful for pulse design.

The \textbf{pSWAP} gate simply consists two \textbf{Hop+U$^1$} pulses, with idling steps interleaved between them [Fig.~\ref{fig:more_pulse_sequences}(a)]. On the 1-p and 3-p sectors, this has the effect of performing a $\pi-$pulse, with an overall phase factor of $i$. This phase could potentially be removed by the use of complex-valued hopping strengths~\cite{aidelsburger2013realization}, but we will not pursue this route here. The 3-p sector will also gain an additional factor of $-1$ from the interaction term $U$. 

On the 2-p sector, each \textbf{Hop+U$^1$} pulse swaps the states $|\text{sing}\rangle$ and $|d\phi^+\rangle$, and two such pulses return these states to themselves. Interleaving idling times after each \textbf{Hop+U$^1$} pulse, however, allows one to imprint arbitrary phases $\exp(i\xi_1)$ and $\exp(i\xi_2)$ on the $|\text{sing}\rangle$ and $|d\phi^+\rangle$. These pulses have the additional property that the $|d\phi^-\rangle$ state, which does not otherwise participate in the dynamics, gains the phase $\exp(i(\xi_1+\xi_2))$. This allows us to engineer the phases $-1,+1,$ and $-1$ on  $|\text{sing}\rangle, |d\phi^+\rangle$ and $|d\phi^-\rangle$ respectively. The phase $-1$ on $|d\phi^-\rangle$, which arises solely due to idling, is the same additional $-1$ phase on the 3-p sector above, and one can verify that the 0-p and 4-p sectors accumulate phase factors of $+1$.

\subsubsection{Bond-bond correlations with opposite spin}
There are two terms involving the opposite spin species, each term being of the form:
\begin{equation}
h^{(2)}_{ijk,(\sigma\bar{\sigma})} \equiv (c^\dagger_{i,\sigma} c^{}_{j,\sigma}+c^\dagger_{j,\sigma} c^{}_{i,\sigma})(c^\dagger_{j,\bar{\sigma}} c^{}_{k,\bar{\sigma}}+c^\dagger_{k,\bar{\sigma}} c^{}_{j,\bar{\sigma}})
\label{eq:cross_spin_terms}
\end{equation}
where $\bar{\sigma}$ is the opposite spin of $\sigma$. This operator conserves particle number, indeed conserves each spin species separately, and is nontrivial when there are $(1,1), (2,1), (1,2)$, and $(2,2)$ $\uparrow$ and $\downarrow$ particles on sites $i,j$ and $k$. In the simplest $(1,1)$ case, the eigenstates and eigenvalues of Eq.~\eqref{eq:cross_spin_terms} are
\begin{align}
    &|\!\uparrow,\downarrow,\phi\rangle \pm |\phi,\uparrow,\downarrow\rangle_{ijk} \label{eq:overlapping_dimers_mixed_spin_eigst_1}\\
     &|\!\uparrow,\phi,\downarrow\rangle \pm |\phi,d,\phi\rangle_{ijk}~,
     \label{eq:overlapping_dimers_mixed_spin_eigst_2}
\end{align}
The sector $(2,1)$ has eigenstates
\begin{align}
    &|\!\uparrow, \downarrow, \uparrow\rangle \pm |\phi, \uparrow, d\rangle_{ijk}~, \label{eq:overlapping_dimers_mixed_spin_eigst_2_1_sector}\\
    &|\!\uparrow, \phi, d\rangle \pm |\phi, d, \uparrow\rangle_{ijk}~,
\end{align}
and the eigenstates in the $(1,2)$ sector are related by exchanging $\uparrow ~\leftrightarrow ~\downarrow$. Finally, the sector $(2,2)$ has eigenstates
\begin{align}
    &|d, \downarrow, \uparrow\rangle \pm |\!\downarrow, \uparrow, d\rangle_{ijk}~,\\
    &|d, \phi, d\rangle \pm |\!\downarrow, d, \uparrow\rangle_{ijk}~,
\label{eq:overlapping_dimers_mixed_spin_eigst_2_2_sector}
\end{align}
In all cases, it is easy to verify the following property: if one performs a \textbf{SWAP} on dimer $\langle i,j\rangle$ on the eigenstates (\ref{eq:overlapping_dimers_mixed_spin_eigst_1}-\ref{eq:overlapping_dimers_mixed_spin_eigst_2_2_sector}), \textit{only} when there are an odd number (i.e.~1 or 3) of particles, then in all cases site $i$ becomes disentangled from sites $j,k$. That is, the only non-identity mappings of \textbf{odd-SWAP} are:
\begin{align}
     |\sigma, \phi \rangle \mapsto i |\phi,\sigma\rangle,~&~|\phi,\sigma\rangle \mapsto i |\sigma,\phi\rangle, \label{eq:oddSWAP_1}\\
    |\sigma, d \rangle \mapsto i |d,\sigma\rangle,~&~|d,\sigma \rangle \mapsto i |\sigma,d\rangle, \label{eq:oddSWAP_2}
\end{align}

Fortunately, this transformation, which we call the \textbf{odd-SWAP} gate, is a simple modification of the \textbf{pSWAP} gate above [Fig.~\ref{fig:more_pulse_sequences}(a)], where we instead imprint the phase $+1$ on all states $|\text{sing}\rangle, |d\phi^+\rangle$ and $|d\phi^-\rangle$. This has the net effect of performing the SWAP along with a phase factor of $i$ on both the 1-p and 3-p sectors, while acting as the identity, with phase $+1$ on the 0-p, 2-p and 4-p sectors.

As an illustrative example, the odd-SWAP gate applied to Eq.~\eqref{eq:overlapping_dimers_mixed_spin_eigst_1} gives
\begin{equation} |\!\uparrow,\downarrow,\phi\rangle_{ijk} \pm |\phi,\uparrow,\downarrow\rangle_{ijk} \overset{(\textbf{odd-SWAP})_{ij}}{\mapsto} |\!\uparrow\rangle_i \otimes\left(|\!\downarrow,\phi\rangle_{jk} \pm i|\phi,\downarrow\rangle_{jk}\right), 
\end{equation}
while when applied to Eq.~\eqref{eq:overlapping_dimers_mixed_spin_eigst_2} gives 
\begin{equation}
    |\!\uparrow,\phi,\downarrow\rangle_{ijk} \pm |\phi,d,\phi\rangle_{ijk} \overset{(\textbf{odd-SWAP})_{ij}}{\mapsto} |\phi\rangle_i \otimes\left(i|\!\uparrow,\downarrow\rangle \pm |d,\phi\rangle\right)_{jk} 
\end{equation}
It is easy to check that \textbf{odd-SWAP} decouples the $i$ site when applied to all other eigenstates (\ref{eq:overlapping_dimers_mixed_spin_eigst_2_1_sector}-\ref{eq:overlapping_dimers_mixed_spin_eigst_2_2_sector}).
\subsubsection{Pulse sequence to measure spin-resolved current}
\label{app:spin-resolved_current}
 In fact, after the \textbf{odd-SWAP} gate, the remaining states on $\langle j,k\rangle $ have a common property: they are the eigenstates of the spin-resolved current 
 \begin{equation}
 J^{\downarrow}_{jk} = i(c^\dagger_{j,\downarrow} c^{}_{k,\downarrow} -c^\dagger_{k,\downarrow} c^{}_{j,\downarrow})
 \label{eq:spin_resolved_current}
 \end{equation}
 Specifically, $J^{\downarrow}_{jk}$ has eigenstates:
\begin{align} &|\!\downarrow,\phi\rangle\pm i|\phi,\downarrow\rangle_{jk}\\
&|d,\phi\rangle \pm i|\!\uparrow,\downarrow\rangle_{jk}\\
&|\!\downarrow,\uparrow\rangle \pm i |\phi,d\rangle_{jk}\\
&|d,\uparrow\rangle \pm i|\!\uparrow,d\rangle_{jk}
\end{align}
In this section, we develop a pulse sequence to measure the spin-resolved current. In addition to its use for estimating the three-body term $h^{(2)}_{ijk,(\sigma\bar{\sigma})}$, one can also verify that a variant of this pulse can be used to estimate the kinetic-interaction energy cross term $\{(c^\dagger_{i,\sigma}c^{}_{j,\sigma} +c^\dagger_{j,\sigma}c^{}_{i,\sigma}), \hat{n}_{j,\sigma} \hat{n}_{j,\bar{\sigma}} \}$, where $\{\cdot , \cdot\}$ is the anticommutator.

As usual, we seek a pulse on dimer $\langle j,k \rangle$ that maps these states into readout states, in the 1-p, 2-p and 3-p sectors. Unlike the pulse sequences on $\langle i,j\rangle $, since we measure immediately after this pulse, we do not have to keep track of the phases on each particle number sector. We design the pulse sequence with the following steps [Fig.~\ref{fig:more_pulse_sequences}(b)]:
\begin{align}
    &|d,\phi\rangle \pm i|\!\uparrow,\downarrow\rangle = \frac{|d\phi^+\rangle + |d\phi^-\rangle \pm i (|\text{trip}\rangle+|\text{sing}\rangle)}{\sqrt{2}} \overset{\textbf{Hop+U$^{2}$}}{\mapsto}
   e^{i\xi_1}|\text{sing}/d\phi^+\rangle + \frac{|d\phi^-\rangle \pm i |\text{trip}\rangle}{\sqrt{2}}\overset{\text{STO}(\pi)+\text{Tilt}(\pi)}{\mapsto} \label{eq:spin_resolved_current}\\
   &i e^{i\xi_1}|\text{trip}/d\phi^-\rangle + \frac{i |d\phi^+\rangle \mp |\text{sing}\rangle}{\sqrt{2}} \overset{\textbf{Hop+U$^2$}}{\mapsto} ie^{i\xi_1}|\text{trip}/d\phi^-\rangle + e^{i\xi_2}|\text{sing}/d\phi^+\rangle \overset{\textbf{Hop+U$^3$}\text{+Idle+}\textbf{Hop+U$^3$}}{\mapsto} |\!\uparrow,\downarrow\rangle/|d,\phi\rangle \nonumber
\end{align}
The same pulse sequence also maps the other set of 2-p eigenstates $|\phi,d\rangle \pm i |\!\downarrow,\uparrow\rangle  \mapsto |\!\downarrow,\uparrow\rangle/|\phi,d\rangle$. In the 1-p and 3-p sectors, this pulse sequence realizes the maps
\begin{align}
    |\sigma\rangle_\text{sp}\otimes|Y^\pm\rangle_\text{pos} \overset{\textbf{Hop+U$^2$}}{\mapsto}
   |\sigma\rangle_\text{sp}\otimes|Z^\pm\rangle_\text{pos} \overset{\text{STO}(\pi)+\text{Tilt}(\pi)}{\mapsto}e^{i\phi}|\sigma\rangle_\text{sp} &\otimes|Z^\pm\rangle_\text{pos} \overset{\textbf{Hop+U$^2$}}{\mapsto} e^{i\phi}|\sigma\rangle_\text{sp} \otimes|Y^\mp\rangle_\text{pos} \\
   \overset{\textbf{Hop+U$^3$}\text{+Idle+}\textbf{Hop+U$^3$}}{\mapsto} e^{i\phi'}|\sigma\rangle_\text{sp}&\otimes|Z^\pm\rangle_\text{pos} \nonumber
\end{align}
where we have accumulated (spin-dependent) phases which will be irrelevant to the measurement outcomes.

Here, we introduce two new hopping pulse sequences, \textbf{Hop+U$^2$} and \textbf{Hop+U$^3$}. \textbf{Hop+U$^2$} is a pulse sequence designed to perform [Fig.~\ref{fig:more_pulse_sequences}(d)]: (i) a $\pi/2$ rotation about the X-axis in the 1-p and 3-p sectors, and (ii) on the 2-p sector, the maps
\begin{align}
 |d\phi^+\rangle + i |\text{sing} \rangle &\mapsto |d\phi^+\rangle \\
 |d\phi^+\rangle - i |\text{sing} \rangle &\mapsto |\text{sing}\rangle
\end{align}
To design a four-pulse version of this sequence, we follow the analysis of Appendix~\ref{app:four_pulse_sequence}, and simply use a different starting point $\vec{r}_i = (0,1,0)^T$, this modifies $\vec{r}_{+}$ in Eq.~\eqref{eq:v1_m} to
\begin{equation}
    \vec{r}_{+} = \begin{pmatrix}
        -\cos\theta \sin \phi^{(+)}_\text{hop}\\
        \cos \phi^{(+)}_\text{hop}\\
        \sin\theta \sin \phi^{(+)}_\text{hop}
    \end{pmatrix}~,
\end{equation}
where $\cot \theta = u/4$. Matching the $z-$coordinates as in Eq.~\eqref{eq:hopping_times_angles} gives the equation
\begin{equation}
\sin\theta \sin \phi^{(+)}_\text{hop} = -\cos^2\theta + \sin^2 \theta \cos \phi^{(-)}_\text{hop}~.
\end{equation}
Substituting the 1-p constraint $\phi^{(+)}_\text{hop} + \phi^{(-)}_\text{hop} = 2 \sqrt{1+ \frac{u^2}{16}} \frac{\pi}{2}$ [Eq.~\eqref{eq:matching_hopping_times}] allows $\phi^{(+)}_\text{hop} - \phi^{(-)}_\text{hop}$ to be numerically solved for.

\textbf{Hop+U$^3$} is a pulse sequence designed to perform [Fig.~\ref{fig:more_pulse_sequences}(e)]: (i) a $\pi/4$ rotation about the X-axis in the 1-p and 3-p sectors, and (ii) the map $|d\phi^+\rangle \leftrightarrow |\text{sing}\rangle$ on the 2-p sector. We perform two \textbf{Hop+U$^3$} sequences, interleaved by an idling step, to correct for the phase difference $i \exp[i(\xi_1 - \xi_2)]$ accumulated by the previous steps in Eq.~\eqref{eq:spin_resolved_current}. This final step also rotates the 1-p and 3-p states into the readout basis. The parameters of \textbf{Hop+U$^3$} directly follow from the analysis in Appendix~\ref{app:four_pulse_sequence}, with $\frac{m\pi}{2}$ in Eq.~\eqref{eq:matching_hopping_times} replaced by $\frac{\pi}{4}$.

\subsubsection{Same-dimer energy correlations}
For completeness, we also describe a pulse sequence to measure the energy correlations on the \textit{same} dimer, i.e. $h_{ij} h_{ij}$. This introduces a new non-trivial term
\begin{equation}
    h^{(2)}_{ij,(\sigma\bar{\sigma})} \equiv (c^\dagger_{i,\sigma} c^{}_{j,\sigma}+c^\dagger_{j,\sigma} c^{}_{i,\sigma})(c^\dagger_{i,\bar{\sigma}} c^{}_{j,\bar{\sigma}}+c^\dagger_{j,\bar{\sigma}} c^{}_{i,\bar{\sigma}})
\label{eq:same_dimer_cross_spin}
\end{equation}
This operator is non-trivial only in the 2-p sector, and its eigenstates are precisely the control basis states $|d\phi^+ \rangle,|d\phi^- \rangle, |\text{sing}\rangle, |\text{trip}\rangle$.

Our approach to map these states into the readout basis will be to exploit dimer tilt and the singlet-triplet oscillation (STO) to admix $|\text{sing}\rangle/|\text{trip}\rangle$ and $|d\phi^{\pm}\rangle$. This admixing on $|\text{sing}\rangle/|\text{trip}\rangle$ is simply the last four steps of our energy-density pulse sequence [Fig.~\ref{fig:pulse_sequences}(c)], and we simply prepend a STO step to also mix $|d\phi^{\pm}\rangle$. This pulse sequence has the added effect of measuring the 1-p and 3-p sectors in the computational basis. Finally, there are also kinetic-interaction energy cross-terms on the same dimer. Measuring this quantity requires precisely the same pulse sequence as measuring the spin-resolved current, modified to account for the lack of the factor of $i$.

\subsection{Application: converting parity readout to spin- and charge-resolved measurements}
As a final application of the techniques we develop, we demonstrate how our pulse sequences can be used to enhance the readout capabilities of quantum gas microscopes. While in this work we had assumed the capability to perform full spin- and charge-resolved readout, many quantum gas microscopes in fact do not have this capability. Instead, they perform \textit{parity readout}, in which measurements are able to distinguish whether there is one particle on a site (without resolving $\uparrow$ vs. $\downarrow$), or either zero or two particles. In other words, it measures the \textit{parity} (odd or even) of the number of particles on the site. This is due to the presence of light-assisted collisions, in which two particles occupying a lattice site collide and are lost~\cite{parsons2016site}. We note that there are experimental techniques to perform full charge- and spin- measurements by applying a spin-dependent potential which separates $\uparrow$ and $\downarrow$ particles into different imaging planes~\cite{koepsell2020robust,hartke2020doublon}. However, here we analyze the more standard parity measurement scheme and discuss how it can be extended using the pulse sequences developed in this work. 

Our results are as follows: with only parity-readout and doublon-readout, the pulse sequences as described in the rest of this work are not sufficient to measure $d$-wave pairing or other quantities discussed in this work. However, additional pulse sequences can be performed right before parity-readout, which together emulate full spin- and charge-resolution. Even without doublon-readout, which is more challenging than parity-readout, a large class of observables, including all those considered in this work, can be measured just with parity-readout augmented by the pulses sequences developed here.

The challenge we seek to address is that parity measurements do not give full spin- and charge-resolved information. These measurements provide experimental probabilities $P$ of a set of outcomes, and our goal is to infer the probability distribution $\vec{p}$ of every element of the readout basis $p_{\phi,\phi},p_{\uparrow,\phi},p_{\downarrow,\phi},\dots$, where each probability $p_{a,b} \equiv \langle a,b|\rho|a,b\rangle$ is the population of the relevant state. Being able to recover $\vec{p}$ would emulate full spin- and charge-resolved measurements on a state, at least for the purpose of observables on dimers. Parity measurements on a dimer give four outcomes: $(0,0),(0,1),(1,0),(1,1)$, corresponding to the parity on each site. The probability $P^{(p)}$ of each outcome is a sum of the probabilities $p_{ab}$:
\begin{align}
    P^{(p)}(0,0) &= p_{\phi,\phi} + p_{\phi,d} + p_{d,\phi} + p_{d,d}, \label{eq:parity_1}\\
    P^{(p)}(0,1) &= p_{\phi,\uparrow} + p_{\phi,\downarrow} + p_{d,\uparrow} + p_{d,\downarrow},\\
    P^{(p)}(1,0) &= p_{\uparrow,\phi} + p_{\uparrow,d} + p_{\downarrow,\phi}  + p_{\downarrow,d},\\
    P^{(p)}(1,1) &= p_{\uparrow,\uparrow} + p_{\uparrow,\downarrow} + p_{\downarrow,\uparrow} + p_{\downarrow,\downarrow},\label{eq:parity_4}
\end{align}
i.e.~parity readout groups the 16 spin- and charge-resolved readout basis states into four groups, and cannot distinguish between states within a group. Parity measurement can be augmented by selectively removing particles of a certain spin species, as has been demonstrated in Ref.~\cite{parsons2016site}. For example, by first removing the $\downarrow$ spins and then measuring, we instead obtain the following outcome probabilities
\begin{align}
    P^{(\backslash \downarrow)}(0,0) &= p_{\phi,\phi} + p_{\phi,\downarrow} + p_{\downarrow,\phi}  + p_{\downarrow,\downarrow},\label{eq:parity_no_down_1}\\
    P^{(\backslash \downarrow)}(0,1) &= p_{\phi,\uparrow} + p_{\phi,d} + p_{\downarrow,\uparrow}  + p_{\downarrow,d},\\
    P^{(\backslash \downarrow)}(1,0) &= p_{\uparrow,\phi} + p_{\uparrow,\downarrow} + p_{d,\phi} + p_{d,\downarrow},\\
    P^{(\backslash \downarrow)}(1,1) &= p_{\uparrow,\uparrow} + p_{\uparrow,d} + p_{d,\uparrow} + p_{d,d}.\label{eq:parity_no_down_4}
\end{align}
Similar probability relations $P^{(\backslash \uparrow)}$ are obtained when the $\uparrow$ particles are removed before measuring parity. Finally, the doublons can be directly imaged by first removing all single particles $\uparrow, \downarrow$ and then specifically measuring for doublons on a site.
\begin{align}
    P^{(d)}(0,0) &= p_{\phi,\phi} + p_{\phi,\uparrow} + p_{\phi,\downarrow}  + p_{\uparrow,\phi} + p_{\uparrow,\uparrow} + p_{\uparrow,\downarrow} + p_{\downarrow,\phi} + p_{\downarrow,\uparrow} + p_{\downarrow,\downarrow} ,\label{eq:doublon_meas_1} \\
    P^{(d)}(0,1) &= p_{\phi,d} + p_{\uparrow,d}  + p_{\downarrow,d},\\
    P^{(d)}(1,0) &= p_{d,\phi} + p_{d,\uparrow}  + p_{d,\downarrow},\\
    P^{(d)}(1,1) &= p_{d,d}. \label{eq:doublon_meas_4}
\end{align}
This doublon-imaging step is experimentally more challenging. As we shall see, while it is required in principle for full spin- and charge-resolution, in practice it can often be avoided. To summarize, a conventional measurement scheme on a quantum gas microscope is to perform the following measurements:
\begin{enumerate}
    \item Parity-projection:
    \begin{enumerate}
    \item measuring the parity of occupation number on each site.
    \item removing $\downarrow$ particles and measuring site occupation.
    \item removing $\uparrow$ particles and measuring site occupation.
  \end{enumerate}
  \item Removing single $\uparrow$ and $\downarrow$ particles, then imaging the doublons.
\end{enumerate}

We now analyze what quantities can and cannot be estimated with these measurements. Simply by counting, one might expect that these four sets of measurements of $P$ to be sufficient to reconstruct all $p_{ab}$, since we have a system of sixteen linear equations for sixteen unknown parameters. This is not true, not just because there are normalization constraints which reduce this to a system of just 12 independent equations. To see this, one can re-write Eq.~(\ref{eq:parity_1}-\ref{eq:parity_4}) as a matrix equation
\begin{equation}
\begin{pmatrix}
    P^{(p)}(0,0)\\
    P^{(p)}(0,1)\\
    P^{(p)}(1,0)\\
    P^{(p)}(1,1)\\
\end{pmatrix}    = \left(\begin{array}{c c c c c c c c c c c c c c c c c c c c}
    1 & & 0 & 0 & 0 & 0 & & 1 & 0 & 0 & 0 & 0 & 1 & & 0 & 0 & 0 & 0 & & 1\\
    0 & & 0 & 0 & 1 & 1 & & 0 & 0 & 0 & 0 & 0 & 0 & & 0 & 0 & 1 & 1 & & 0\\
    0 & & 1 & 1 & 0 & 0 & & 0 & 0 & 0 & 0 & 0 & 0 & & 1 & 1 & 0 & 0 & & 0\\
    0 & & 0 & 0 & 0 & 0 & & 0 & 1 & 1 & 1 & 1 & 0 & & 0 & 0 & 0 & 0 & & 0\\
\end{array}\right) \vec{p}
\equiv A^{(p)} \vec{p},
\end{equation}
where $\vec{p} \equiv (p_{\phi,\phi},~p_{\uparrow,\phi},p_{\downarrow,\phi},p_{\phi,\uparrow},p_{\phi,\downarrow},~p_{d,\phi},p_{\uparrow,\uparrow},p_{\uparrow,\downarrow},p_{\downarrow,\uparrow},p_{\downarrow,\downarrow},p_{\phi,d},~p_{\uparrow,d},p_{\downarrow,d},p_{d,\uparrow},p_{d,\downarrow},~p_{d,d})^T$ is the vector of spin- and charge- resolved probabilities, and where we have separated the columns by total particle number on the dimer for clarity. We can define similar matrix equations for Eq.~(\ref{eq:parity_no_down_1}-\ref{eq:doublon_meas_4}) with matrices $A^{(\backslash \downarrow)}, A^{(\backslash \uparrow)}, A^{(d)}$ and combine them into a matrix equation
\begin{equation}
    \vec{P} = \left(\begin{array}{c}
       A^{(p)}  \\
       A^{(\backslash \downarrow)}\\
       A^{(\backslash \uparrow)}\\
       A^{(d)}
    \end{array}\right) \vec{p} \equiv A \vec{p},
\end{equation}
where $\vec{P} = (P^{(p)}(0,0),\dots, P^{(d)}(1,1))^T$ represents the measured probabilities. The goal is to infer the probabilities in the standard basis $\vec{p}$ : this is possible if and only if the matrix $A$ has a left-inverse $B$ such that $BA = \mathbb{I}_{16}$, the $16\times 16$ identity matrix. Unfortunately, this is not the case: the matrix $A$ has a five-dimensional null space $\text{ker}(A)$ (i.e.~such that $A\vec{q} = \vec{0}$ when $\vec{q}\in \text{ker}(A)$), which implies that a left-inverse $B$ cannot exist. The standard-basis probabilities $\vec{p}$ can only be estimated modulo addition  in the null space: for any vector $\vec{q} \in \text{ker}(A)$ in the null space and coefficient $\lambda$, the measured probabilities $A\vec{p}$ and $A(\vec{p} + \lambda \vec{q})$ are equal, and therefore $\vec{p}$ and $\vec{p} + \lambda \vec{q}$ are indistinguishable from the measurement outcomes $\vec{P}$. As an example, an element of the null space of $A$ is the vector $\vec{q} = (0,~1,-1,-1,1,~0,0,-1,1,0,0,~0,0,0,0,~0)^T$. 

The null space determines what quantities can and cannot be estimated from the measurement outcomes $\vec{P}$. For any observable $O = \text{diag}(o_{ab})$ which is diagonal in the readout basis, its expectation value $\langle  O \rangle \equiv \text{tr}(O\rho)$ is equal to the inner product of $o_{ab}$ with the probabilities $p_{ab}$:
\begin{equation}
    \langle O  \rangle = \sum_{ab} o_{ab} p_{ab} = \vec{o}\cdot \vec{p}
\end{equation}
$\langle O \rangle$ can be estimated only if the inner products $\vec{o}\cdot \vec{p} = \vec{o}\cdot (\vec{p}+\lambda \vec{q})$ are equal, which is in turn true if and only if $\vec{o}$ is orthogonal to the nullspace $\text{ker}(A)$.

As an example, in the measurement protocol discussed in Appendix~\ref{app:dwave_hopping_pulse}, estimating the $d-$wave pairing correlator corresponds to processing the probabilities $\vec{p}$ by taking its inner product with the vector 
\begin{equation}
\vec{o}_{d\text{-wave}} = (0,~1,-1,-1,1,~0,0,-2,2,0,0,~1,-1,-1,1,~0)^T.    
\end{equation}
One can check that $\vec{o}$ in fact lies in the null space $\text{ker}(A)$ and therefore $d$-wave pairing \textit{cannot} be estimated using just measurements 1.abc and 2. However, as we will see below, by appending additional pulse sequences before parity measurements, we can enhance readout capabilities to estimate $d$-wave pairing, along with all other observables considered in this work.

Before discussing the pulse-sequence-enhanced measurements, we first briefly comment on the ability to fully resolve spin- and charge- on a \textit{single-site} vs.~on two sites. Measurements 1(abc) suffice to resolve the spin and charge densities $(p_\phi, p_\uparrow,p_\downarrow,p_d)$ on a single site. Here, each measurement has two measurement outcomes, $0$ or $1$, and the $A$ matrix is
\begin{equation}
    \begin{pmatrix}
        P^{(p)}(0)\\ 
        P^{(p)}(1)\\ 
        P^{(\backslash\downarrow)}(0)\\ 
        P^{(\backslash\downarrow)}(1)\\ 
        P^{(\backslash\uparrow)}(0)\\ 
        P^{(\backslash\uparrow)}(1)\\ 
    \end{pmatrix}
     = \begin{pmatrix}
        1 & & 0 & 0 && 1\\
        0 & & 1 & 1 && 0\\
        1 & & 0 & 1 && 0\\
        0 & & 1 & 0 && 1\\
        1 & & 1 & 0 && 0\\
        0 & & 0 & 1 && 1\\
    \end{pmatrix}
    \begin{pmatrix}
        p_\phi\\ 
        p_\uparrow\\ 
        p_\downarrow\\ 
        p_d\\
    \end{pmatrix}
\end{equation}
This matrix is invertible, which indicates that measurements 1(abc) suffice to reconstruct all on-site densities. However, the two-site quantities of interest correspond to nearest-neighbor \textit{spin- and charge-correlations} and cannot be inferred just from on-site densities: these require more measurement settings which we discuss below.

We seek to enhance the capabilities of parity-readout using our pulse sequences. Three sets of additional pulse sequences will suffice to extend parity- and doublon-readout into full spin- and charge-resolved readout. Our approach is to use pulse sequences to permute the readout basis states. Each permutation will result in a new $A^{(\cdot)}$ matrix, which when concatenated results in an invertible matrix equation $\vec{P} = A' \vec{p}$. 

Even though our pulse sequences offer a great deal of flexibility, only a few might be useful here. All our pulse sequences are particle-number conserving and equivalent on the 1-p and 3-p sectors. There are 4! total permutations on the 1-/3-p sectors, but the most easily implemented are global spin-flip (microwave) and exchanging sites (SWAP) by hopping. A global spin-flip would not provide any new information, since it merely exchanges the $\backslash\!\uparrow$ and $\backslash\!\downarrow$ measurement outcomes. However, a spin-flip on \textit{half the sites}, with a $\delta B_X$ magnetic field gradient (which can be implemented with a $\delta B_X$ gradient, or a sequence of $\delta B_Z$, $\bar{B}_X$ and $\bar{B}_Z$ pulses) gives additional information, outlined below. Analogously, swapping sites over \textit{all} number sectors would simply exchange the $(0,1)$ and $(1,0)$ outcomes. However, the \textbf{odd-SWAP} gate allows for a SWAP only on the 1- and 3-p sectors, which gives new information when combined with $\backslash\!\uparrow$ and $\backslash\!\downarrow$ measurements. Finally, we can combine these two operations. These three measurements will turn out to extract as much information as possible and additional permutations (for example, arbitrary permutations on the 2-p sector) do not help. In summary, the additional measurements we consider are:
\begin{enumerate}
    \setcounter{enumi}{2}
    \item Exchanging $\uparrow~\leftrightarrow~\downarrow$ on \textit{every other site} with a $\delta B_X$ magnetic field gradient, then
    \begin{enumerate}
    \item removing $\downarrow$ spins and measuring parity.
    \item removing $\uparrow$ spins and measuring parity.
  \end{enumerate}
    \item Applying \textbf{odd-SWAP} on the dimer [with the hopping and idling pulse in Fig.~\ref{fig:more_pulse_sequences}(a)], then
    \begin{enumerate}
    \item removing $\downarrow$ spins and measuring parity.
    \item removing $\uparrow$ spins and measuring parity.
  \end{enumerate}
\item Exchanging $\uparrow~\leftrightarrow~\downarrow$ on every other site and applying \textbf{odd-SWAP}. In this case, it suffices to only
\begin{enumerate}
    \item remove $\downarrow$ spins and measure parity.
  \end{enumerate}
\end{enumerate}
These result in matrices $A^{(\delta B_X, \backslash \downarrow)}, A^{(\delta B_X, \backslash \uparrow)}, A^{(\textbf{odd-SWAP}, \backslash \downarrow)}, A^{(\textbf{odd-SWAP}, \backslash \uparrow)}, A^{(\delta B_X,~\textbf{odd-SWAP}, \backslash \downarrow)}$. For example, measurement 3(a) results in the matrix
\begin{equation}
\vec{P}^{(\delta B_X, \backslash \downarrow)}    = \left(\begin{array}{c c c c c c c c c c c c c c c c c c c c}
    1 & & 0 & 1 & 1 & 0 & & 0 & 0 & 0 & 1 & 0 & 0 & & 0 & 0 & 0 & 0 & & 0\\
    0 & & 1 & 0 & 0 & 0 & & 1 & 1 & 0 & 0 & 0 & 0 & & 0 & 0 & 1 & 0 & & 0\\
    0 & & 0 & 0 & 0 & 1 & & 0 & 0 & 0 & 0 & 1 & 1 & & 0 & 1 & 0 & 0 & & 0\\
    0 & & 0 & 0 & 0 & 0 & & 0 & 0 & 1 & 0 & 0 & 0 & & 1 & 0 & 0 & 1 & & 1\\
\end{array}\right) \vec{p}
\equiv A^{(\delta B_X, \backslash \downarrow)} \vec{p}.
\end{equation}

Each of the five additional steps reduces the dimension of the nullspace by one, and hence the matrix $A'$, in which all $A^{(\cdot)}$ matrices are concatenated, has no null-space. Therefore, measurements 1-5 together suffice to reconstruct all $p_{ab}$, mimicking full spin- and charge-resolved readout. In turn, our universality result implies that this enables the expectation values $\langle O\rangle$ of any number-conserving observable $O$ to be measured with the protocol in this work. 

Finally, we remark on the role of the doublon-occupation measurement (measurement 2). While it is necessary to completely reconstruct $\vec{p}$, in many cases the information it provides is not necessary for the quantity in question: indeed, in all applications considered here it is not required. Without measurement 2 (i.e.~only measurements 1,3,4, and 5), the concatenated matrix $A'$ has a one-dimensional nullspace: the vector 
\begin{equation}
 \vec{q}_\text{no doub} = (-2,~1,1,1,1,~0,0,0,0,0,0,~-1,-1,-1,-1,~2)^T.   
\end{equation}
Any (diagonal) observable $O$ which is orthogonal to $\vec{q}_\text{no doub}$ can be measured without doublon readout, i.e.~by a combination of pulse sequences and parity measurements. This is the case for all observables considered in this work. In addition to the $d$-wave pairing correlator, the kinetic energy~\eqref{eq:kinetic_energy}, energy density~\eqref{eq:energy_density}, and spin-resolved current $J^{\downarrow}$~\eqref{eq:spin_resolved_current} have data-processing vectors
\begin{align}
    \vec{o}_\text{kin} &= -t(0, ~-1,-1,1,1,~2,0,0,0,0,-2,~1,1,-1,-1,~0)^T\\
    \vec{o}_\text{en.} &= (0, ~-t,-t,t,t,~U/8+\lambda,0,U/4,0,0,U/8-\lambda,~U/4+t,U/4+t,U/4-t,U/4-t,~U/2)^T\\
    \vec{o}_{J^{\downarrow}}&= (0, ~0,-1,0,1,~1,0,1,-1,0,-1,~0,1,0,-1,~0)^T 
\end{align}
with $\lambda= \sqrt{4t^2+U^2/64}$. It is easy to verify that all three vectors are orthogonal to $\vec{q}_\text{no doub}$ above, hence can be estimated only by a combination of pulse sequences and parity readout.  

In fact, any of the below conditions are sufficient for an observable to have a data-processing vector that is orthogonal to $\vec{q}_\text{no doub}$:
\begin{enumerate}[label=(\roman *)]
    \item The observable is symmetric under hole-doublon exchange (e.g.~$\vec{o}_{d\text{-wave}}$)
    \item The observable has trace zero within each particle sector (e.g.~$\vec{o}_\text{kin}$ and $\vec{o}_{J^{\downarrow}}$)
    \item The observable has trace within each particle sector which is orthogonal to the vector $(-2,1,0,-1,2)$ [e.g.~$\vec{o}_\text{en.}$ has traces $(0,0,U/2,U,U/2)$]. Note that (i) and (ii) are special cases of this condition, which is both necessary and sufficient for orthogonality to $\vec{q}_\text{no doub}$.
\end{enumerate}
We conclude that a large class of observables can be estimated without the doublon-readout measurement. A simple example of an observable which \textit{does} require doublon-readout measurements is the density of doublons, which has traces $(0,0,2,4,1)$. Finally, we remark that since $\vec{q}_\text{no doub}$ is proportional to the identity on each particle-number sector, any further number-preserving operations will leave the sum $\vec{p}\cdot\vec{q}_\text{no doub} = -2 p_\text{0-p} + p_\text{1-p} - p_\text{3-p} + 2 p_\text{4-p}$ invariant, and hence not add additional information. Therefore, doublon measurement (measurement 2) cannot be replaced by further pulse sequences followed by parity measurement, and we expect doublon measurement to be essential to fully reconstruct $\vec{p}$. 
\end{document}